\documentclass[
    reprint, 
    amsmath,
    amssymb, 
    aps, 
    twoside, 
    superscriptaddress,
]{revtex4-1}

\usepackage[T1]{fontenc} 
\usepackage{physics}
\usepackage{dsfont} 
\usepackage{color,newfloat}
\usepackage{graphicx}
\usepackage{dcolumn}
\usepackage{bm}
\usepackage{bbm}
\usepackage{amsmath,amssymb}
\usepackage{subfiles}

\usepackage[nonumberlist, acronym, toc]{glossaries}
\usepackage[inline]{enumitem}
\usepackage[ampersand]{easylist}

\usepackage{array}  
\usepackage{hhline}
\usepackage{multirow}
\usepackage{afterpage}
\usepackage[normalem]{ulem}
\usepackage{float}
\usepackage{nomencl}
\makenomenclature
\usepackage{etoolbox}

\usepackage[UKenglish]{babel}
\usepackage{variables}

\usepackage{todonotes}
\usepackage{soul}

\usepackage{siunitx}
\usepackage[ruled]{algorithm2e} 
\usepackage{newfloat,algcompatible} 
\usepackage[pdftex,colorlinks=true,bookmarks=false,citecolor=blue,urlcolor=blue]{hyperref}

\def\({\left(}
\def\){\right)}
\def\[{\left[}
\def\]{\right]}

\usepackage{mathtools}
\DeclarePairedDelimiter\ceil{\lceil}{\rceil}

\newcommand{\ttt}[1]{\texttt{#1}}
\newcommand{\italicbold}[1]{\textbf{\emph{#1}}}

\begin{document}

\title{
Learning models of quantum systems from experiments
}

\preprint{APS/123-QED}
\author{Antonio A. Gentile}
\email{These authors contributed equally}
\affiliation{Quantum Engineering Technology Labs, University of Bristol, BS8 1FD, Bristol, UK}
\author{Brian Flynn}
\email{These authors contributed equally}
\affiliation{Quantum Engineering Technology Labs, University of Bristol, BS8 1FD, Bristol, UK}
\affiliation{Quantum Engineering Centre for Doctoral Training, University of Bristol, Bristol BS8 1FD, UK}
\author{Sebastian Knauer}
\email{These authors contributed equally}
\affiliation{Quantum Engineering Technology Labs, University of Bristol, BS8 1FD, Bristol, UK}
\affiliation{Centre of Excellence for Quantum Computation and Communication Technology, School of Physics, University of New South Wales, Sydney, NSW 2052, Australia (current address).}
\author{Nathan Wiebe}
\affiliation{Department of Physics, University of Washington, Seattle, Washington 98105, USA}
\affiliation{Pacific Northwest National Laboratory, Richland, Washington 99354, USA}
\author{Stefano Paesani}
\affiliation{Quantum Engineering Technology Labs, University of Bristol, BS8 1FD, Bristol, UK}
\author{Christopher E. Granade}
\affiliation{Quantum Architectures and Computation Group, Microsoft Research, Redmond, Washington 98052, USA}
\author{John G. Rarity}
\affiliation{Quantum Engineering Technology Labs, University of Bristol, BS8 1FD, Bristol, UK}
\author{Raffaele Santagati}
\email{raffaele.santagati@gmail.com}
\affiliation{Quantum Engineering Technology Labs, University of Bristol, BS8 1FD, Bristol, UK}
\author{Anthony Laing}
\affiliation{Quantum Engineering Technology Labs, University of Bristol, BS8 1FD, Bristol, UK}

\date{\today}
\newacronym{qmd}{QMD}{Quantum Model Development}
\newacronym{qml}{QML}{Quantum Model Learning}
\newacronym{qhl}{QHL}{Quantum Hamiltonian Learning}
\newacronym{bf}{BF}{Bayes Factors}
\newacronym{nv}{NV}{Nitrogen---Vacancy}
\newacronym{hmf}{HMF}{Hamiltonian Model Form}
\newacronym{qmc}{QMC}{Quantum Model Comparison}
\newacronym{qmla}{QMLA}{Quantum Model Learning Agent}
\newacronym{gr}{GR}{Growth Rule}
\newacronym{sr}{SR}{Success}
\newacronym{dag}{DAG}{Directed Acyclic Graph}
\newacronym{cdag}{cDAG}{comparative Directed Acyclic Graph}
\newacronym{sdag}{sDAG}{structural Directed Acyclic Graph}
\newacronym{cle}{CLE}{Classical Likelihood Estimation}
\newacronym{qle}{QLE}{Quantum Likelihood Estimation}
\newacronym{smc}{SMC}{Sequential Monte Carlo}
\newacronym{mha}{MHA}{Metropolis Hastings Algorithm}


\begin{abstract}

An isolated system of interacting quantum particles is described by a Hamiltonian operator.
Hamiltonian models underpin the study and analysis of physical and chemical processes throughout science and industry, so  it is crucial they are faithful to the system they represent.
However, formulating and testing Hamiltonian models of quantum systems from experimental data is difficult because it is impossible to directly observe which interactions the quantum system is subject to. 
Here, we propose and demonstrate an approach to retrieving a Hamiltonian model from experiments, using unsupervised machine learning.
We test our methods experimentally on an electron spin in a nitrogen-vacancy interacting with its spin bath environment, and numerically, finding success rates up to 86\%.
By building agents capable of learning science, which recover meaningful representations, we can gain further insight on the physics of quantum systems.

\end{abstract}

\maketitle


Distilling models of quantum systems from experimental data in an intelligible form, without excessive over-fitting, is a core challenge with far-reaching implications ~\cite{waltz2009automating, nautrup2020operationally, Ghahramani2015probML, kirsch2011inverseprobs}.  
Great strides towards automating the discovery process have been made in learning classical dynamics. 
In particular, optimal parameterisation of models with known structure~\cite{crutchfield1987eq_motion} has been performed, as well as searches for models that exploit nontrivial conservation laws~\cite{Schmidt2009autoconslaws, crutchfield2014theodreams}.
Other methods invoke innovative neural network architectures for modelling physical systems~\cite{Toth2019, Greydanus2019, carrasquilla2017machine, iten2018}.
However, for quantum systems, these methodologies face new challenges, such as the inherent fragility of quantum states, the computational complexity of simulating quantum systems and the need to provide interpretable models for the underlying dynamics. We overcome these challenges here by introducing a new agent-based approach to Bayesian learning that we show can successfully automate model learning for quantum systems.

Reliable characterisation protocols are the lynchpin of our protocol, as well a plethora of other approaches~\cite{Stenberg2016, hills2015algorithm, Carleo2019}. 
Machine learning methods have recently been applied to learn efficient representations for quantum states 
~\cite{Carleo2017, Dunjko_2018}, to encode physical parameters~\cite{iten2018}, and to perform quantum error correction~\cite{Fosel2018, Cong2019,Nautrup2019} and sensing~\cite{Santagati2019, Lumino2018, Aharon2019, Liu2019}.
Supervised learning techniques have also been applied to the automated design of quantum gates~\cite{banchi2016gatenet}, tuning quantum devices~\cite{kalantre2019machine,darulova2019autonomous} and experimental state preparation~\cite{krenn2016automated}.

Several methods for the characterisation of quantum systems with known models have been demonstrated. 
For example, when a parameterised Hamiltonian model $\hat{H}_0(\vec{x})=\ho$ is known in advance, the parameters $\vec{x}_0$ that best describe the observed dynamics can be efficiently learned via \gls{qhl}~\cite{Granade:2012kj, Wiebe:2014qhl, evans2019scalable}. 
However, these protocols cannot be used in cases with unknown Hamiltonian models: the description and modification of \ho \ are left to the user \cite{wang2017qhlexp, hincks2018nvlearn} which is a major hurdle in automating the discovery process.

Here, we present a \gls{qmla} to provide approximate, automated solutions to the inverse problem of inferring a Hamiltonian model from experimental data.
We test the \gls{qmla} in numerical simulations and apply it to the experimental study of the open-system dynamics of an electron spin in a \gls{nv} centre in diamond.  
We show that it is possible to bootstrap \gls{qmla} for larger systems, by investigating the spin-bath dynamics of an electron-spin, finding the number of interacting particles required to explain observations. 

The \gls{qmla} is designed to represent models as a combination of independent terms which map directly to physical interactions. 
Therefore the output of the procedure is easily interpretable, offering insight 
on the system under study, in particular by understanding its dominant interactions and their relative strengths.
We describe our method as a learning agent~\cite{russell2016ai} because it uses its cumulative knowledge to design experiments to perform on the quantum system under study, while also capable of generating new descriptive models to test against the experimental evidence. 
The agency of the protocol, combined with the interpretability of its outputs, provides a powerful application for quantum computers, supporting the characterisation and development of quantum technologies, as well as enhancing understanding of quantum processes.

\begin{figure*}[h!t]
    \centering
    \includegraphics[width=0.89\textwidth]{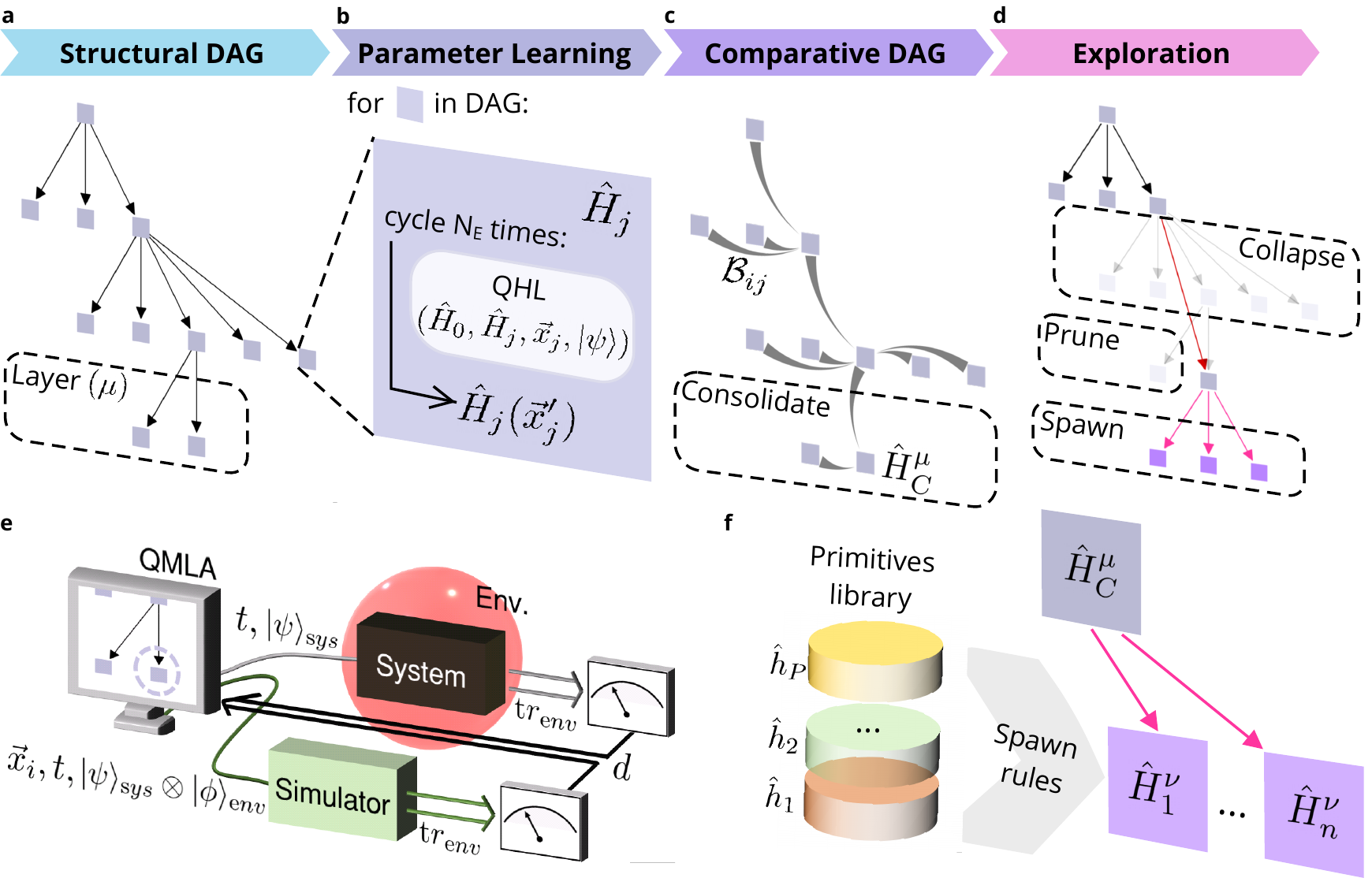}
    \caption{
    \textbf{a-d}, Schematic representation of a single iteration of a QMLA instance. 
    \textbf{a}, Individual models $\hat{H}_j$ are held in nodes of an \gls{sdag}, and models are grouped in layers $\mu$. 
    \textbf{b}, Each model $\hat{H}_j$ within a layer undergoes Quantum Hamiltonian Learning (QHL) to optimise the parameterisation $\vec{x}_j$. 
    \textbf{c}, Within a \gls{cdag}, $\mu$ is consolidated: pairwise Bayes factors \bij are computed for all pairs $\hi,\hj \in\mu$.
    A layer champion \huc \ is chosen.
    \textbf{d}, \huc \ is compared with its parent. This can result in the collapse of the parent layer if \huc \ is superior. 
    Other models in $\mu$ are pruned, and a new layer is spawned from \huc. 
    \textbf{e}, Schematic of QHL used in \textbf{b}. A (quantum) simulator is used as a subroutine of a Bayesian inference protocol to learn the parameters of the considered Hamiltonian model. QHL chooses random probe states $\ket{\psi}_{sys}$ and adaptive evolution times $t$ as in \cite{Wiebe:2014qhl}, which are used both for the time evolution of the system and simulator. 
    This retrieves the optimised parameterisation $\vec{x}_i^{\prime}$.
    \textbf{f}, Spawn rules combine the previous layer champion, \huc, with the primitives library to determine models to constitute the next later $\nu$.
    }
    \label{fig:Fig1}
\end{figure*}

\noindent{\textbf{Quantum Model Learning Agent}}

The overarching idea of the \gls{qmla} is that, to find an approximate model of a system of interest, a series of models are tested against experimental evidence gathered from the system.
This is achieved by training individual models on the data, iteratively constructing new models of increasing complexity, and finally selecting the best model, i.e. of the models tested, the one which best replicates \ho. 
In this section, we describe the steps of the \gls{qmla} algorithm and introduce the concepts involved therein; a glossary of terms used throughout this section is given in Supplementary~\S~\ref{supp_concepts}. 

\gls{qmla}'s model search occurs across a \gls{dag} with two components: structural (sDAG) and comparative (cDAG). 
Each node of the \gls{dag} represents a single candidate model, \hj. 
The \gls{dag} is composed of \emph{layers} $\mu$, each containing a set of similar models, typically of the same Hilbert space dimension and/or number of parameters, Fig.~\ref{fig:Fig1}a. 
For each \hj in $\mu$, QHL is used to find the best parameterisation to approximate the system Hamiltonian \ho, yielding $\hj \rightarrow \hjp$, Fig.~\ref{fig:Fig1}b.
The \gls{qhl} update used for learning parameters is computed using the quantum likelihood estimation protocol presented in~\cite{Wiebe:2014qhl}, and depicted in Fig.~\ref{fig:Fig1}e; additional details in Supplementary~\S~\ref{supp_qhl}.  Classical likelihood evaluation can also be performed if the dimension of the Hilbert space is sufficiently small.

Once all $\hj \in \mu$ are trained, $\mu$ is \emph{consolidated} by systematically comparing models through \gls{bf} analysis, a measure for comparing the predictive power of different models that incorporates a form of Occam's razor to penalise models with more degrees of freedom.
Such comparisons are stored in the \gls{cdag}, Fig.\ref{fig:Fig1}c.
The \gls{bf}, denoted $\mathcal{B}$ penalises models which impose higher structure, naturally limiting over--fitting ~\cite{kass1995bayes}. 
This allows for the selection of a \emph{layer champion} $\huc$, the model within $\mu$ which best reproduces the dynamics of \ho. 
Other models within $\mu$ are then deactivated (or \emph{pruned} from the DAG), and \huc \ is compared with its parent.  If the Bayes factor for \huc is sufficiently greater than that of its parent, our protocol deactivates the parent (and therefore its entire layer) as a cost-saving measure. 
\huc \ is used, along with a set of \emph{primitive} terms, and according to user-defined \emph{spawn rules}, to construct a new set of models, which constitute the next layer of the \gls{dag}, $\nu$, Fig.~\ref{fig:Fig1}f. 
This search over model space is termed \emph{exploration}, Fig.~\ref{fig:Fig1}d.
The procedure iterates until QMLA decides, according to predetermined criteria, that the search should terminate. 
Together, the termination criteria, the spawn rules and the choice of primitive terms are called the \gls{gr}, specified by the user.
The \gls{gr} is flexible, allowing the user to modify \gls{qmla} to learn from any quantum system; the \gls{gr} used in this work is listed in Supplementary~\S~\ref{supp_concepts}.
\par 

A quantum computer/simulator can be used for the efficient learning of each $\hj$ and in the \gls{bf} calculations, (Fig.\ref{fig:Fig1}b,c), as in~\cite{Wiebe:2014qhl, wang2017qhlexp, Santagati2019}.
After termination is confirmed, the set of surviving layer champions $\mathbb{H}_C$ is consolidated to select the global \emph{champion model}, $\hat{H}^{\prime}$. 
This is the model \gls{qmla} determines, based on the experimental evidence observed, to most reliably reproduce the dynamics of \ho.

\noindent{\textbf{Experimental set-up}}

We applied \gls{qmla} to the experimental study of the dynamics of an NV-centre electron spin ($^{14}\text{N}$) in diamond, interacting with the surrounding nuclei of its spin bath, according to an expected Hamiltonian model:
\begin{equation}
 \hat{H}_{\mathrm{full}} = \mu_B \mathbf{g} \cdot \mathbf{B} \cdot \hat{S} + \hat{S} \cdot \mathbf{A} \cdot \hat{I},
 \label{eq:full_Hamiltonian}
\end{equation} 
where $\hat{S}$ and $\hat{I}$ are the total electron and nuclear spin operators respectively,  $\mu_B$ the Bohr magneton and $\mathbf{B}$ the magnetic field, $\mathbf{A}$ and $\mathbf{g}$ the hyperfine and electron gyromagnetic tensors~\cite{Kalb2018}. 
In a simplified picture, we represent the NV centre system in a two-qubit  scheme: one qubit acts as the electron spin, while the entire environment (i.e. open system contribution) is encoded in the second qubit~\cite{Sweke2014}, see Methods for details of this mapping. 
We relate Eqn.~\ref{eq:full_Hamiltonian} to a simulatable model by introducing terms in three forms: Pauli rotation acting on the electron spin $\hat{S}_i = \hat{\sigma}_i$; hyperfine coupling of the spin with the environmental qubit along field axes, $A_{j} = \hat{\sigma}_j\otimes\hat{\sigma}_j$, and non-axial coupling of the spin with the environmental qubit, transverse terms $\hat{T}_{kl}=\hat{\sigma}_k\otimes\hat{\sigma}_l$ . 
We thus yield a Hamiltonian,
\begin{equation}
    \label{formulated_hamiltonian}
    \begin{split}
    \hat{H}_0 \sim \hat{S}_{xyz} + \hat{A}_{xyz} + \hat{T}_{xy, xz, yz}
    \end{split},
\end{equation}
where 
$\hat{S}_{xyz} = \sum_{i\in\{x,y,z\}} \alpha_{i} \hat{S}_i$,
$\hat{A}_{xyz} = \sum_{j\in\{x,y,z\}} \alpha_{j} \hat{A}_j$ 
and 
$\hat{T}_{xy, xz, yz} = \sum_{k,l\in\{x,y,z\}} \alpha_{k,l} \hat{T}_{kl}$. 
In this compact representation, $\alpha$'s are the parameters learned during \gls{qhl}. 
We expect the system under study to be represented by a subset of terms in Eqn.~\ref{formulated_hamiltonian}. 

\begin{figure*}[ht!]
    \centering
    \includegraphics[width=0.95\textwidth]{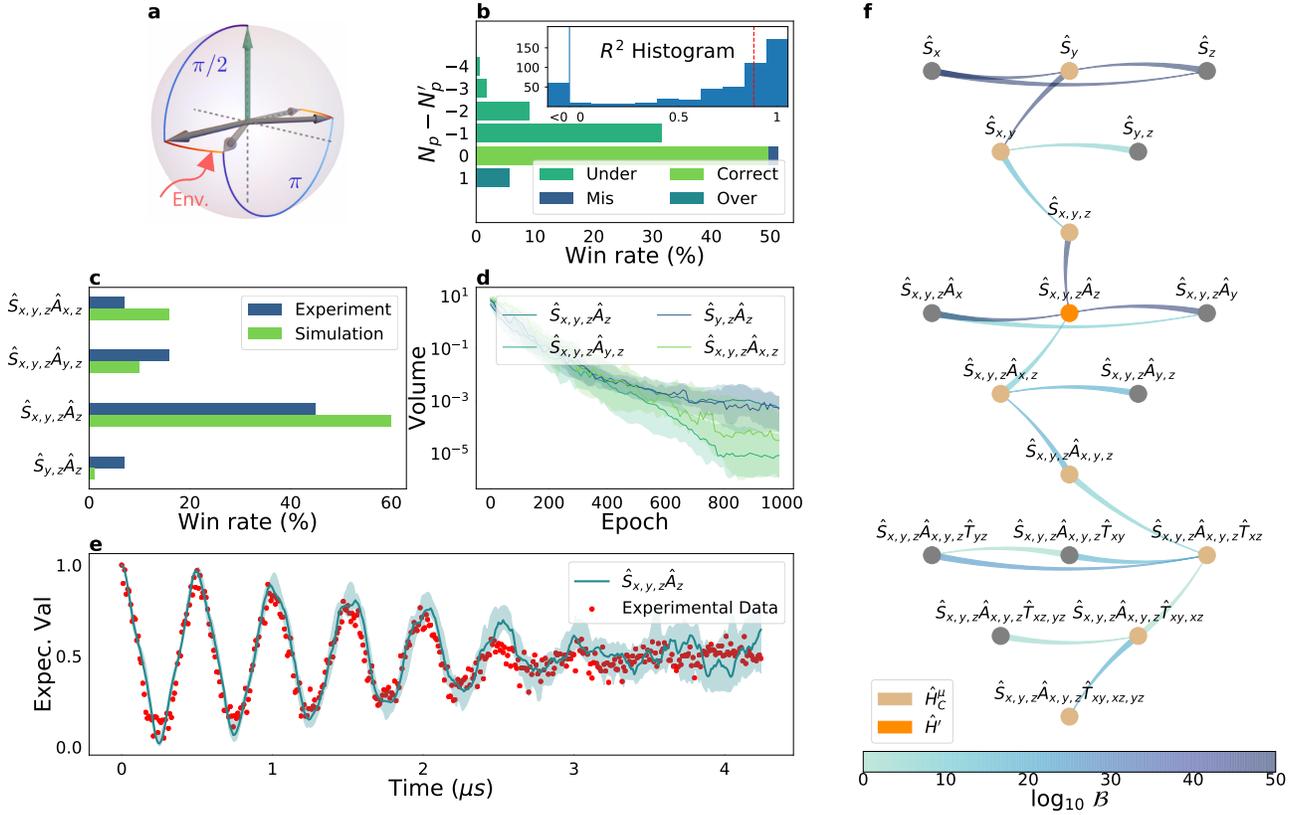}
    \caption{
    \textbf{a}, Time evolution of the electron spin quantum state (represented as a qubit in a Bloch sphere) during the pulses for the Hahn echo sequences. 
    \textbf{b}, Simulation of 500 independent QMLA instances where $\hat{H}_0$ is chosen randomly. We report the win rate as a function of the difference $(N_{p}-N^{\prime}_p)$ between the specific number of parameters in the winning and true models. The models presented are divided in four classes: \textit{Under-parameterised and Over-parameterised} (respectively models with more and less parameters than \ho), \textit{Correct} (true model \ho  found exactly) and \textit{Mis-parameterised} (same parameters cardinality as $\hat{H}_0$, but with a different set of parameters). \textbf{Inset}, Histogram of $R^2$ values, representing predictive power, of the champion models of the 500 QMLA instances, with median $R^2=0.84$ (red dotted line). 
    \textbf{c}, Win rates of credible models (see main text), for 100 instances of QMLA each for experimental and simulated data.  
    Experimentally, the credible models are collectively found in  \experimentalCredibleSuccessRate of instances.
    In simulation for $\ho = \SxyzAz$, an exact success rate of \simulationExactSuccessRate is seen, rising to \simulationCredibleSuccessRate for the same set of credible models. 
    \textbf{d}, Total volume of parameter space for the models in \textbf{c}. Shaded areas indicate $66\%$ credible regions. 
    \textbf{e}, Expectation values, $\abs{\bra{+}e^{-iHt}\ket{+}}^2$, reproduced by the model with the highest win rate, $\hat{S}_{x,y,z}\hat{A}_{z}$ (turquoise), compared with NV-centre system experimental data (red-dots), with median $R^2=0.82$.
    \textbf{f}, A single QMLA instance depicted as a \gls{cdag}, for experimental data. Individual models are represented as nodes with pairwise Bayes Factors ($\mathcal{B}$) connecting nodes. 
    The thin end of the edge point to the stronger model; the colour of the edges reflect $log_{10}\mathcal{B}$, and can be interpreted as the relative strength of the winning model. 
    Layer champions \huc \ (light orange) are determined for each layer, which generates the next layer of models through spawn rules. Consolidation of layer champions determines the global champion, \hp \ (dark orange).
    }
    \label{fig:Fig2}       
\end{figure*}

The experiment design has two main controls: the choice of the evolution time for the system of study, and the initial quantum state of the electron spin, called the \emph{probe state} $\ket{\psi}$~\cite{wiebe2014qhlpra}.
The data we use are obtained from Hahn-echo sequences 
(Fig.~\ref{fig:Fig2}a), collected using the confocal set-up and microwave (MW) pulses that control the dynamics of the electron spin in diamond, at room temperature. 

Echo sequences attempt to decouple the electron spin dynamics from its nuclear bath \cite{charnock2001hf, Rowan1965, Childress2006, Blok2014manip}, making it an ideal study case where QMLA can learn residual contributions from the environment.
For more details on the experimental set-up, the chosen system Hamiltonian and the measurements performed, see Methods and Supplementary~\S~\ref{supp_experimental}.

\noindent{\textbf{Analysis}} 


To assess the \gls{qmla}'s performance we run tests for three distinct cases:
\begin{enumerate}[label=(\roman*)]
    \item simulated NV systems, varying $\hat{H}_0$ and the probe state,
    \item simulated data, mimicking our experiment, with   $\hat{H}_0=\hat{S}_{xyz}\hat{A}_z$ and $\ket{\psi}=\ket{++^{\prime}} = \ket{+} \frac{\ket{0} + e^{i\phi}\ket{1}}{\sqrt{2}}$ (with $\phi$ random),
    \item experimental data obtained with $\ket{\psi}=\ket{++^{\prime}}$ and unknown \ho.
\end{enumerate}

An  assessment of 
\gls{qmla} performance is obtained running (i) for $500$ independent instances.
Each instance chooses $\hat{H}_0$ randomly from a subset of models within the scope of the model search:
$\{ \hat{S}_{xyz}, \ \hat{S}_{xyz}\hat{A}_{z},  \ \hat{S}_{xyz}\hat{A}_{xy}, \ \hat{S}_{xyz}\hat{A}_{xyz}, \ \hat{S}_{xyz}\hat{A}_{xyz}\hat{T}_{xz} \}$. 
In Fig. \ref{fig:Fig2}b we report the cumulative \emph{win rate} as a function of the difference between the cardinality of \ho, i.e. the number of terms in the parameterisation $\vec{x}_0$, $N_p$, and that of the champion model for each instance, $N^{\prime}_p$. 
Defining the \emph{success rate}, $SR$, as the fraction of QMLA instances for which the correct (known) model is deemed champion, we observe $SR = 50 \pm 0.5 \%$.
The winning models show similar performance with respect to predictive power, with median $R^2=0.84$ across all instances (inset, Fig.~\ref{fig:Fig2}b). 

For case (ii), we again simulate \ho, restricting \gls{qmla} by fixing the probe state, to reflect experimental limitations. 
Here we introduce a set of four roughly equivalent models, termed  \emph{credible models}: \credibleModels. 
These models are in good agreement with those successfully used in ab-initio calculations for this kind of physical system~\cite{Gali2008, Gali2009b}, so we assert that \gls{qmla} is successful when these are found.
In simulations, the exact model is found in \simulationExactSuccessRate of instances, while the set of credible models are found in \simulationCredibleSuccessRate, Fig.~\ref{fig:Fig2}c. 

Finally, for case (iii), \gls{qmla} is tested against experimental data, where the underlying model is unknown, and Eq.~\ref{formulated_hamiltonian} is expected to provide a strong approximation to the system. 
In this case, the credible models are found collectively in \experimentalCredibleSuccessRate of cases (Fig.~\ref{fig:Fig2}c). 
In Fig.~\ref{fig:Fig2}d we report the total volume of the parameter space across the epochs of the \gls{qmla} training, for the credible models.
The volume gives an estimate of the width of the posterior distribution at each epoch, with its exponential decrease indicating a successful convergence towards the final parameterisation (further details in  Supplementary~\S~\ref{supp_qhl}).
In Fig. \ref{fig:Fig2}e we provide the normalised photoluminescence from the NV centre during a typical Hahn experiment, against the one predicted by the model selected most often by QMLA, \hp=\SxyzAz. 
Here $H^{\prime}$ reproduces the dynamics of $\ho$ with
median $R^2 = \experimentalSxyzAzRsquared$. 

Fig.~\ref{fig:Fig2}f shows a representative example of the cDAG built during \gls{qmla} for the experimental data. 
Adopting $\{\Sx, \Sy, \Sz\}$ as primitives, QMLA builds progressively up to \SxyzAxyzTxyxzyx.

\noindent{\textbf{Analysis of the bath}} 
\begin{figure}[ht!]
    \centering
    \includegraphics[width=0.45\textwidth]
    {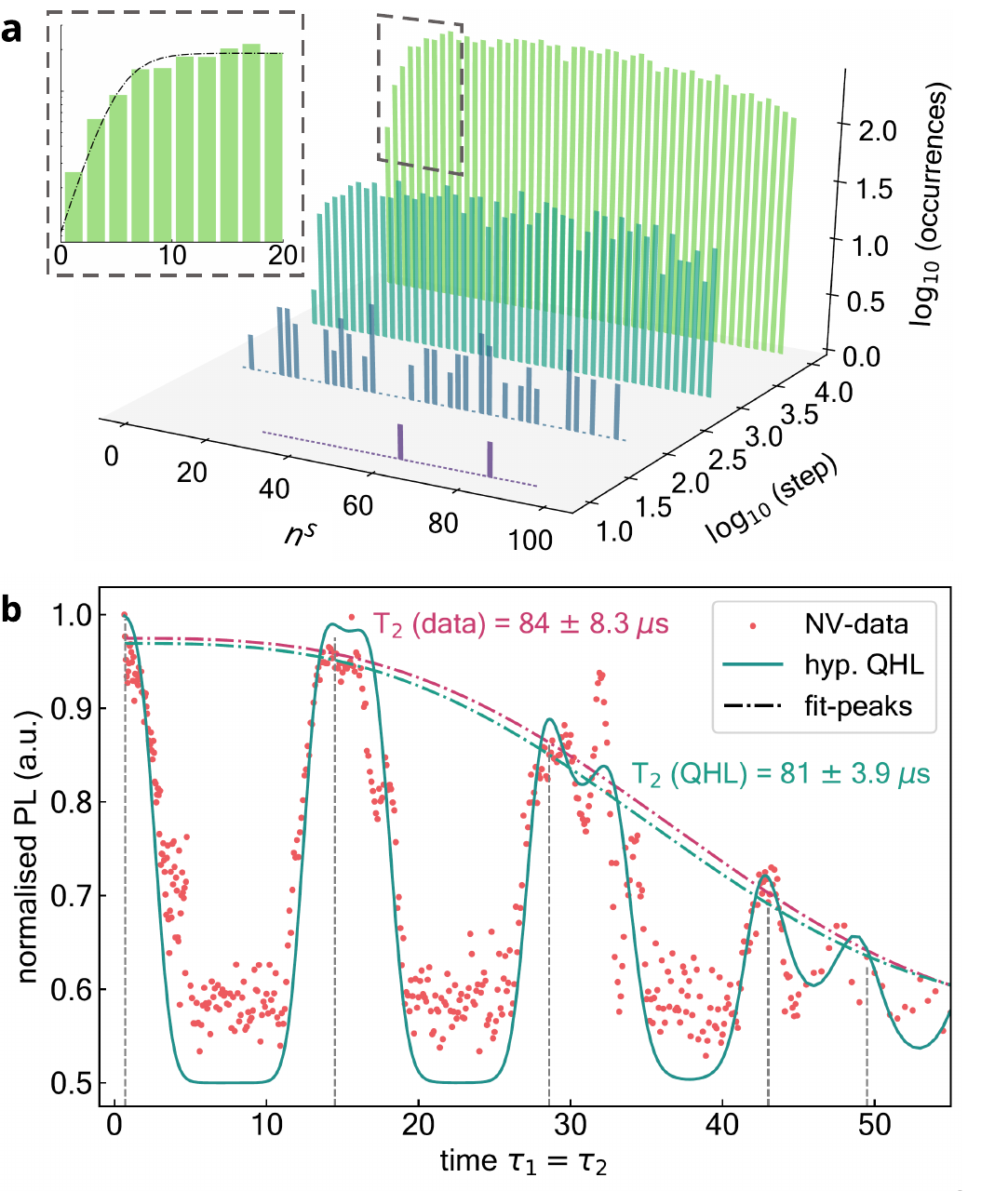}   
    \caption{
    \textbf{a}, Outcome of Metropolis-Hastings algorithm (MHA) sampling from a distribution proportional to the absolute log-likelihood $|\ell(n^{\textrm{s}})|$, with $n^{\textrm{s}}$ the number of environmental spins 
    Exemplary steps out of a single MHA run up to 10000 steps are displayed. 
    \textbf{Inset}, Detail of the distribution behaviour for $n^{\textrm{s}} \leq 20$. A dash-dotted black line indicates the fit of the logistic mock function to the output sample from MHA.  
    \textbf{b}, Normalised photo--luminescence (PL) response for Hahn echo experiments, against evolution time $\tau = \tau'$. Red dots represent experimental data, whereas the green line is the expected PL obtained from the model with $n^{\textrm{s}} = 20$, whose parameters are the average outcome of 100 independent QHL runs. 
    Estimates for the decoherence time $T_2$ obtained from fits on the experimental data, and from QHL, are also reported.
    }
    \label{fig:Fig3}
\end{figure}

The analysis so far has left aside an in-depth study of the spin bath in which the electron spin resides, which has been mapped to a single environmental qubit.
Such an approximation is valid for the rapid dynamics of the fast--decay (such as in Fig.~\ref{fig:Fig2}e), however, the effects of a finite--size bath consisting of $n^{\textrm{s}}$ spins are expected to be evident at the longer times probed by Hahn echo experiments \cite{Childress2006, breuer2002}.

It is infeasible to apply \gls{qmla} to characterise the bath through a complete parameterisation, without a large scale quantum processor, as \gls{qmla} relies on the complete quantum evolution of the global system.  
Instead, preliminary analysis on the system of interest can inform the model space which \gls{qmla} ought to explore. 
Here this is accomplished by using an analytical likelihood function, modelling the interaction of the NV centre with the bath \cite{Rowan1965}, instead of capturing the dynamics of the global system.

Our ad-hoc method obtains a probability distribution $P(n^{\textrm{s}})$ via a \gls{mha}, whereby at each step of the algorithm, an $n^{\textrm{s}}$ is sampled, and a trial form of the analytical likelihood is trained via \gls{cle}. 
The sample acceptance depends upon the value of  $\ell (D|n^{\textrm{s}})$, i.e. the likelihood that a set of observations $D$ is observed with a value $n^{\textrm{s}}$. 
Further details about this procedure are given in Methods.

In Fig. \ref{fig:Fig3}a we show the \gls{mha} outcome after up to 10000 steps. 
In the inset, we emphasise how the approximated distribution starts to plateau for $n^{\textrm{s}} \sim 13$, so that for this system, there is no compelling evidence to consider many additional spins in the bath.
Interestingly, our estimate is well below the number of nuclear sites employed in initial simulations of Hahn-echo experiments with NV-centres~\cite{Childress2006} but agrees in order of magnitude with the number of \textsuperscript{13}C in the \textit{first-shell}, known to be hyperpolarisable~\cite{alvarez2015hyperpol, Hou2019}. 
Finally, we show in Fig.~\ref{fig:Fig3}b the expected normalised photoluminescence (PL) signal, estimated via \gls{cle} from the same hyperparameterised model, with $n^{\textrm{s}} = 20$ 
, together with experimental data. 
Simulated PL accurately reproduces experimental findings, including the revival peak positions, allowing an independent estimate of the decoherence time for this system, from the envelope of the revived PL signals: $T_2 = 81 \pm 3.9$ $\mu \textrm{s}$.
This analysis can be used to bootstrap a QMLA procedure adopting a Hamiltonian formalism, towards models $\hat{H}_j$ of appropriate dimensionality, to describe the system dynamics at longer times.

\noindent{\textbf{Discussion} }

Identifying interaction terms within a Hamiltonian
is fundamental to
understanding the physics of the quantum system it describes.
In the context of quantum technologies,
\gls{qmla} could be used to improve our knowledge of decoherence processes
and design adaptive methodologies to counteract them~\cite{Bouganne2020}.
To efficiently tackle long-time dynamics, more complex interactions, and a deeper study of decoherence, we envisage that future work will extend \gls{qmla} to address open systems, exploiting alternative descriptions, such as Lindblad operators~\cite{Reiter2012}.

The interpretability of \gls{qmla}'s outputs gives users a unique insight into the processes within their system of study. 
This can be helpful in diagnosing imperfect experiments and devices, aiding quantum engineers' efforts towards reliable quantum technologies. 
Moreover, we believe that the potential to deepen our understanding of physics provides a powerful application of noisy intermediate-scale quantum devices.

\clearpage
\bibliography{biblio}

\begin{thebibliography}{71}%
\makeatletter
\providecommand \@ifxundefined [1]{%
 \@ifx{#1\undefined}
}%
\providecommand \@ifnum [1]{%
 \ifnum #1\expandafter \@firstoftwo
 \else \expandafter \@secondoftwo
 \fi
}%
\providecommand \@ifx [1]{%
 \ifx #1\expandafter \@firstoftwo
 \else \expandafter \@secondoftwo
 \fi
}%
\providecommand \natexlab [1]{#1}%
\providecommand \enquote  [1]{``#1''}%
\providecommand \bibnamefont  [1]{#1}%
\providecommand \bibfnamefont [1]{#1}%
\providecommand \citenamefont [1]{#1}%
\providecommand \href@noop [0]{\@secondoftwo}%
\providecommand \href [0]{\begingroup \@sanitize@url \@href}%
\providecommand \@href[1]{\@@startlink{#1}\@@href}%
\providecommand \@@href[1]{\endgroup#1\@@endlink}%
\providecommand \@sanitize@url [0]{\catcode `\\12\catcode `\$12\catcode
  `\&12\catcode `\#12\catcode `\^12\catcode `\_12\catcode `\%12\relax}%
\providecommand \@@startlink[1]{}%
\providecommand \@@endlink[0]{}%
\providecommand \url  [0]{\begingroup\@sanitize@url \@url }%
\providecommand \@url [1]{\endgroup\@href {#1}{\urlprefix }}%
\providecommand \urlprefix  [0]{URL }%
\providecommand \Eprint [0]{\href }%
\providecommand \doibase [0]{http://dx.doi.org/}%
\providecommand \selectlanguage [0]{\@gobble}%
\providecommand \bibinfo  [0]{\@secondoftwo}%
\providecommand \bibfield  [0]{\@secondoftwo}%
\providecommand \translation [1]{[#1]}%
\providecommand \BibitemOpen [0]{}%
\providecommand \bibitemStop [0]{}%
\providecommand \bibitemNoStop [0]{.\EOS\space}%
\providecommand \EOS [0]{\spacefactor3000\relax}%
\providecommand \BibitemShut  [1]{\csname bibitem#1\endcsname}%
\let\auto@bib@innerbib\@empty
\bibitem [{\citenamefont {Waltz}\ and\ \citenamefont
  {Buchanan}(2009)}]{waltz2009automating}%
  \BibitemOpen
  \bibfield  {author} {\bibinfo {author} {\bibfnamefont {D.}~\bibnamefont
  {Waltz}}\ and\ \bibinfo {author} {\bibfnamefont {B.~G.}\ \bibnamefont
  {Buchanan}},\ }\href@noop {} {\bibfield  {journal} {\bibinfo  {journal}
  {Science}\ }\textbf {\bibinfo {volume} {324}},\ \bibinfo {pages} {43}
  (\bibinfo {year} {2009})}\BibitemShut {NoStop}%
\bibitem [{\citenamefont {Nautrup}\ \emph {et~al.}(2020)\citenamefont
  {Nautrup}, \citenamefont {Metger}, \citenamefont {Iten}, \citenamefont
  {Jerbi}, \citenamefont {Trenkwalder}, \citenamefont {Wilming}, \citenamefont
  {Briegel},\ and\ \citenamefont {Renner}}]{nautrup2020operationally}%
  \BibitemOpen
  \bibfield  {author} {\bibinfo {author} {\bibfnamefont {H.~P.}\ \bibnamefont
  {Nautrup}}, \bibinfo {author} {\bibfnamefont {T.}~\bibnamefont {Metger}},
  \bibinfo {author} {\bibfnamefont {R.}~\bibnamefont {Iten}}, \bibinfo {author}
  {\bibfnamefont {S.}~\bibnamefont {Jerbi}}, \bibinfo {author} {\bibfnamefont
  {L.~M.}\ \bibnamefont {Trenkwalder}}, \bibinfo {author} {\bibfnamefont
  {H.}~\bibnamefont {Wilming}}, \bibinfo {author} {\bibfnamefont {H.~J.}\
  \bibnamefont {Briegel}}, \ and\ \bibinfo {author} {\bibfnamefont
  {R.}~\bibnamefont {Renner}},\ }\href@noop {} {\bibfield  {journal} {\bibinfo
  {journal} {arXiv preprint arXiv:2001.00593}\ } (\bibinfo {year}
  {2020})}\BibitemShut {NoStop}%
\bibitem [{\citenamefont {Ghahramani}(2015)}]{Ghahramani2015probML}%
  \BibitemOpen
  \bibfield  {author} {\bibinfo {author} {\bibfnamefont {Z.}~\bibnamefont
  {Ghahramani}},\ }\href {\doibase 10.1038/nature14541} {\bibfield  {journal}
  {\bibinfo  {journal} {Nature}\ }\textbf {\bibinfo {volume} {521}},\ \bibinfo
  {pages} {452} (\bibinfo {year} {2015})}\BibitemShut {NoStop}%
\bibitem [{\citenamefont {Kirsch}(2011)}]{kirsch2011inverseprobs}%
  \BibitemOpen
  \bibfield  {author} {\bibinfo {author} {\bibfnamefont {A.}~\bibnamefont
  {Kirsch}},\ }\href@noop {} {\emph {\bibinfo {title} {An introduction to the
  mathematical theory of inverse problems}}},\ Vol.\ \bibinfo {volume} {120}\
  (\bibinfo  {publisher} {Springer Science \& Business Media},\ \bibinfo {year}
  {2011})\BibitemShut {NoStop}%
\bibitem [{\citenamefont {Crutchfield}\ and\ \citenamefont
  {McNamara}(1987)}]{crutchfield1987eq_motion}%
  \BibitemOpen
  \bibfield  {author} {\bibinfo {author} {\bibfnamefont {J.~P.}\ \bibnamefont
  {Crutchfield}}\ and\ \bibinfo {author} {\bibfnamefont {B.~S.}\ \bibnamefont
  {McNamara}},\ }\href@noop {} {\bibfield  {journal} {\bibinfo  {journal}
  {Complex systems}\ }\textbf {\bibinfo {volume} {1}},\ \bibinfo {pages} {121}
  (\bibinfo {year} {1987})}\BibitemShut {NoStop}%
\bibitem [{\citenamefont {Schmidt}\ and\ \citenamefont
  {Lipson}(2009)}]{Schmidt2009autoconslaws}%
  \BibitemOpen
  \bibfield  {author} {\bibinfo {author} {\bibfnamefont {M.}~\bibnamefont
  {Schmidt}}\ and\ \bibinfo {author} {\bibfnamefont {H.}~\bibnamefont
  {Lipson}},\ }\href {\doibase 10.1126/science.1165893} {\bibfield  {journal}
  {\bibinfo  {journal} {Science}\ }\textbf {\bibinfo {volume} {324}},\ \bibinfo
  {pages} {81} (\bibinfo {year} {2009})}\BibitemShut {NoStop}%
\bibitem [{\citenamefont {Crutchfield}(2014)}]{crutchfield2014theodreams}%
  \BibitemOpen
  \bibfield  {author} {\bibinfo {author} {\bibfnamefont {J.~P.}\ \bibnamefont
  {Crutchfield}},\ }\href@noop {} {\bibfield  {journal} {\bibinfo  {journal}
  {Wiley Interdisciplinary Reviews: Computational Statistics}\ }\textbf
  {\bibinfo {volume} {6}},\ \bibinfo {pages} {75} (\bibinfo {year}
  {2014})}\BibitemShut {NoStop}%
\bibitem [{\citenamefont {Toth}\ \emph {et~al.}(2019)\citenamefont {Toth},
  \citenamefont {Rezende}, \citenamefont {Jaegle}, \citenamefont {Racaniere},
  \citenamefont {Botev},\ and\ \citenamefont {Higgins}}]{Toth2019}%
  \BibitemOpen
  \bibfield  {author} {\bibinfo {author} {\bibfnamefont {P.}~\bibnamefont
  {Toth}}, \bibinfo {author} {\bibfnamefont {D.~J.}\ \bibnamefont {Rezende}},
  \bibinfo {author} {\bibfnamefont {A.}~\bibnamefont {Jaegle}}, \bibinfo
  {author} {\bibfnamefont {S.}~\bibnamefont {Racaniere}}, \bibinfo {author}
  {\bibfnamefont {A.}~\bibnamefont {Botev}}, \ and\ \bibinfo {author}
  {\bibfnamefont {I.}~\bibnamefont {Higgins}},\ }\href@noop {} {\bibfield
  {journal} {\bibinfo  {journal} {arXiv}\ } (\bibinfo {year} {2019})},\ \Eprint
  {http://arxiv.org/abs/1909.13789} {arXiv:1909.13789 [cs.LG]} \BibitemShut
  {NoStop}%
\bibitem [{\citenamefont {Greydanus}\ \emph {et~al.}(2019)\citenamefont
  {Greydanus}, \citenamefont {Dzamba},\ and\ \citenamefont
  {Yosinski}}]{Greydanus2019}%
  \BibitemOpen
  \bibfield  {author} {\bibinfo {author} {\bibfnamefont {S.}~\bibnamefont
  {Greydanus}}, \bibinfo {author} {\bibfnamefont {M.}~\bibnamefont {Dzamba}}, \
  and\ \bibinfo {author} {\bibfnamefont {J.}~\bibnamefont {Yosinski}},\
  }\href@noop {} {\bibfield  {journal} {\bibinfo  {journal} {arXiv}\ }
  (\bibinfo {year} {2019})},\ \Eprint {http://arxiv.org/abs/1906.01563}
  {arXiv:1906.01563 [cs.NE]} \BibitemShut {NoStop}%
\bibitem [{\citenamefont {Carrasquilla}\ and\ \citenamefont
  {Melko}(2017)}]{carrasquilla2017machine}%
  \BibitemOpen
  \bibfield  {author} {\bibinfo {author} {\bibfnamefont {J.}~\bibnamefont
  {Carrasquilla}}\ and\ \bibinfo {author} {\bibfnamefont {R.~G.}\ \bibnamefont
  {Melko}},\ }\href@noop {} {\bibfield  {journal} {\bibinfo  {journal} {Nature
  Physics}\ }\textbf {\bibinfo {volume} {13}},\ \bibinfo {pages} {431}
  (\bibinfo {year} {2017})}\BibitemShut {NoStop}%
\bibitem [{\citenamefont {Iten}\ \emph {et~al.}(2018)\citenamefont {Iten},
  \citenamefont {Metger}, \citenamefont {Wilming}, \citenamefont {del Rio},\
  and\ \citenamefont {Renner}}]{iten2018}%
  \BibitemOpen
  \bibfield  {author} {\bibinfo {author} {\bibfnamefont {R.}~\bibnamefont
  {Iten}}, \bibinfo {author} {\bibfnamefont {T.}~\bibnamefont {Metger}},
  \bibinfo {author} {\bibfnamefont {H.}~\bibnamefont {Wilming}}, \bibinfo
  {author} {\bibfnamefont {L.}~\bibnamefont {del Rio}}, \ and\ \bibinfo
  {author} {\bibfnamefont {R.}~\bibnamefont {Renner}},\ }\href@noop {}
  {\bibfield  {journal} {\bibinfo  {journal} {Phys. Rev. Lett. arXiv preprint
  arXiv:1807.10300}\ } (\bibinfo {year} {2018})}\BibitemShut {NoStop}%
\bibitem [{\citenamefont {Stenberg}\ \emph {et~al.}(2016)\citenamefont
  {Stenberg}, \citenamefont {K\"ohn},\ and\ \citenamefont
  {Wilhelm}}]{Stenberg2016}%
  \BibitemOpen
  \bibfield  {author} {\bibinfo {author} {\bibfnamefont {M.~P.~V.}\
  \bibnamefont {Stenberg}}, \bibinfo {author} {\bibfnamefont {O.}~\bibnamefont
  {K\"ohn}}, \ and\ \bibinfo {author} {\bibfnamefont {F.~K.}\ \bibnamefont
  {Wilhelm}},\ }\href {\doibase 10.1103/PhysRevA.93.012122} {\bibfield
  {journal} {\bibinfo  {journal} {Phys. Rev. A}\ }\textbf {\bibinfo {volume}
  {93}},\ \bibinfo {pages} {012122} (\bibinfo {year} {2016})}\BibitemShut
  {NoStop}%
\bibitem [{\citenamefont {Hills}\ \emph {et~al.}(2015)\citenamefont {Hills},
  \citenamefont {Gr{\"u}tter},\ and\ \citenamefont
  {Hudson}}]{hills2015algorithm}%
  \BibitemOpen
  \bibfield  {author} {\bibinfo {author} {\bibfnamefont {D.~J.}\ \bibnamefont
  {Hills}}, \bibinfo {author} {\bibfnamefont {A.~M.}\ \bibnamefont
  {Gr{\"u}tter}}, \ and\ \bibinfo {author} {\bibfnamefont {J.~J.}\ \bibnamefont
  {Hudson}},\ }\href@noop {} {\bibfield  {journal} {\bibinfo  {journal} {PeerJ
  Computer Science}\ }\textbf {\bibinfo {volume} {1}},\ \bibinfo {pages} {e31}
  (\bibinfo {year} {2015})}\BibitemShut {NoStop}%
\bibitem [{\citenamefont {Carleo}\ \emph {et~al.}(2019)\citenamefont {Carleo},
  \citenamefont {Cirac}, \citenamefont {Cranmer}, \citenamefont {Daudet},
  \citenamefont {Schuld}, \citenamefont {Tishby}, \citenamefont
  {Vogt-Maranto},\ and\ \citenamefont {Zdeborov\'a}}]{Carleo2019}%
  \BibitemOpen
  \bibfield  {author} {\bibinfo {author} {\bibfnamefont {G.}~\bibnamefont
  {Carleo}}, \bibinfo {author} {\bibfnamefont {I.}~\bibnamefont {Cirac}},
  \bibinfo {author} {\bibfnamefont {K.}~\bibnamefont {Cranmer}}, \bibinfo
  {author} {\bibfnamefont {L.}~\bibnamefont {Daudet}}, \bibinfo {author}
  {\bibfnamefont {M.}~\bibnamefont {Schuld}}, \bibinfo {author} {\bibfnamefont
  {N.}~\bibnamefont {Tishby}}, \bibinfo {author} {\bibfnamefont
  {L.}~\bibnamefont {Vogt-Maranto}}, \ and\ \bibinfo {author} {\bibfnamefont
  {L.}~\bibnamefont {Zdeborov\'a}},\ }\href {\doibase
  10.1103/RevModPhys.91.045002} {\bibfield  {journal} {\bibinfo  {journal}
  {Rev. Mod. Phys.}\ }\textbf {\bibinfo {volume} {91}},\ \bibinfo {pages}
  {045002} (\bibinfo {year} {2019})}\BibitemShut {NoStop}%
\bibitem [{\citenamefont {Carleo}\ and\ \citenamefont
  {Troyer}(2017)}]{Carleo2017}%
  \BibitemOpen
  \bibfield  {author} {\bibinfo {author} {\bibfnamefont {G.}~\bibnamefont
  {Carleo}}\ and\ \bibinfo {author} {\bibfnamefont {M.}~\bibnamefont
  {Troyer}},\ }\href {\doibase 10.1126/science.aag2302} {\bibfield  {journal}
  {\bibinfo  {journal} {Science}\ }\textbf {\bibinfo {volume} {355}},\ \bibinfo
  {pages} {602} (\bibinfo {year} {2017})}\BibitemShut {NoStop}%
\bibitem [{\citenamefont {Dunjko}\ and\ \citenamefont
  {Briegel}(2018)}]{Dunjko_2018}%
  \BibitemOpen
  \bibfield  {author} {\bibinfo {author} {\bibfnamefont {V.}~\bibnamefont
  {Dunjko}}\ and\ \bibinfo {author} {\bibfnamefont {H.~J.}\ \bibnamefont
  {Briegel}},\ }\href {\doibase 10.1088/1361-6633/aab406} {\bibfield  {journal}
  {\bibinfo  {journal} {Reports on Progress in Physics}\ }\textbf {\bibinfo
  {volume} {81}},\ \bibinfo {pages} {074001} (\bibinfo {year}
  {2018})}\BibitemShut {NoStop}%
\bibitem [{\citenamefont {F\"osel}\ \emph {et~al.}(2018)\citenamefont
  {F\"osel}, \citenamefont {Tighineanu}, \citenamefont {Weiss},\ and\
  \citenamefont {Marquardt}}]{Fosel2018}%
  \BibitemOpen
  \bibfield  {author} {\bibinfo {author} {\bibfnamefont {T.}~\bibnamefont
  {F\"osel}}, \bibinfo {author} {\bibfnamefont {P.}~\bibnamefont {Tighineanu}},
  \bibinfo {author} {\bibfnamefont {T.}~\bibnamefont {Weiss}}, \ and\ \bibinfo
  {author} {\bibfnamefont {F.}~\bibnamefont {Marquardt}},\ }\href {\doibase
  10.1103/PhysRevX.8.031084} {\bibfield  {journal} {\bibinfo  {journal} {Phys.
  Rev. X}\ }\textbf {\bibinfo {volume} {8}},\ \bibinfo {pages} {031084}
  (\bibinfo {year} {2018})}\BibitemShut {NoStop}%
\bibitem [{\citenamefont {Cong}\ \emph {et~al.}(2019)\citenamefont {Cong},
  \citenamefont {Choi},\ and\ \citenamefont {Lukin}}]{Cong2019}%
  \BibitemOpen
  \bibfield  {author} {\bibinfo {author} {\bibfnamefont {I.}~\bibnamefont
  {Cong}}, \bibinfo {author} {\bibfnamefont {S.}~\bibnamefont {Choi}}, \ and\
  \bibinfo {author} {\bibfnamefont {M.~D.}\ \bibnamefont {Lukin}},\ }\href
  {\doibase 10.1038/s41567-019-0648-8} {\bibfield  {journal} {\bibinfo
  {journal} {Nature Physics}\ }\textbf {\bibinfo {volume} {15}},\ \bibinfo
  {pages} {1273} (\bibinfo {year} {2019})}\BibitemShut {NoStop}%
\bibitem [{\citenamefont {Poulsen~Nautrup}\ \emph {et~al.}(2019)\citenamefont
  {Poulsen~Nautrup}, \citenamefont {Delfosse}, \citenamefont {Dunjko},
  \citenamefont {Briegel},\ and\ \citenamefont {Friis}}]{Nautrup2019}%
  \BibitemOpen
  \bibfield  {author} {\bibinfo {author} {\bibfnamefont {H.}~\bibnamefont
  {Poulsen~Nautrup}}, \bibinfo {author} {\bibfnamefont {N.}~\bibnamefont
  {Delfosse}}, \bibinfo {author} {\bibfnamefont {V.}~\bibnamefont {Dunjko}},
  \bibinfo {author} {\bibfnamefont {H.~J.}\ \bibnamefont {Briegel}}, \ and\
  \bibinfo {author} {\bibfnamefont {N.}~\bibnamefont {Friis}},\ }\href
  {\doibase 10.22331/q-2019-12-16-215} {\bibfield  {journal} {\bibinfo
  {journal} {Quantum}\ }\textbf {\bibinfo {volume} {3}},\ \bibinfo {pages}
  {215} (\bibinfo {year} {2019})}\BibitemShut {NoStop}%
\bibitem [{\citenamefont {Santagati}\ \emph {et~al.}(2019)\citenamefont
  {Santagati}, \citenamefont {Gentile}, \citenamefont {Knauer}, \citenamefont
  {Schmitt}, \citenamefont {Paesani}, \citenamefont {Granade}, \citenamefont
  {Wiebe}, \citenamefont {Osterkamp}, \citenamefont {McGuinness}, \citenamefont
  {Wang}, \citenamefont {Thompson}, \citenamefont {Rarity}, \citenamefont
  {Jelezko},\ and\ \citenamefont {Laing}}]{Santagati2019}%
  \BibitemOpen
  \bibfield  {author} {\bibinfo {author} {\bibfnamefont {R.}~\bibnamefont
  {Santagati}}, \bibinfo {author} {\bibfnamefont {A.~A.}\ \bibnamefont
  {Gentile}}, \bibinfo {author} {\bibfnamefont {S.}~\bibnamefont {Knauer}},
  \bibinfo {author} {\bibfnamefont {S.}~\bibnamefont {Schmitt}}, \bibinfo
  {author} {\bibfnamefont {S.}~\bibnamefont {Paesani}}, \bibinfo {author}
  {\bibfnamefont {C.}~\bibnamefont {Granade}}, \bibinfo {author} {\bibfnamefont
  {N.}~\bibnamefont {Wiebe}}, \bibinfo {author} {\bibfnamefont
  {C.}~\bibnamefont {Osterkamp}}, \bibinfo {author} {\bibfnamefont {L.~P.}\
  \bibnamefont {McGuinness}}, \bibinfo {author} {\bibfnamefont
  {J.}~\bibnamefont {Wang}}, \bibinfo {author} {\bibfnamefont {M.~G.}\
  \bibnamefont {Thompson}}, \bibinfo {author} {\bibfnamefont {J.~G.}\
  \bibnamefont {Rarity}}, \bibinfo {author} {\bibfnamefont {F.}~\bibnamefont
  {Jelezko}}, \ and\ \bibinfo {author} {\bibfnamefont {A.}~\bibnamefont
  {Laing}},\ }\href {\doibase 10.1103/PhysRevX.9.021019} {\bibfield  {journal}
  {\bibinfo  {journal} {Phys. Rev. X}\ }\textbf {\bibinfo {volume} {9}},\
  \bibinfo {pages} {021019} (\bibinfo {year} {2019})}\BibitemShut {NoStop}%
\bibitem [{\citenamefont {Lumino}\ \emph {et~al.}(2018)\citenamefont {Lumino},
  \citenamefont {Polino}, \citenamefont {Rab}, \citenamefont {Milani},
  \citenamefont {Spagnolo}, \citenamefont {Wiebe},\ and\ \citenamefont
  {Sciarrino}}]{Lumino2018}%
  \BibitemOpen
  \bibfield  {author} {\bibinfo {author} {\bibfnamefont {A.}~\bibnamefont
  {Lumino}}, \bibinfo {author} {\bibfnamefont {E.}~\bibnamefont {Polino}},
  \bibinfo {author} {\bibfnamefont {A.~S.}\ \bibnamefont {Rab}}, \bibinfo
  {author} {\bibfnamefont {G.}~\bibnamefont {Milani}}, \bibinfo {author}
  {\bibfnamefont {N.}~\bibnamefont {Spagnolo}}, \bibinfo {author}
  {\bibfnamefont {N.}~\bibnamefont {Wiebe}}, \ and\ \bibinfo {author}
  {\bibfnamefont {F.}~\bibnamefont {Sciarrino}},\ }\href {\doibase
  10.1103/PhysRevApplied.10.044033} {\bibfield  {journal} {\bibinfo  {journal}
  {Phys. Rev. Applied}\ }\textbf {\bibinfo {volume} {10}},\ \bibinfo {pages}
  {044033} (\bibinfo {year} {2018})}\BibitemShut {NoStop}%
\bibitem [{\citenamefont {Aharon}\ \emph {et~al.}(2019)\citenamefont {Aharon},
  \citenamefont {Rotem}, \citenamefont {McGuinness}, \citenamefont {Jelezko},
  \citenamefont {Retzker},\ and\ \citenamefont {Ringel}}]{Aharon2019}%
  \BibitemOpen
  \bibfield  {author} {\bibinfo {author} {\bibfnamefont {N.}~\bibnamefont
  {Aharon}}, \bibinfo {author} {\bibfnamefont {A.}~\bibnamefont {Rotem}},
  \bibinfo {author} {\bibfnamefont {L.~P.}\ \bibnamefont {McGuinness}},
  \bibinfo {author} {\bibfnamefont {F.}~\bibnamefont {Jelezko}}, \bibinfo
  {author} {\bibfnamefont {A.}~\bibnamefont {Retzker}}, \ and\ \bibinfo
  {author} {\bibfnamefont {Z.}~\bibnamefont {Ringel}},\ }\href {\doibase
  10.1038/s41598-019-54119-9} {\bibfield  {journal} {\bibinfo  {journal}
  {Scientific Reports}\ }\textbf {\bibinfo {volume} {9}},\ \bibinfo {pages}
  {17802} (\bibinfo {year} {2019})}\BibitemShut {NoStop}%
\bibitem [{\citenamefont {Liu}\ \emph {et~al.}(2019)\citenamefont {Liu},
  \citenamefont {Chen}, \citenamefont {Liu}, \citenamefont {Layden},\ and\
  \citenamefont {Cappellaro}}]{Liu2019}%
  \BibitemOpen
  \bibfield  {author} {\bibinfo {author} {\bibfnamefont {G.}~\bibnamefont
  {Liu}}, \bibinfo {author} {\bibfnamefont {M.}~\bibnamefont {Chen}}, \bibinfo
  {author} {\bibfnamefont {Y.~X.}\ \bibnamefont {Liu}}, \bibinfo {author}
  {\bibfnamefont {D.}~\bibnamefont {Layden}}, \ and\ \bibinfo {author}
  {\bibfnamefont {P.}~\bibnamefont {Cappellaro}},\ }\href
  {https://arxiv.org/abs/1907.11947} {\bibfield  {journal} {\bibinfo  {journal}
  {arXiv}\ }\textbf {\bibinfo {volume} {abs/1907.11947}} (\bibinfo {year}
  {2019})},\ \Eprint {http://arxiv.org/abs/1907.11947} {arXiv:1907.11947}
  \BibitemShut {NoStop}%
\bibitem [{\citenamefont {Banchi}\ \emph {et~al.}(2016)\citenamefont {Banchi},
  \citenamefont {Pancotti},\ and\ \citenamefont {Bose}}]{banchi2016gatenet}%
  \BibitemOpen
  \bibfield  {author} {\bibinfo {author} {\bibfnamefont {L.}~\bibnamefont
  {Banchi}}, \bibinfo {author} {\bibfnamefont {N.}~\bibnamefont {Pancotti}}, \
  and\ \bibinfo {author} {\bibfnamefont {S.}~\bibnamefont {Bose}},\ }\href@noop
  {} {\bibfield  {journal} {\bibinfo  {journal} {npj Quantum Information}\
  }\textbf {\bibinfo {volume} {2}},\ \bibinfo {pages} {16019} (\bibinfo {year}
  {2016})}\BibitemShut {NoStop}%
\bibitem [{\citenamefont {Kalantre}\ \emph {et~al.}(2019)\citenamefont
  {Kalantre}, \citenamefont {Zwolak}, \citenamefont {Ragole}, \citenamefont
  {Wu}, \citenamefont {Zimmerman}, \citenamefont {Stewart},\ and\ \citenamefont
  {Taylor}}]{kalantre2019machine}%
  \BibitemOpen
  \bibfield  {author} {\bibinfo {author} {\bibfnamefont {S.~S.}\ \bibnamefont
  {Kalantre}}, \bibinfo {author} {\bibfnamefont {J.~P.}\ \bibnamefont
  {Zwolak}}, \bibinfo {author} {\bibfnamefont {S.}~\bibnamefont {Ragole}},
  \bibinfo {author} {\bibfnamefont {X.}~\bibnamefont {Wu}}, \bibinfo {author}
  {\bibfnamefont {N.~M.}\ \bibnamefont {Zimmerman}}, \bibinfo {author}
  {\bibfnamefont {M.~D.}\ \bibnamefont {Stewart}}, \ and\ \bibinfo {author}
  {\bibfnamefont {J.~M.}\ \bibnamefont {Taylor}},\ }\href@noop {} {\bibfield
  {journal} {\bibinfo  {journal} {npj Quantum Information}\ }\textbf {\bibinfo
  {volume} {5}},\ \bibinfo {pages} {1} (\bibinfo {year} {2019})}\BibitemShut
  {NoStop}%
\bibitem [{\citenamefont {Darulov{\'a}}\ \emph {et~al.}(2019)\citenamefont
  {Darulov{\'a}}, \citenamefont {Pauka}, \citenamefont {Wiebe}, \citenamefont
  {Chan}, \citenamefont {Cassidy},\ and\ \citenamefont
  {Troyer}}]{darulova2019autonomous}%
  \BibitemOpen
  \bibfield  {author} {\bibinfo {author} {\bibfnamefont {J.}~\bibnamefont
  {Darulov{\'a}}}, \bibinfo {author} {\bibfnamefont {S.}~\bibnamefont {Pauka}},
  \bibinfo {author} {\bibfnamefont {N.}~\bibnamefont {Wiebe}}, \bibinfo
  {author} {\bibfnamefont {K.}~\bibnamefont {Chan}}, \bibinfo {author}
  {\bibfnamefont {M.}~\bibnamefont {Cassidy}}, \ and\ \bibinfo {author}
  {\bibfnamefont {M.}~\bibnamefont {Troyer}},\ }\href@noop {} {\bibfield
  {journal} {\bibinfo  {journal} {arXiv preprint arXiv:1911.10709}\ } (\bibinfo
  {year} {2019})}\BibitemShut {NoStop}%
\bibitem [{\citenamefont {Krenn}\ \emph {et~al.}(2016)\citenamefont {Krenn},
  \citenamefont {Malik}, \citenamefont {Fickler}, \citenamefont {Lapkiewicz},\
  and\ \citenamefont {Zeilinger}}]{krenn2016automated}%
  \BibitemOpen
  \bibfield  {author} {\bibinfo {author} {\bibfnamefont {M.}~\bibnamefont
  {Krenn}}, \bibinfo {author} {\bibfnamefont {M.}~\bibnamefont {Malik}},
  \bibinfo {author} {\bibfnamefont {R.}~\bibnamefont {Fickler}}, \bibinfo
  {author} {\bibfnamefont {R.}~\bibnamefont {Lapkiewicz}}, \ and\ \bibinfo
  {author} {\bibfnamefont {A.}~\bibnamefont {Zeilinger}},\ }\href {\doibase
  10.1103/PhysRevLett.116.090405} {\bibfield  {journal} {\bibinfo  {journal}
  {Physical review letters}\ }\textbf {\bibinfo {volume} {116}},\ \bibinfo
  {pages} {090405} (\bibinfo {year} {2016})}\BibitemShut {NoStop}%
\bibitem [{\citenamefont {Granade}\ \emph {et~al.}(2012)\citenamefont
  {Granade}, \citenamefont {Ferrie}, \citenamefont {Wiebe},\ and\ \citenamefont
  {Cory}}]{Granade:2012kj}%
  \BibitemOpen
  \bibfield  {author} {\bibinfo {author} {\bibfnamefont {C.~E.}\ \bibnamefont
  {Granade}}, \bibinfo {author} {\bibfnamefont {C.}~\bibnamefont {Ferrie}},
  \bibinfo {author} {\bibfnamefont {N.}~\bibnamefont {Wiebe}}, \ and\ \bibinfo
  {author} {\bibfnamefont {D.~G.}\ \bibnamefont {Cory}},\ }\href {\doibase
  10.1088/1367-2630/14/10/103013} {\bibfield  {journal} {\bibinfo  {journal}
  {New Journal of Physics}\ }\textbf {\bibinfo {volume} {14}},\ \bibinfo
  {pages} {103013} (\bibinfo {year} {2012})}\BibitemShut {NoStop}%
\bibitem [{\citenamefont {Wiebe}\ \emph
  {et~al.}(2014{\natexlab{a}})\citenamefont {Wiebe}, \citenamefont {Granade},
  \citenamefont {Ferrie},\ and\ \citenamefont {Cory}}]{Wiebe:2014qhl}%
  \BibitemOpen
  \bibfield  {author} {\bibinfo {author} {\bibfnamefont {N.}~\bibnamefont
  {Wiebe}}, \bibinfo {author} {\bibfnamefont {C.}~\bibnamefont {Granade}},
  \bibinfo {author} {\bibfnamefont {C.}~\bibnamefont {Ferrie}}, \ and\ \bibinfo
  {author} {\bibfnamefont {D.~G.}\ \bibnamefont {Cory}},\ }\href {\doibase
  10.1103/PhysRevLett.112.190501} {\bibfield  {journal} {\bibinfo  {journal}
  {Physical Review Letters}\ }\textbf {\bibinfo {volume} {112}},\ \bibinfo
  {pages} {190501} (\bibinfo {year} {2014}{\natexlab{a}})}\BibitemShut
  {NoStop}%
\bibitem [{\citenamefont {Evans}\ \emph {et~al.}(2019)\citenamefont {Evans},
  \citenamefont {Harper},\ and\ \citenamefont {Flammina}}]{evans2019scalable}%
  \BibitemOpen
  \bibfield  {author} {\bibinfo {author} {\bibfnamefont {T.}~\bibnamefont
  {Evans}}, \bibinfo {author} {\bibfnamefont {R.}~\bibnamefont {Harper}}, \
  and\ \bibinfo {author} {\bibfnamefont {S.}~\bibnamefont {Flammina}},\ }\href
  {http://arxiv.org/abs/1912.07636} {\bibfield  {journal} {\bibinfo  {journal}
  {arXiv}\ }\textbf {\bibinfo {volume} {abs/1912.07636}} (\bibinfo {year}
  {2019})},\ \Eprint {http://arxiv.org/abs/1912.07636} {arXiv:1912.07636}
  \BibitemShut {NoStop}%
\bibitem [{\citenamefont {Wang}\ \emph {et~al.}(2017)\citenamefont {Wang},
  \citenamefont {Paesani}, \citenamefont {Santagati}, \citenamefont {Knauer},
  \citenamefont {Gentile}, \citenamefont {Wiebe}, \citenamefont {Petruzzella},
  \citenamefont {O'Brien}, \citenamefont {Rarity}, \citenamefont {Laing} \emph
  {et~al.}}]{wang2017qhlexp}%
  \BibitemOpen
  \bibfield  {author} {\bibinfo {author} {\bibfnamefont {J.}~\bibnamefont
  {Wang}}, \bibinfo {author} {\bibfnamefont {S.}~\bibnamefont {Paesani}},
  \bibinfo {author} {\bibfnamefont {R.}~\bibnamefont {Santagati}}, \bibinfo
  {author} {\bibfnamefont {S.}~\bibnamefont {Knauer}}, \bibinfo {author}
  {\bibfnamefont {A.~A.}\ \bibnamefont {Gentile}}, \bibinfo {author}
  {\bibfnamefont {N.}~\bibnamefont {Wiebe}}, \bibinfo {author} {\bibfnamefont
  {M.}~\bibnamefont {Petruzzella}}, \bibinfo {author} {\bibfnamefont {J.~L.}\
  \bibnamefont {O'Brien}}, \bibinfo {author} {\bibfnamefont {J.~G.}\
  \bibnamefont {Rarity}}, \bibinfo {author} {\bibfnamefont {A.}~\bibnamefont
  {Laing}},  \emph {et~al.},\ }\href@noop {} {\bibfield  {journal} {\bibinfo
  {journal} {Nature Physics}\ }\textbf {\bibinfo {volume} {13}},\ \bibinfo
  {pages} {551} (\bibinfo {year} {2017})}\BibitemShut {NoStop}%
\bibitem [{\citenamefont {Hincks}\ \emph {et~al.}(2018)\citenamefont {Hincks},
  \citenamefont {Granade},\ and\ \citenamefont {Cory}}]{hincks2018nvlearn}%
  \BibitemOpen
  \bibfield  {author} {\bibinfo {author} {\bibfnamefont {I.}~\bibnamefont
  {Hincks}}, \bibinfo {author} {\bibfnamefont {C.}~\bibnamefont {Granade}}, \
  and\ \bibinfo {author} {\bibfnamefont {D.~G.}\ \bibnamefont {Cory}},\
  }\href@noop {} {\bibfield  {journal} {\bibinfo  {journal} {New Journal of
  Physics}\ }\textbf {\bibinfo {volume} {20}},\ \bibinfo {pages} {013022}
  (\bibinfo {year} {2018})}\BibitemShut {NoStop}%
\bibitem [{\citenamefont {Russell}\ and\ \citenamefont
  {Norvig}(2016)}]{russell2016ai}%
  \BibitemOpen
  \bibfield  {author} {\bibinfo {author} {\bibfnamefont {S.~J.}\ \bibnamefont
  {Russell}}\ and\ \bibinfo {author} {\bibfnamefont {P.}~\bibnamefont
  {Norvig}},\ }\href@noop {} {\emph {\bibinfo {title} {Artificial intelligence:
  a modern approach}}}\ (\bibinfo  {publisher} {Pearson Education Limited,},\
  \bibinfo {year} {2016})\BibitemShut {NoStop}%
\bibitem [{\citenamefont {Kass}\ and\ \citenamefont
  {Raftery}(1995)}]{kass1995bayes}%
  \BibitemOpen
  \bibfield  {author} {\bibinfo {author} {\bibfnamefont {R.~E.}\ \bibnamefont
  {Kass}}\ and\ \bibinfo {author} {\bibfnamefont {A.~E.}\ \bibnamefont
  {Raftery}},\ }\href@noop {} {\bibfield  {journal} {\bibinfo  {journal}
  {Journal of the american statistical association}\ }\textbf {\bibinfo
  {volume} {90}},\ \bibinfo {pages} {773} (\bibinfo {year} {1995})}\BibitemShut
  {NoStop}%
\bibitem [{\citenamefont {Kalb}\ \emph {et~al.}(2018)\citenamefont {Kalb},
  \citenamefont {Humphreys}, \citenamefont {Slim},\ and\ \citenamefont
  {Hanson}}]{Kalb2018}%
  \BibitemOpen
  \bibfield  {author} {\bibinfo {author} {\bibfnamefont {N.}~\bibnamefont
  {Kalb}}, \bibinfo {author} {\bibfnamefont {P.~C.}\ \bibnamefont {Humphreys}},
  \bibinfo {author} {\bibfnamefont {J.~J.}\ \bibnamefont {Slim}}, \ and\
  \bibinfo {author} {\bibfnamefont {R.}~\bibnamefont {Hanson}},\ }\href
  {\doibase 10.1103/PhysRevA.97.062330} {\bibfield  {journal} {\bibinfo
  {journal} {Phys. Rev. A}\ }\textbf {\bibinfo {volume} {97}},\ \bibinfo
  {pages} {062330} (\bibinfo {year} {2018})}\BibitemShut {NoStop}%
\bibitem [{\citenamefont {Sweke}\ \emph {et~al.}(2014)\citenamefont {Sweke},
  \citenamefont {Sinayskiy},\ and\ \citenamefont {Petruccione}}]{Sweke2014}%
  \BibitemOpen
  \bibfield  {author} {\bibinfo {author} {\bibfnamefont {R.}~\bibnamefont
  {Sweke}}, \bibinfo {author} {\bibfnamefont {I.}~\bibnamefont {Sinayskiy}}, \
  and\ \bibinfo {author} {\bibfnamefont {F.}~\bibnamefont {Petruccione}},\
  }\href {\doibase 10.1103/PhysRevA.90.022331} {\bibfield  {journal} {\bibinfo
  {journal} {Phys. Rev. A}\ }\textbf {\bibinfo {volume} {90}},\ \bibinfo
  {pages} {022331} (\bibinfo {year} {2014})}\BibitemShut {NoStop}%
\bibitem [{\citenamefont {Wiebe}\ \emph
  {et~al.}(2014{\natexlab{b}})\citenamefont {Wiebe}, \citenamefont {Granade},
  \citenamefont {Ferrie},\ and\ \citenamefont {Cory}}]{wiebe2014qhlpra}%
  \BibitemOpen
  \bibfield  {author} {\bibinfo {author} {\bibfnamefont {N.}~\bibnamefont
  {Wiebe}}, \bibinfo {author} {\bibfnamefont {C.}~\bibnamefont {Granade}},
  \bibinfo {author} {\bibfnamefont {C.}~\bibnamefont {Ferrie}}, \ and\ \bibinfo
  {author} {\bibfnamefont {D.}~\bibnamefont {Cory}},\ }\href@noop {} {\bibfield
   {journal} {\bibinfo  {journal} {Physical Review A}\ }\textbf {\bibinfo
  {volume} {89}},\ \bibinfo {pages} {042314} (\bibinfo {year}
  {2014}{\natexlab{b}})}\BibitemShut {NoStop}%
\bibitem [{\citenamefont {Charnock}\ and\ \citenamefont
  {Kennedy}(2001)}]{charnock2001hf}%
  \BibitemOpen
  \bibfield  {author} {\bibinfo {author} {\bibfnamefont {F.~T.}\ \bibnamefont
  {Charnock}}\ and\ \bibinfo {author} {\bibfnamefont {T.}~\bibnamefont
  {Kennedy}},\ }\href@noop {} {\bibfield  {journal} {\bibinfo  {journal}
  {Physical Review B}\ }\textbf {\bibinfo {volume} {64}},\ \bibinfo {pages}
  {041201} (\bibinfo {year} {2001})}\BibitemShut {NoStop}%
\bibitem [{\citenamefont {Rowan}\ \emph {et~al.}(1965)\citenamefont {Rowan},
  \citenamefont {Hahn},\ and\ \citenamefont {Mims}}]{Rowan1965}%
  \BibitemOpen
  \bibfield  {author} {\bibinfo {author} {\bibfnamefont {L.~G.}\ \bibnamefont
  {Rowan}}, \bibinfo {author} {\bibfnamefont {E.~L.}\ \bibnamefont {Hahn}}, \
  and\ \bibinfo {author} {\bibfnamefont {W.~B.}\ \bibnamefont {Mims}},\ }\href
  {\doibase 10.1103/PhysRev.137.A61} {\bibfield  {journal} {\bibinfo  {journal}
  {Phys. Rev.}\ }\textbf {\bibinfo {volume} {137}},\ \bibinfo {pages} {A61}
  (\bibinfo {year} {1965})}\BibitemShut {NoStop}%
\bibitem [{\citenamefont {Childress}\ \emph {et~al.}(2006)\citenamefont
  {Childress}, \citenamefont {Gurudev~Dutt}, \citenamefont {Taylor},
  \citenamefont {Zibrov}, \citenamefont {Jelezko}, \citenamefont {Wrachtrup},
  \citenamefont {Hemmer},\ and\ \citenamefont {Lukin}}]{Childress2006}%
  \BibitemOpen
  \bibfield  {author} {\bibinfo {author} {\bibfnamefont {L.}~\bibnamefont
  {Childress}}, \bibinfo {author} {\bibfnamefont {M.~V.}\ \bibnamefont
  {Gurudev~Dutt}}, \bibinfo {author} {\bibfnamefont {J.~M.}\ \bibnamefont
  {Taylor}}, \bibinfo {author} {\bibfnamefont {A.~S.}\ \bibnamefont {Zibrov}},
  \bibinfo {author} {\bibfnamefont {F.}~\bibnamefont {Jelezko}}, \bibinfo
  {author} {\bibfnamefont {J.}~\bibnamefont {Wrachtrup}}, \bibinfo {author}
  {\bibfnamefont {P.~R.}\ \bibnamefont {Hemmer}}, \ and\ \bibinfo {author}
  {\bibfnamefont {M.~D.}\ \bibnamefont {Lukin}},\ }\href {\doibase
  10.1126/science.1131871} {\bibfield  {journal} {\bibinfo  {journal}
  {Science}\ }\textbf {\bibinfo {volume} {314}},\ \bibinfo {pages} {281}
  (\bibinfo {year} {2006})}\BibitemShut {NoStop}%
\bibitem [{\citenamefont {Blok}\ \emph {et~al.}(2014)\citenamefont {Blok},
  \citenamefont {Bonato}, \citenamefont {Markham}, \citenamefont {Twitchen},
  \citenamefont {Dobrovitski},\ and\ \citenamefont {Hanson}}]{Blok2014manip}%
  \BibitemOpen
  \bibfield  {author} {\bibinfo {author} {\bibfnamefont {M.}~\bibnamefont
  {Blok}}, \bibinfo {author} {\bibfnamefont {C.}~\bibnamefont {Bonato}},
  \bibinfo {author} {\bibfnamefont {M.}~\bibnamefont {Markham}}, \bibinfo
  {author} {\bibfnamefont {D.}~\bibnamefont {Twitchen}}, \bibinfo {author}
  {\bibfnamefont {V.}~\bibnamefont {Dobrovitski}}, \ and\ \bibinfo {author}
  {\bibfnamefont {R.}~\bibnamefont {Hanson}},\ }\href@noop {} {\bibfield
  {journal} {\bibinfo  {journal} {Nature Physics}\ }\textbf {\bibinfo {volume}
  {10}},\ \bibinfo {pages} {189} (\bibinfo {year} {2014})}\BibitemShut
  {NoStop}%
\bibitem [{\citenamefont {Gali}\ \emph {et~al.}(2008)\citenamefont {Gali},
  \citenamefont {Fyta},\ and\ \citenamefont {Kaxiras}}]{Gali2008}%
  \BibitemOpen
  \bibfield  {author} {\bibinfo {author} {\bibfnamefont {A.}~\bibnamefont
  {Gali}}, \bibinfo {author} {\bibfnamefont {M.}~\bibnamefont {Fyta}}, \ and\
  \bibinfo {author} {\bibfnamefont {E.}~\bibnamefont {Kaxiras}},\ }\href
  {\doibase 10.1103/PhysRevB.77.155206} {\bibfield  {journal} {\bibinfo
  {journal} {Phys. Rev. B}\ }\textbf {\bibinfo {volume} {77}},\ \bibinfo
  {pages} {1} (\bibinfo {year} {2008})}\BibitemShut {NoStop}%
\bibitem [{\citenamefont {Gali}(2009)}]{Gali2009b}%
  \BibitemOpen
  \bibfield  {author} {\bibinfo {author} {\bibfnamefont {A.}~\bibnamefont
  {Gali}},\ }\href {\doibase 10.1103/PhysRevB.80.241204} {\bibfield  {journal}
  {\bibinfo  {journal} {Phys. Rev. B}\ }\textbf {\bibinfo {volume} {80}},\
  \bibinfo {pages} {241204} (\bibinfo {year} {2009})}\BibitemShut {NoStop}%
\bibitem [{\citenamefont {Breuer}\ and\ \citenamefont
  {Petruccione}(2002)}]{breuer2002}%
  \BibitemOpen
  \bibfield  {author} {\bibinfo {author} {\bibfnamefont {H.-P.}\ \bibnamefont
  {Breuer}}\ and\ \bibinfo {author} {\bibfnamefont {F.}~\bibnamefont
  {Petruccione}},\ }\href@noop {} {\emph {\bibinfo {title} {The theory of open
  quantum systems}}}\ (\bibinfo  {publisher} {Oxford University Press on
  Demand},\ \bibinfo {year} {2002})\BibitemShut {NoStop}%
\bibitem [{\citenamefont {{\'A}lvarez}\ \emph {et~al.}(2015)\citenamefont
  {{\'A}lvarez}, \citenamefont {Bretschneider}, \citenamefont {Fischer},
  \citenamefont {London}, \citenamefont {Kanda}, \citenamefont {Onoda},
  \citenamefont {Isoya}, \citenamefont {Gershoni},\ and\ \citenamefont
  {Frydman}}]{alvarez2015hyperpol}%
  \BibitemOpen
  \bibfield  {author} {\bibinfo {author} {\bibfnamefont {G.~A.}\ \bibnamefont
  {{\'A}lvarez}}, \bibinfo {author} {\bibfnamefont {C.~O.}\ \bibnamefont
  {Bretschneider}}, \bibinfo {author} {\bibfnamefont {R.}~\bibnamefont
  {Fischer}}, \bibinfo {author} {\bibfnamefont {P.}~\bibnamefont {London}},
  \bibinfo {author} {\bibfnamefont {H.}~\bibnamefont {Kanda}}, \bibinfo
  {author} {\bibfnamefont {S.}~\bibnamefont {Onoda}}, \bibinfo {author}
  {\bibfnamefont {J.}~\bibnamefont {Isoya}}, \bibinfo {author} {\bibfnamefont
  {D.}~\bibnamefont {Gershoni}}, \ and\ \bibinfo {author} {\bibfnamefont
  {L.}~\bibnamefont {Frydman}},\ }\href@noop {} {\bibfield  {journal} {\bibinfo
   {journal} {Nature communications}\ }\textbf {\bibinfo {volume} {6}},\
  \bibinfo {pages} {8456} (\bibinfo {year} {2015})}\BibitemShut {NoStop}%
\bibitem [{\citenamefont {Hou}\ \emph {et~al.}(2019)\citenamefont {Hou},
  \citenamefont {He}, \citenamefont {Wang}, \citenamefont {Huang},
  \citenamefont {Zhang}, \citenamefont {Ouyang}, \citenamefont {Wang},
  \citenamefont {Lian}, \citenamefont {Chang},\ and\ \citenamefont
  {Duan}}]{Hou2019}%
  \BibitemOpen
  \bibfield  {author} {\bibinfo {author} {\bibfnamefont {P.-Y.}\ \bibnamefont
  {Hou}}, \bibinfo {author} {\bibfnamefont {L.}~\bibnamefont {He}}, \bibinfo
  {author} {\bibfnamefont {F.}~\bibnamefont {Wang}}, \bibinfo {author}
  {\bibfnamefont {X.-Z.}\ \bibnamefont {Huang}}, \bibinfo {author}
  {\bibfnamefont {W.-G.}\ \bibnamefont {Zhang}}, \bibinfo {author}
  {\bibfnamefont {X.-L.}\ \bibnamefont {Ouyang}}, \bibinfo {author}
  {\bibfnamefont {X.}~\bibnamefont {Wang}}, \bibinfo {author} {\bibfnamefont
  {W.-Q.}\ \bibnamefont {Lian}}, \bibinfo {author} {\bibfnamefont {X.-Y.}\
  \bibnamefont {Chang}}, \ and\ \bibinfo {author} {\bibfnamefont {L.-M.}\
  \bibnamefont {Duan}},\ }\href {\doibase 10.1088/0256-307x/36/10/100303}
  {\bibfield  {journal} {\bibinfo  {journal} {Chinese Physics Letters}\
  }\textbf {\bibinfo {volume} {36}},\ \bibinfo {pages} {100303} (\bibinfo
  {year} {2019})}\BibitemShut {NoStop}%
\bibitem [{\citenamefont {Bouganne}\ \emph {et~al.}(2020)\citenamefont
  {Bouganne}, \citenamefont {Bosch~Aguilera}, \citenamefont {Ghermaoui},
  \citenamefont {Beugnon},\ and\ \citenamefont {Gerbier}}]{Bouganne2020}%
  \BibitemOpen
  \bibfield  {author} {\bibinfo {author} {\bibfnamefont {R.}~\bibnamefont
  {Bouganne}}, \bibinfo {author} {\bibfnamefont {M.}~\bibnamefont
  {Bosch~Aguilera}}, \bibinfo {author} {\bibfnamefont {A.}~\bibnamefont
  {Ghermaoui}}, \bibinfo {author} {\bibfnamefont {J.}~\bibnamefont {Beugnon}},
  \ and\ \bibinfo {author} {\bibfnamefont {F.}~\bibnamefont {Gerbier}},\ }\href
  {\doibase 10.1038/s41567-019-0678-2} {\bibfield  {journal} {\bibinfo
  {journal} {Nature Physics}\ }\textbf {\bibinfo {volume} {16}},\ \bibinfo
  {pages} {21} (\bibinfo {year} {2020})}\BibitemShut {NoStop}%
\bibitem [{\citenamefont {Reiter}\ and\ \citenamefont
  {S\o{}rensen}(2012)}]{Reiter2012}%
  \BibitemOpen
  \bibfield  {author} {\bibinfo {author} {\bibfnamefont {F.}~\bibnamefont
  {Reiter}}\ and\ \bibinfo {author} {\bibfnamefont {A.~S.}\ \bibnamefont
  {S\o{}rensen}},\ }\href {\doibase 10.1103/PhysRevA.85.032111} {\bibfield
  {journal} {\bibinfo  {journal} {Phys. Rev. A}\ }\textbf {\bibinfo {volume}
  {85}},\ \bibinfo {pages} {032111} (\bibinfo {year} {2012})}\BibitemShut
  {NoStop}%
\bibitem [{\citenamefont {Smeltzer}\ \emph {et~al.}(2011)\citenamefont
  {Smeltzer}, \citenamefont {Childress},\ and\ \citenamefont
  {Gali}}]{smeltzer2011hf}%
  \BibitemOpen
  \bibfield  {author} {\bibinfo {author} {\bibfnamefont {B.}~\bibnamefont
  {Smeltzer}}, \bibinfo {author} {\bibfnamefont {L.}~\bibnamefont {Childress}},
  \ and\ \bibinfo {author} {\bibfnamefont {A.}~\bibnamefont {Gali}},\
  }\href@noop {} {\bibfield  {journal} {\bibinfo  {journal} {New Journal of
  Physics}\ }\textbf {\bibinfo {volume} {13}},\ \bibinfo {pages} {025021}
  (\bibinfo {year} {2011})}\BibitemShut {NoStop}%
\bibitem [{\citenamefont {Granade}\ \emph {et~al.}(2017)\citenamefont
  {Granade}, \citenamefont {Ferrie}, \citenamefont {Hincks}, \citenamefont
  {Casagrande}, \citenamefont {Alexander}, \citenamefont {Gross}, \citenamefont
  {Kononenko},\ and\ \citenamefont {Sanders}}]{granade2017qinfer}%
  \BibitemOpen
  \bibfield  {author} {\bibinfo {author} {\bibfnamefont {C.}~\bibnamefont
  {Granade}}, \bibinfo {author} {\bibfnamefont {C.}~\bibnamefont {Ferrie}},
  \bibinfo {author} {\bibfnamefont {I.}~\bibnamefont {Hincks}}, \bibinfo
  {author} {\bibfnamefont {S.}~\bibnamefont {Casagrande}}, \bibinfo {author}
  {\bibfnamefont {T.}~\bibnamefont {Alexander}}, \bibinfo {author}
  {\bibfnamefont {J.}~\bibnamefont {Gross}}, \bibinfo {author} {\bibfnamefont
  {M.}~\bibnamefont {Kononenko}}, \ and\ \bibinfo {author} {\bibfnamefont
  {Y.}~\bibnamefont {Sanders}},\ }\href@noop {} {\bibfield  {journal} {\bibinfo
   {journal} {Quantum}\ }\textbf {\bibinfo {volume} {1}},\ \bibinfo {pages} {5}
  (\bibinfo {year} {2017})}\BibitemShut {NoStop}%
\bibitem [{\citenamefont {Liu}\ and\ \citenamefont {West}(2001)}]{Liu:2001cz}%
  \BibitemOpen
  \bibfield  {author} {\bibinfo {author} {\bibfnamefont {J.}~\bibnamefont
  {Liu}}\ and\ \bibinfo {author} {\bibfnamefont {M.}~\bibnamefont {West}},\
  }in\ \href {\doibase 10.1007/978-1-4757-3437-9_10} {\emph {\bibinfo
  {booktitle} {Sequential Monte Carlo algorithms for a class of outer
  measures}}}\ (\bibinfo  {publisher} {Springer New York},\ \bibinfo {address}
  {New York, NY},\ \bibinfo {year} {2001})\ pp.\ \bibinfo {pages}
  {197--223}\BibitemShut {NoStop}%
\bibitem [{\citenamefont {Granade}\ and\ \citenamefont
  {Wiebe}(2017)}]{Granade:2016txa}%
  \BibitemOpen
  \bibfield  {author} {\bibinfo {author} {\bibfnamefont {C.}~\bibnamefont
  {Granade}}\ and\ \bibinfo {author} {\bibfnamefont {N.}~\bibnamefont
  {Wiebe}},\ }\href {\doibase 10.1088/1367-2630/aa77cf} {\bibfield  {journal}
  {\bibinfo  {journal} {New Journal of Physics}\ }\textbf {\bibinfo {volume}
  {19}},\ \bibinfo {pages} {1} (\bibinfo {year} {2017})}\BibitemShut {NoStop}%
\bibitem [{\citenamefont {Balian}\ \emph {et~al.}(2014)\citenamefont {Balian},
  \citenamefont {Wolfowicz}, \citenamefont {Morton},\ and\ \citenamefont
  {Monteiro}}]{Balian2014}%
  \BibitemOpen
  \bibfield  {author} {\bibinfo {author} {\bibfnamefont {S.~J.}\ \bibnamefont
  {Balian}}, \bibinfo {author} {\bibfnamefont {G.}~\bibnamefont {Wolfowicz}},
  \bibinfo {author} {\bibfnamefont {J.~J.~L.}\ \bibnamefont {Morton}}, \ and\
  \bibinfo {author} {\bibfnamefont {T.~S.}\ \bibnamefont {Monteiro}},\ }\href
  {\doibase 10.1103/PhysRevB.89.045403} {\bibfield  {journal} {\bibinfo
  {journal} {Phys. Rev. B}\ }\textbf {\bibinfo {volume} {89}},\ \bibinfo
  {pages} {045403} (\bibinfo {year} {2014})}\BibitemShut {NoStop}%
\bibitem [{\citenamefont {{Meinhardt}}\ \emph {et~al.}(2017)\citenamefont
  {{Meinhardt}}, \citenamefont {{Moeller}}, \citenamefont {{Hazirbas}},\ and\
  \citenamefont {{Cremers}}}]{Meinhardt2017}%
  \BibitemOpen
  \bibfield  {author} {\bibinfo {author} {\bibfnamefont {T.}~\bibnamefont
  {{Meinhardt}}}, \bibinfo {author} {\bibfnamefont {M.}~\bibnamefont
  {{Moeller}}}, \bibinfo {author} {\bibfnamefont {C.}~\bibnamefont
  {{Hazirbas}}}, \ and\ \bibinfo {author} {\bibfnamefont {D.}~\bibnamefont
  {{Cremers}}},\ }in\ \href {\doibase 10.1109/ICCV.2017.198} {\emph {\bibinfo
  {booktitle} {2017 IEEE International Conference on Computer Vision (ICCV)}}}\
  (\bibinfo {year} {2017})\ pp.\ \bibinfo {pages} {1799--1808}\BibitemShut
  {NoStop}%
\bibitem [{\citenamefont {Brunner}\ \emph {et~al.}(2008)\citenamefont
  {Brunner}, \citenamefont {Pironio}, \citenamefont {Acin}, \citenamefont
  {Gisin}, \citenamefont {M\'ethot},\ and\ \citenamefont
  {Scarani}}]{Brunner2008}%
  \BibitemOpen
  \bibfield  {author} {\bibinfo {author} {\bibfnamefont {N.}~\bibnamefont
  {Brunner}}, \bibinfo {author} {\bibfnamefont {S.}~\bibnamefont {Pironio}},
  \bibinfo {author} {\bibfnamefont {A.}~\bibnamefont {Acin}}, \bibinfo {author}
  {\bibfnamefont {N.}~\bibnamefont {Gisin}}, \bibinfo {author} {\bibfnamefont
  {A.~A.}\ \bibnamefont {M\'ethot}}, \ and\ \bibinfo {author} {\bibfnamefont
  {V.}~\bibnamefont {Scarani}},\ }\href {\doibase
  10.1103/PhysRevLett.100.210503} {\bibfield  {journal} {\bibinfo  {journal}
  {Phys. Rev. Lett.}\ }\textbf {\bibinfo {volume} {100}},\ \bibinfo {pages}
  {210503} (\bibinfo {year} {2008})}\BibitemShut {NoStop}%
\bibitem [{\citenamefont {Hutson}(2018)}]{Hutson2018}%
  \BibitemOpen
  \bibfield  {author} {\bibinfo {author} {\bibfnamefont {M.}~\bibnamefont
  {Hutson}},\ }\href {\doibase 10.1126/science.359.6377.725} {\bibfield
  {journal} {\bibinfo  {journal} {Science}\ }\textbf {\bibinfo {volume}
  {359}},\ \bibinfo {pages} {725} (\bibinfo {year} {2018})}\BibitemShut
  {NoStop}%
\bibitem [{\citenamefont {Bruus}\ and\ \citenamefont
  {Flensberg}(2004)}]{Bruus2004}%
  \BibitemOpen
  \bibfield  {author} {\bibinfo {author} {\bibfnamefont {H.}~\bibnamefont
  {Bruus}}\ and\ \bibinfo {author} {\bibfnamefont {K.}~\bibnamefont
  {Flensberg}},\ }\href@noop {} {\emph {\bibinfo {title} {Many-Body Quantum
  Theory in Condensed Matter Physics An Introduction}}}\ (\bibinfo  {publisher}
  {Oxford University Press},\ \bibinfo {year} {2004})\BibitemShut {NoStop}%
\bibitem [{\citenamefont {Berry}\ \emph {et~al.}(2007)\citenamefont {Berry},
  \citenamefont {Ahokas}, \citenamefont {Cleve},\ and\ \citenamefont
  {Sanders}}]{Berry2007}%
  \BibitemOpen
  \bibfield  {author} {\bibinfo {author} {\bibfnamefont {D.~W.}\ \bibnamefont
  {Berry}}, \bibinfo {author} {\bibfnamefont {G.}~\bibnamefont {Ahokas}},
  \bibinfo {author} {\bibfnamefont {R.}~\bibnamefont {Cleve}}, \ and\ \bibinfo
  {author} {\bibfnamefont {B.~C.}\ \bibnamefont {Sanders}},\ }\href {\doibase
  10.1007/s00220-006-0150-x} {\bibfield  {journal} {\bibinfo  {journal}
  {Communications in Mathematical Physics}\ }\textbf {\bibinfo {volume}
  {270}},\ \bibinfo {pages} {359} (\bibinfo {year} {2007})}\BibitemShut
  {NoStop}%
\bibitem [{\citenamefont {Childs}\ and\ \citenamefont
  {Kothari}(2011)}]{Childs2011}%
  \BibitemOpen
  \bibfield  {author} {\bibinfo {author} {\bibfnamefont {A.~M.}\ \bibnamefont
  {Childs}}\ and\ \bibinfo {author} {\bibfnamefont {R.}~\bibnamefont
  {Kothari}},\ }in\ \href@noop {} {\emph {\bibinfo {booktitle} {Theory of
  Quantum Computation, Communication, and Cryptography}}}\ (\bibinfo
  {publisher} {Springer Berlin Heidelberg},\ \bibinfo {address} {Berlin,
  Heidelberg},\ \bibinfo {year} {2011})\ pp.\ \bibinfo {pages}
  {94--103}\BibitemShut {NoStop}%
\bibitem [{\citenamefont {Peruzzo}\ \emph {et~al.}(2014)\citenamefont
  {Peruzzo}, \citenamefont {McClean}, \citenamefont {Shadbolt}, \citenamefont
  {Yung}, \citenamefont {Zhou}, \citenamefont {Love}, \citenamefont
  {Aspuru-Guzik},\ and\ \citenamefont {O'Brien}}]{Peruzzo2014}%
  \BibitemOpen
  \bibfield  {author} {\bibinfo {author} {\bibfnamefont {A.}~\bibnamefont
  {Peruzzo}}, \bibinfo {author} {\bibfnamefont {J.}~\bibnamefont {McClean}},
  \bibinfo {author} {\bibfnamefont {P.}~\bibnamefont {Shadbolt}}, \bibinfo
  {author} {\bibfnamefont {M.-H.}\ \bibnamefont {Yung}}, \bibinfo {author}
  {\bibfnamefont {X.-Q.}\ \bibnamefont {Zhou}}, \bibinfo {author}
  {\bibfnamefont {P.~J.}\ \bibnamefont {Love}}, \bibinfo {author}
  {\bibfnamefont {A.}~\bibnamefont {Aspuru-Guzik}}, \ and\ \bibinfo {author}
  {\bibfnamefont {J.~L.}\ \bibnamefont {O'Brien}},\ }\href {\doibase
  10.1038/ncomms5213} {\bibfield  {journal} {\bibinfo  {journal} {Nature
  Communications}\ }\textbf {\bibinfo {volume} {5}},\ \bibinfo {pages} {4213}
  (\bibinfo {year} {2014})}\BibitemShut {NoStop}%
\bibitem [{\citenamefont {Sra}\ \emph {et~al.}(2012)\citenamefont {Sra},
  \citenamefont {Nowozin},\ and\ \citenamefont {Wright}}]{sra2012}%
  \BibitemOpen
  \bibfield  {author} {\bibinfo {author} {\bibfnamefont {S.}~\bibnamefont
  {Sra}}, \bibinfo {author} {\bibfnamefont {S.}~\bibnamefont {Nowozin}}, \ and\
  \bibinfo {author} {\bibfnamefont {S.~J.}\ \bibnamefont {Wright}},\
  }\href@noop {} {\emph {\bibinfo {title} {Optimization for Machine
  Learning}}}\ (\bibinfo  {publisher} {{MIT Press}},\ \bibinfo {year}
  {2012})\BibitemShut {NoStop}%
\bibitem [{\citenamefont {B.}(1998)}]{Bollobas1998}%
  \BibitemOpen
  \bibfield  {author} {\bibinfo {author} {\bibfnamefont {B.}~\bibnamefont
  {B.}},\ }\href@noop {} {\emph {\bibinfo {title} {Connectivity and Matching.
  In: Modern Graph Theory. Graduate Texts in Mathematics}}},\ Vol.\ \bibinfo
  {volume} {184}\ (\bibinfo  {publisher} {Springer},\ \bibinfo {year}
  {1998})\BibitemShut {NoStop}%
\bibitem [{\citenamefont {Murphy}(2012)}]{Murphy2012}%
  \BibitemOpen
  \bibfield  {author} {\bibinfo {author} {\bibfnamefont {K.~P.}\ \bibnamefont
  {Murphy}},\ }\href@noop {} {\emph {\bibinfo {title} {Machine Learning: A
  Probabilistic Perspective}}}\ (\bibinfo  {publisher} {The MIT Press},\
  \bibinfo {year} {2012})\BibitemShut {NoStop}%
\bibitem [{\citenamefont {Bishop}(2006)}]{bishop2006}%
  \BibitemOpen
  \bibfield  {author} {\bibinfo {author} {\bibfnamefont {C.}~\bibnamefont
  {Bishop}},\ }\href {https://books.google.co.in/books?id=kTNoQgAACAAJ} {\emph
  {\bibinfo {title} {Pattern Recognition and Machine Learning}}},\ Information
  Science and Statistics\ (\bibinfo  {publisher} {Springer},\ \bibinfo {year}
  {2006})\BibitemShut {NoStop}%
\bibitem [{\citenamefont {Ferrie}\ \emph {et~al.}(2012)\citenamefont {Ferrie},
  \citenamefont {Granade},\ and\ \citenamefont {Cory}}]{Ferrie:2012ip}%
  \BibitemOpen
  \bibfield  {author} {\bibinfo {author} {\bibfnamefont {C.}~\bibnamefont
  {Ferrie}}, \bibinfo {author} {\bibfnamefont {C.~E.}\ \bibnamefont {Granade}},
  \ and\ \bibinfo {author} {\bibfnamefont {D.~G.}\ \bibnamefont {Cory}},\
  }\href {\doibase 10.1007/s11128-012-0407-6} {\bibfield  {journal} {\bibinfo
  {journal} {Quantum Information Processing}\ }\textbf {\bibinfo {volume}
  {12}},\ \bibinfo {pages} {611} (\bibinfo {year} {2012})}\BibitemShut
  {NoStop}%
\bibitem [{\citenamefont {Itano}\ \emph {et~al.}(1993)\citenamefont {Itano},
  \citenamefont {Bergquist}, \citenamefont {Bollinger}, \citenamefont
  {Gilligan}, \citenamefont {Heinzen}, \citenamefont {Moore}, \citenamefont
  {Raizen},\ and\ \citenamefont {Wineland}}]{Itano1993}%
  \BibitemOpen
  \bibfield  {author} {\bibinfo {author} {\bibfnamefont {W.~M.}\ \bibnamefont
  {Itano}}, \bibinfo {author} {\bibfnamefont {J.~C.}\ \bibnamefont
  {Bergquist}}, \bibinfo {author} {\bibfnamefont {J.~J.}\ \bibnamefont
  {Bollinger}}, \bibinfo {author} {\bibfnamefont {J.~M.}\ \bibnamefont
  {Gilligan}}, \bibinfo {author} {\bibfnamefont {D.~J.}\ \bibnamefont
  {Heinzen}}, \bibinfo {author} {\bibfnamefont {F.~L.}\ \bibnamefont {Moore}},
  \bibinfo {author} {\bibfnamefont {M.~G.}\ \bibnamefont {Raizen}}, \ and\
  \bibinfo {author} {\bibfnamefont {D.~J.}\ \bibnamefont {Wineland}},\ }\href
  {\doibase 10.1103/PhysRevA.47.3554} {\bibfield  {journal} {\bibinfo
  {journal} {Phys. Rev. A}\ }\textbf {\bibinfo {volume} {47}},\ \bibinfo
  {pages} {3554} (\bibinfo {year} {1993})}\BibitemShut {NoStop}%
\bibitem [{\citenamefont {Zhao}\ \emph {et~al.}(2012)\citenamefont {Zhao},
  \citenamefont {Honert}, \citenamefont {Schmid}, \citenamefont {Klas},
  \citenamefont {Isoya}, \citenamefont {Markham}, \citenamefont {Twitchen},
  \citenamefont {Jelezko}, \citenamefont {Liu}, \citenamefont {Fedder} \emph
  {et~al.}}]{zhao2012nucspins}%
  \BibitemOpen
  \bibfield  {author} {\bibinfo {author} {\bibfnamefont {N.}~\bibnamefont
  {Zhao}}, \bibinfo {author} {\bibfnamefont {J.}~\bibnamefont {Honert}},
  \bibinfo {author} {\bibfnamefont {B.}~\bibnamefont {Schmid}}, \bibinfo
  {author} {\bibfnamefont {M.}~\bibnamefont {Klas}}, \bibinfo {author}
  {\bibfnamefont {J.}~\bibnamefont {Isoya}}, \bibinfo {author} {\bibfnamefont
  {M.}~\bibnamefont {Markham}}, \bibinfo {author} {\bibfnamefont
  {D.}~\bibnamefont {Twitchen}}, \bibinfo {author} {\bibfnamefont
  {F.}~\bibnamefont {Jelezko}}, \bibinfo {author} {\bibfnamefont {R.-B.}\
  \bibnamefont {Liu}}, \bibinfo {author} {\bibfnamefont {H.}~\bibnamefont
  {Fedder}},  \emph {et~al.},\ }\href@noop {} {\bibfield  {journal} {\bibinfo
  {journal} {Nature nanotechnology}\ }\textbf {\bibinfo {volume} {7}},\
  \bibinfo {pages} {657} (\bibinfo {year} {2012})}\BibitemShut {NoStop}%
\bibitem [{\citenamefont {Schlosshauer}(2005)}]{schlosshauer2005}%
  \BibitemOpen
  \bibfield  {author} {\bibinfo {author} {\bibfnamefont {M.}~\bibnamefont
  {Schlosshauer}},\ }\href {\doibase 10.1103/RevModPhys.76.1267} {\bibfield
  {journal} {\bibinfo  {journal} {Rev. Mod. Phys.}\ }\textbf {\bibinfo {volume}
  {76}},\ \bibinfo {pages} {1267} (\bibinfo {year} {2005})}\BibitemShut
  {NoStop}%
\bibitem [{\citenamefont {Weiss}(2012)}]{Weiss2012}%
  \BibitemOpen
  \bibfield  {author} {\bibinfo {author} {\bibfnamefont {U.}~\bibnamefont
  {Weiss}},\ }\href {\doibase 10.1142/8334} {\emph {\bibinfo {title} {Quantum
  Dissipative Systems}}},\ \bibinfo {edition} {4th}\ ed.\ (\bibinfo
  {publisher} {WORLD SCIENTIFIC},\ \bibinfo {year} {2012})\ \Eprint
  {http://arxiv.org/abs/https://www.worldscientific.com/doi/pdf/10.1142/8334}
  {https://www.worldscientific.com/doi/pdf/10.1142/8334} \BibitemShut {NoStop}%
\bibitem [{\citenamefont {Chirolli}\ and\ \citenamefont
  {Burkard}(2008)}]{chirolli2008}%
  \BibitemOpen
  \bibfield  {author} {\bibinfo {author} {\bibfnamefont {L.}~\bibnamefont
  {Chirolli}}\ and\ \bibinfo {author} {\bibfnamefont {G.}~\bibnamefont
  {Burkard}},\ }\href {\doibase 10.1080/00018730802218067} {\bibfield
  {journal} {\bibinfo  {journal} {Advances in Physics}\ }\textbf {\bibinfo
  {volume} {57}},\ \bibinfo {pages} {225} (\bibinfo {year} {2008})}\BibitemShut
  {NoStop}%
\bibitem [{\citenamefont {Deng}\ \emph {et~al.}(2005)\citenamefont {Deng},
  \citenamefont {Porras},\ and\ \citenamefont {Cirac}}]{Deng2005}%
  \BibitemOpen
  \bibfield  {author} {\bibinfo {author} {\bibfnamefont {X.-L.}\ \bibnamefont
  {Deng}}, \bibinfo {author} {\bibfnamefont {D.}~\bibnamefont {Porras}}, \ and\
  \bibinfo {author} {\bibfnamefont {J.~I.}\ \bibnamefont {Cirac}},\ }\href
  {\doibase 10.1103/PhysRevA.72.063407} {\bibfield  {journal} {\bibinfo
  {journal} {Phys. Rev. A}\ }\textbf {\bibinfo {volume} {72}},\ \bibinfo
  {pages} {063407} (\bibinfo {year} {2005})}\BibitemShut {NoStop}%
\end{thebibliography}%


\newpage
\section*{Methods}
{\small
\noindent\textbf{Experimental set-up details --}
To focus the investigation on the hyperfine interaction of the NV centre with its environment we omit in Eqn.~\ref{eq:full_Hamiltonian} the zero-field splitting ($G_{\text{gs}} S_z^2$) and contributions from the quadrupole interaction ($P I_z^2$)
as constant shifts on the fine states and a splitting on the nitrogen hyperfine states respectively.
%
The hyperfine interaction term $\hat{S} \cdot \mathbf{A} \cdot \hat{I}$ in Eqn.\ref{eq:full_Hamiltonian} replaces its expansion $\hat{S} \cdot \sum_{X, \chi} \mathbf{A}_{X,\chi} \cdot \hat{I}_{X,\chi}$ in contributions due to each atom species $X$ and lattice site $\chi$ \cite{smeltzer2011hf}.
Thus, all nuclear contributions to the electron spin dynamics are mapped to a single \textit{environmental} qubit.
We analyse the Hamiltonian describing the action of an external  magnetic field $\mathbf{B}$, and the hyperfine  coupling described via the term
\begin{equation}
\mathbf{S} \cdot \mathbf{A} \cdot \mathbf{I} \simeq A_{\parallel} S_z I_z + A_{\perp} (S_x I_x + S_y I_y),
\label{eq:appr_Hamiltonian}
\end{equation}
by the (non-)axial tensor parameters $A_{\parallel}$ ($A_{\perp}$)~\cite{charnock2001hf}.

Hahn signal experiments can be interpreted according to the hyperfine interaction~\cite{Rowan1965}, describing the coupling of the electronic NV spin, in state $m_s$, to the $j^{th}$ nuclear \textsuperscript{13}C spin, combined in a four-level system $\mathcal{S}$. 
Here we summarise this interaction in terms of an effective magnetic field $\textbf{B}^j_{m_{s}}$, whereby the ground states of $\mathcal{S}$ precess at a rate $\omega_{j,0}$, whereas the excited states incur in a splitting $\omega_{j,1}$ ~\cite{Childress2006}.
In free evolution, after the initial $\pi/2$ pulse, the nuclear and electron spin become progressively more entangled at a rate dictated by the hyperfine interaction, getting maximally correlated at times $\tau \propto \pi/A$~\cite{Blok2014manip}, where the Hahn signal is weakest. 
When the two spins get disentangled again, revivals can be observed in the experimental Hahn-echo signal (see Fig.~\ref{fig:Fig3}).
The unitary evolution originated from such simplified dipole Hamiltonian can be analytically solved~\cite{Rowan1965} to obtain the Hahn-echo signal:
\begin{equation}
  \rm{Pr}(1| \tau ; \{ \mathbf{B}^j \}, \{ \omega_j \}) =  \big( \prod_j \; (S_j) + 1 \big) /2, 
  \label{eq:prhahnsignal_methods}
 \end{equation}
 via the pseudospins:
 \begin{equation}
   S_j =  1 - \frac{|\mathbf{B}_0 \times \mathbf{B}_1^j|^2}{|\mathbf{B}_0|^2 |\mathbf{B}_1^j|^2} 
    \sin^2 (\omega_{0} \; \tau /2)  \sin^2 (\omega_{j,1} \; \tau /2), 
\label{eq:pseudospins_methods}
\end{equation}
with $\mathbf{B}_0$ the external magnetic field and $\omega_{0}$ the bare Larmor frequency. 
Observing Eqn.~\ref{eq:pseudospins_methods}, decays and revivals in the PL signal can then be interpreted respectively as beatings among the modulation frequencies $\omega_{j,1}$, and re-synchronisation when $\tau = 2 \pi / \omega_0$. 
See Supplementary \S~\ref{supp_experimental} for further details.

\noindent\textbf{Quantum Hamiltonian Learning --}
For each explored model $\hat H_j(\vec{x}_j)$, \gls{qhl} is performed by updating a distribution of $N_P$ particles for $N_E$ epochs, where each epoch corresponds to a single experiment performed on the NV centre electron spin. 
The experiment design here is left to a heuristic rule within \gls{qhl}, which can easily be replaced by more sophisticated methods when required.
Each experiment involves preparing the electron spin in the chosen probe state $\ket{\psi}$, evolving the spin for the chosen evolution time $t$, and measuring. 
In practice, for a given $t$, we repeat $10^6$ times to obtain a reliable estimate of the expectation value $\left| \bra{\psi} e^{-i\ho t}\ket{\psi} \right|^2$.
QHL designs experiments to run on the system of interest in order to maximise knowledge gained, and can invoke a trusted (quantum) 
simulator to test and iteratively improve parameterisations, resulting in trained $\vec{x}_j^{\prime}$ (yielding $\hat{H}_j^{\prime}$) ~\cite{Wiebe:2014qhl, wang2017qhlexp, Santagati2019}. 

\gls{qhl} learns the parameters within given models through Bayesian inference. 
Bayesian inference relies on a likelihood function to compare the true model with a particle $\vec{x}_p$ sampled from the current parameter probability distribution.
The likelihood function comes from Bayes' rule,
\begin{equation}
    \label{bayes_rule_eqn}
        Pr(\vec{x}_p | d; t) = \frac{Pr(d| \vec{x}_p; t) \ Pr(\vec{x}_p)}{Pr(d|t)},
\end{equation} 
where $\vec{x}_p$ is a single particle, $Pr(\vec{x}_p)$ is the current probability distribution for particles, $t$ is the time the system evolved for,and  $d$ is a datum, i.e. either 0 or 1 measured from the system. 

For each epoch an experiment is performed resulting in a single datum $d \in \{0,1\}$, so here we use the likelihood function $\mathrm{Pr}(d | \hat{H}_j, t) = \ev{\tr_{\mathrm{env}} [  \rho_{\Psi}(t)]}{d}_{\mathrm{sys}} $, 
where $\lvert d\rangle_{\mathrm{sys}}$ is the chosen measurement basis, and $\rho_{\Psi}(t) $ the global density matrix evolved under each $\hat{H}_j$.
The data used at each epoch are stored as $D_j$.
Complete details for \gls{qhl} are given in Supplementary \S~\ref{supp_qhl}. 

\noindent\textbf{Quantum Model Learning Agent further details --}
During the consolidation phase for layer $\mu$, all models in $\mu$ are compared pairwise according to their \gls{bf}: $\mathcal{B}_{ij} = \exp [\ell(D_{ij}|\hip) - \ell(D_{ij}|\hjp)]$, where $D_{ij} = D_i \cup D_j$ and $\ell(D_{ij}|\hjp) = \sum_{d \in D_{ij}} \log \Pr(d | \hjp)$ is the cumulative log-likelihood \cite{granade2017qinfer}. 
Directed edges between $(\hat{H}_i, \hat{H}_j)$ represent $\mathcal{B}_{ij}$ in the \gls{cdag}, with directionality indicating performance, i.e. $\mathcal{B}_{ij} \gg 1$ ($\bij \ll 1)$ is proof $\hat{H}_i$ (\hj) is superior, Fig. \ref{fig:Fig1}c. 
In a given consolidation among any subset of models $\{\hat{H}_k\}$, the champion is the node at which most edges point, i.e. which wins the most pairwise comparisons. 
In particular, in a given layer $\mu$ we use consolidation to rank models, or to determine a single layer champion $\hat{H}^{\mu}_C$. \par 
An \textit{exploration} stage follows, whereby a new layer $\nu$ is generated, consisting of new models which \emph{inherit} properties from the layer champion (or top-ranked models) of $\mu$, i.e. $\nu = R(\hat{H}^{\mu}_C)$, where $R$ is the unique spawn rule which characterises \gls{qmla}. 
The spawn rule employed in this work is a simple \emph{greedy} search, i.e. at each layer, terms are added individually from a predefined set, with subsequent layer champions informing \gls{qmla} which terms to remove from that set. When the set of available terms is empty, \gls{qmla} either moves to the next stage, or terminates. 
Formally, the spawn rule proceeds as follows. 
\begin{enumerate*}[label=(\roman*)]
    \item Spin rotation terms are adopted as primitives: $\mu^0 = \{ \hat{S}_x, \hat{S}_y, \hat{S}_z\}$;
    \item Terms are added greedily in separate layers until exhausted;
    \item Hilbert space dimension is increased by introducing hyperfine terms $\{\hat{A}_{x}, \hat{A}_y, \hat{A}_z\}$, which are also added individually by layer until exhaustion;
    \item transverse terms $\{\hat{T}_{xy}, \hat{T}_{xz}, \hat{T}_{yz}\}$ are introduced and added greedily until exhaustion.
\end{enumerate*}
Here \gls{qmla} is limited to 9-parameter spin models with specific interactions. 
The termination rule in this case is simply whether all primitive models have been exhausted in greedy search. 
Generated models are considered \emph{children} of the model from which they inherit (\emph{parent}): such relationships are stored in the \gls{sdag}.
A representative example of the \gls{qmla} cDAG is shown in Fig.~\ref{fig:Fig2}f.

\noindent\textbf{Analysis of the spin bath --}
The interaction terms in the Hamiltonian corresponding to the spin bath are hyperfine contributions for each spin, equivalent to those in Eqn. \ref{eq:full_Hamiltonian}; that is, the contributions from each spin are all degenerate. 
Degeneracies in the likelihood function are known to mislead sequential Monte-Carlo algorithms, such as that exploited by \gls{cle}, which are implicitly based on assumptions of unimodality~\cite{Liu:2001cz}.
This problem was recently addressed in \cite{Granade:2016txa}. 

Hahn--echo experiments are designed to minimise the contribution of the bath to the system dynamics, so collective phenomena are expected to dominate individual spin contributions \cite{Balian2014}.
The method starts from the likelihood for the binary outcome of Hahn--echo experiments, 
\begin{equation}
\label{bath_likelihood}
    {Pr}(1| S_j) =  \big( \prod_j \; (S_j) + 1 \big) /2,
\end{equation}
expressed in terms of the pseudospins $S_j (\mathbf{B}_0, \omega_0, \mathbf{B}_j, \omega_j)$, where $\mathbf{B}_0$ ($\mathbf{B}_j$) is the external (effective) magnetic field at site $j$, and $\omega_0$ the bare (modulated) Larmor frequency.

To characterise the spin-bath interaction in the more general case and estimate the total number of interacting spins, $n^s$, producing the observed dynamics, we use a Metropolis-Hasting protocol as follows. 
We construct a hyperparameterisation of the problem, using two normal distributions $\mathcal{N}(B_1, \sigma_B)$ and $\mathcal{N}(\omega_0 + \delta_{\omega}, \sigma_{\omega})$, from which a number $n^{\textrm{s}}_j$ of $\mathbf{B}_1^j$ and $ \omega_{j,1}$ are drawn. In this way, for each tentative $n^{\textrm{s}}_j$ a \gls{cle} iteration can be performed against a reduced hyperparameter set: $ \vec{x} := \{ \mathbf{B}_0, \mathbf{B}_1, \sigma_B, \omega_0, \delta_{\omega}, \sigma_{\omega}   \}$. 
Inferring $n^{\textrm{s}}$ is then left to a Metropolis--Hastings procedure, that approximates the distribution of $P(n^{\textrm{s}})$. 
At each epoch a new tentative $n^{\textrm{s}}_{j+1}$ is sampled; $N_e$ epochs of CLE are performed for each $n^{\textrm{s}}_{j}$. 
The probability of accepting $n^{\textrm{s}}_{j+1}$ as representative of the distribution is taken as $\mathcal{B}_{j+1,j} (D)$, with $D$ the cumulative set of experimental data collected throughout all steps. 
In this way, higher values of $n^{\textrm{s}}_j$ can be considered if they are statistically justified by a better reproduction of the data. 

} 

\section*{Acknowledgements}
{\small
The authors  thank Cristian Bonato, Fedor Jelezko, Florian Marquardt, Th\"{o}mas Fosel, Christopher Woods, Jianwei Wang, Mark G. Thompson, Adeline Paiement and Alessio Cimarelli for useful discussion and feedback. 
BF acknowledges support from Airbus and EPSRC grant code EP/P510427/1.
The authors acknowledge support from the Engineering and Physical Sciences Research Council (EPSRC), Programme Grant EP/L024020/1, from the European project QuCHIP.
A.L. acknowledges fellowship support from EPSRC (EP/N003470/1).
J.G.R. acknowledges support from EPSRC (EP/M024458/1).
This work was carried out using the computational facilities of the Advanced Computing Research Centre, University of Bristol - http://www.bristol.ac.uk/acrc/.

} 

\section*{Author Contribution}
{\small
AAG, BF, SK and RS contributed equally to this work.
RS, AAG, and NW conceived the methodology. BF, AAG, and RS performed simulations with support from NW, SP and CG. SK build the set-up and performed the experiments under guidance of JGR. AAG, RS, BF and SP analysed and interpreted the data with support from NW, SK and CG. RS, AAG, BF, SK, NW, SP, CG and AL wrote the manuscript. RS and AL supervised the project. 
}



\clearpage
\onecolumngrid


\renewcommand \thesection{\arabic{section}}
\renewcommand \thesubsection{\arabic{section}.\arabic{subsection}}
\renewcommand*\thefigure{SM\arabic{figure}}  
\renewcommand*\thetable{SM\arabic{table}}  
\renewcommand{\theequation}{SM\arabic{equation}}
\setcounter{figure}{0}    
\setcounter{equation}{0}

\section*{Supplementary Material}

\section{Additional details about the QMLA protocol}
\subsection{Concepts}\label{supp_concepts}
The concepts referred to in the description of  \gls{qmla} are as follows. 

\begin{easylist}[itemize]
    
    & \italicbold{Models}: individual Hamiltonian models \hj, e.g. $\hat{S}_{xy} = \alpha_x \hat{\sigma}_{x} + \alpha_y \hat{\sigma}_{y}$, $\hat{S}_{y}\hat{A}_x = \alpha_y \hat{\sigma}_{y} + \alpha_{xx}\hat{\sigma}_{x}\hat{\sigma}_x$. Models are trained through \gls{qhl}, described later. 
    
    & \italicbold{Directed Acyclic Graphs}: A structural DAG graph maintains structural information about the \gls{qmla} instance. Nodes represent models after undergoing QHL (\hjp). Edges represent 
    the relationships (parent/child) between those models. 
    A comparitive DAG also represents models as nodes, and edges between nodes $(i,j)$ give the pairwise \gls{bf} \bij.

    & \italicbold{Layers}: In the \gls{sdag}, a group of models are held in a layer $\mu$. Layers usually consist of similar models, e.g. differing only by a single term. In Fig.\ref{fig:Fig2}f (main text), for instance the fourth layer $\mu^4 = \{ \hat{S}_{xyz}\hat{A}_x, \hat{S}_{xyz}\hat{A}_y, \hat{S}_{xyz}\hat{A}_z \}$.
    
    & \italicbold{Consolidation}: Comparison between any set of models $\mathbb{H}_k = \{ \hat{H}_k \}$, achieved by computing the pairwise \gls{bf} $B_{ij} \forall \hi, \hj \in \mathbb{H}_k$. Each pairwise comparison for \hj gains that model a point; after all comparisons models have a score $\hj : s_j$. Consolidation can result in a ranking of models in $\mathbb{H}_k$, or else the highest ranked model alone.  When consolidation is performed on a layer $\mu$, and only the highest ranked model is returned, this results in selection of a \italicbold{layer champion}, \huc.
    
    & \italicbold{Pruning}: disactivation of a model, i.e. pruning it from the graph. Following consolidation within layer $\mu$, all models apart from  $\hat{H}_C^{\mu}$ are pruned. 
    & \italicbold{Exploration} and \italicbold{Spawning}: \gls{qmla} explores the space of models by trying various combination of the primitive terms. In particular this is done by spawning new models when needed, given a \italicbold{seed} model. For instance models to be placed on a new layer $\nu (=\mu +1)$ are spawned where the seed is the previous layer champion $\hat{H}^{\mu}_C$. Spawning follows user defined growth rules, though typically involves \italicbold{inheritance} of the features of the seed, for instance adding a single interaction to the seed. Seed models are considered the \italicbold{parent} of those generated, the \italicbold{child}, and such relationships are recorded in the \gls{cdag}. 
    
    & \italicbold{Primitives}: a set of base models used to generate the models explored by \gls{qmla}. Typically these are some physically interpretable, basic set of operators. These can vary throughout the exploration, depending on the user-defined growth rules. In this work we use $\mathbb{H}_0 = \{ \ssx, \ssy, \ssz \}$. These can be added and tensor-producted together to construct different models.
  
    & \italicbold{Greedy search}: A set of primitive terms, $\{ \hat{h}_0, \hat{h}_1, \dots, \hat{h}_n\}$, are chosen for the exploration stage , each corresponding to an action of (or interaction with) the electron spin. In greedy search, the spawn rule is that the terms present in a given seed model, usually the previous layer champion \huc, are retained, and the new models are $\{ \huc+\hat{h}_0, \huc+\hat{h}_1, \dots, \huc+\hat{h}_n\}$. These models then constitute layer $\nu$, are trained and consolidated, and a layer champion is selected, $\hat{H}^{\nu}_C = \huc+\hat{h}_k$. Then, $\hat{h}_k$ is removed from the available terms, and the process iterates until no terms remain available. In this sense, the algorithm is greedily consuming all the available terms one by one.
    
    & \italicbold{Growth rules}: The precise manner in which seed models and primitives are combined to generate a new set of models, defining which models are proposed at each stage of the exploration. This is defined by the user of the protocol. In this case the \gls{gr} proceeds as follows:
        && Spin rotation terms are adopted as primitives: $\primitives = \mu^0 = \{ \hat{S}_x, \hat{S}_y, \hat{S}_z\}$ (where $\hat{S}_i$ are the Pauli operators).
        && Terms are added \italicbold{greedily} in separate layers until exhausted
        && Hilbert space dimension is increased by introducing hyperfine terms $\{\hat{A}_{x} (=\ssx\otimes\ssx), \hat{A}_y, \hat{A}_z\}$ (also added individually by layer until exhaustion)
        && transverse terms $\{\hat{T}_{xy} (=\ssx\otimes\ssy), \hat{T}_{xz}, \hat{T}_{yz}\}$ are introduced and added greedily until exhaustion.
    & \italicbold{Termination rule}: the \gls{gr} includes a definition of a function to test whether \gls{qmla} should not consider any further models.

    & \italicbold{Layer collapse}: After \gls{qmla} has terminated the exploration stage, so no further models will be entertained, and each layer has selected its layer champion, a preliminary inter-layer championship is conducted. This involves calculation of \gls{bf} for champions of neighbouring layers, i.e. between parent $\mu$ and child $\nu$: $B_{\mu\nu}$. If $B_{\mu\nu} \gg 1 \ (\ll 1)$, i.e. there is strong evidence in favour the parent (child), so the losing model $\nu$ ($\mu$) is pruned. As this was the sole surviving model, $H_C^{\nu}$ ($H_C^{\mu}$) on the corresponding layer, that layer,  $\nu$ ($\mu$) is pruned entirely, i.e. the layer collapses. 
\end{easylist}

\subsection{Protocol Steps}\label{supp_steps}

The steps of the protocol are then, informally,
\begin{easylist}[itemize]
    & Initialise a \gls{sdag}. On the first layer impose the primitives alone, i.e. $\primitives = \mu^0 = \{ \hat{S}_x, \hat{S}_y, \hat{S}_z\}$
    & Iteratively (denoting the most recently added layer as $\mu$):
    && Perform \gls{qhl} on all $\hj \in \mu$
    && Consolidate $\mu$ to prune models and select layer champion $\hat{H}_C^{\mu}$. This involves adding edges in the \gls{cdag} within $\mu$. 
    && Explore: Using $\hat{H}_C^{\mu}$ as the seed model, use the growth rule $R$ and primitives \primitives to generate a new set of models, $R(\hat{H}_C^{\mu})$; assign these to a new layer of the \gls{sdag}. 
    && if the termination function determines that the exploration is complete, terminate iterations. 
    & Collect all layer champions into a champion set $\mathbb{H}_C$. Perform comparisons only between parent/children pairs within $\mathbb{H}_C$, pruning under-performing models (resulting in layer collapse). The result is a reduced $\mathbb{H}_C^{\prime}$.
    & Consolidate $\mathbb{H}_C^{\prime}$ to find the global champion model, \hp. 

\end{easylist}
\par

Primitives are staged in three distinct sets of \emph{sub-primitives}: $\primitives\{ \mathbb{H}_1=\{ \Sx,\Sy\Sz\}, \mathbb{H}_2 = \{ \hat{A}_x, \hat{A}_y, \hat{A}_z \}, \mathbb{H}_3 = \{ \hat{T}_{xy}, \hat{T}_{xz}, \hat{T}_{yz} \}  \}$.
Likewise, the spin terms are explored first: $\mathbb{H} \gets \mathbb{H}_S$.
First layer are the raw electron spin rotation terms, $\mathbb{H}_S \gets \mu$.

The spawn rule $R$ used in this work, for a given \huc, proceeds as follows:
\begin{easylist}
    & For $\hat{h} \in \mathbb{H}$
    && $\hat{H}^{\nu} \gets \huc + \hat{h}$
    && if $\mathbb{H}$ exhausted (i.e. all terms in $\mathbb{H}$ are present in \huc), switch $\mathbb{H}$ to the next set of sub-primitives. 
\end{easylist}
\par

The termination rule, $f_T$ is then simply whether $\mathbb{H}_3$ has been exhausted. 

Collectively, $R$, $f_T$ and \primitives are called the growth rules for a given QMLA instance. These can be decided by the user and altered simply in the QMLA Python framework, to include more complex models and data from any system. 

For formal pseudo code of this protocol, see \S\ref{supp_pseudocode}.

\subsection{Quantum Hamiltonian Learning}\label{supp_qhl}
\gls{qhl} has been studied since 2012, initially proposed by Granade et al.~\cite{Granade:2012kj,Wiebe:2014qhl,wiebe2014qhlpra}, and demostrated experimentally by Wang et al.~\cite{wang2017qhlexp}. Here we reiterate the protocol briefly, but for complete details a reader should see the above references. In particular, \gls{qmla} uses the \gls{smc} updater provided by Qinfer. Pseudocode for this routine is provided in Alg. \ref{qmd_alg}.

Quantum systems dynamics are predicted by their Hamiltonian matrix, $\hat{H}$. In particular, a quantum state, $\ket{\psi}$ of a particular system evolves according to its unitary operator, Eqns \ref{unitary}-\ref{unitary_ev}.

\begin{equation}\label{unitary}
    U = e^{-i\hat{H}t}
\end{equation}

\begin{equation}\label{unitary_ev}
    \ket{\psi} \longrightarrow U\ket{\psi}
\end{equation}

Upon measurement, quantum systems collapse with expected value according to Eqn. \ref{expected_value}

\begin{equation}\label{expected_value}
    E = \left| \bra{\psi} U \ket{\psi} \right|^2
\end{equation}

Quantum Hamiltonian Learning (QHL) is the process of learning Hamiltonian parameters of an unknown quantum system by interfacing with a classical machine learning protocol, Bayesian inference. For example, the spin of an electron from an Nitrogen Vacancy centre, which has been characterised, is known to have a Hamiltonian which depends on its Rabi frequency, $\Omega$, Eqn. \ref{nv_centre_spin}

\begin{equation}\label{nv_centre_spin}
    \hat{H} = \frac{\Omega}{2} \pauliz = \frac{\Omega}{2} \ssz
\end{equation}

Hence by learning Hamiltonians, we infer physical parameters of quantum systems. 

\subsubsection{Likelihood Estimation}
In order to learn Hamiltonian parameters we employ a form of approximate Bayesian Inference known as sequential Monte-Carlo Methods or particle filtering. This process iteratively improves the probability distribution over the parameter space. That is, it starts from some initial knowledge of the system, e.g. uniform probability distribution for all values of $\Omega \in [0,1]$. Call this probability distribution $P$. \par

$P$ is sampled from to obtain a value of $\Omega_1$. This informs a hypothesis, i.e. that the system is governed by $\hat{H} = \Omega_1 \ssz$. This hypothesis is simulated for a time $t$ and some initial state $\ket{\psi}$, determining an expectation value $E_1(t | \ket{\psi})$ according to Eqn. \ref{expected_value}. The quantum system is prepared in the same state $\ket{\psi}$, and let evolve physically for $t$, to determine the true, physical expectation value $E_0(t | \ket{\psi})$. The true and simulated expectation values are compared, computing the likelihood that $\Omega_1=\Omega_0$. 

\begin{figure}
    \centering
    \includegraphics[scale=0.3]{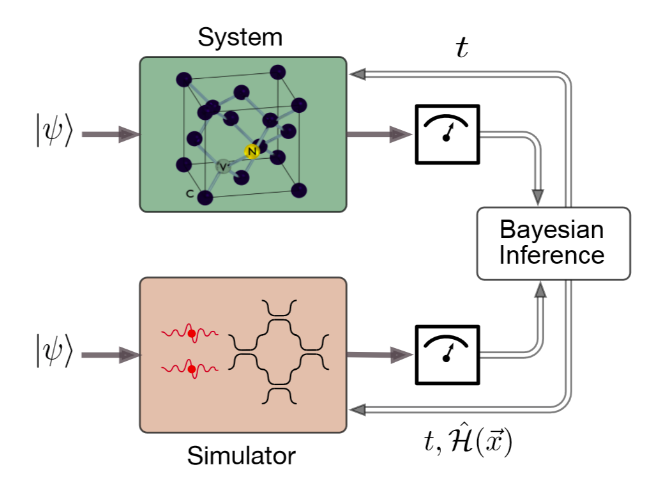}
    \caption{Quantum Hamiltonian Learning: A quantum system (NV--centre electron spin) is evolved, and a proposed Hamiltonian is run on a quantum simulator. The true and simulated outputs are compared using Bayesian inference, leading to improved probability distributions over the parameter space. Image taken from \cite{wang2017qhlexp}.}
    \label{qhl_fig}
\end{figure}

Bayes' Rule, Eqn. \ref{bayes_rule}, describes the probability that a hypothesis is related to an observation. We apply Bayes' rule to relate the expected value of a quantum system to a \emph{likelihood} function, which is used to update probability distributions over parameter space. 

\begin{equation}\label{bayes_rule}
    Pr(\vec{x} | D; t) = \frac{Pr(D| \vec{x}; t) \ Pr(\vec{x})}{Pr(D|t)}    
\end{equation}
In Eqn. \ref{bayes_rule}, $\vec{x}$ is the parameter vector forming the hypothesis (termed $\Omega$ in our example above); $Pr(\vec{x} | D; t)$ is the probability distribution of model parameters given the experimental data, which we can interpret as the probability $\vec{x}=\vec{x_0}$; $Pr(D |\vec{x} ; t)$ is the likelihood function; $ Pr(\vec{x})$ is the prior distribution and $Pr(D|t)$ is a normalisation factor. \par 
Upon sampling from the current parameter probability distribution, we encode the corresponding $H(\vec{x})$, either on a quantum simulator such as a silicon chip, or a simulation of a quantum system (i.e. compute Eqn. \ref{expected_value} analytically). 
This gives us the term $Pr(D|\vec{X}_i; t)$ for a single $\vec{x}_i$. We sub this into Eqn. \ref{bayes_rule} to find the likelihood, $Pr(\vec{x}_i|D;t)$ corresponding to that $\vec{x}_i$. 
The likelihood is a value between 0 and 1 which reflects the likelihood that the chosen $\vec{x}_i$ is the correct value, given the data observed. \par 

Once we have the likelihood of a particular $\vec{x}_i$, we can update the weight associated with that parameter vector. 
Before being chosen as a hypothesis to explain observed data, the vector had some weight, $w_i^{old}$. 
That value is now multiplied by the likelihood to determine its new weighting in the next iteration of the probability distribution over the parameter space:

\begin{equation}\label{weight_likelihood}
    w_i^{new}  = Pr(\vec{x}_i|D;t) \times w_i^{old}
\end{equation}

Each sample drawn is called a \emph{particle} in Bayesian Inference. Drawing particles (i.e. sampling) from an initial parameter distribution $N_p$ times, we perform weight updates on $N_p$ individual values.
We introduce a weight threshold, $a$ and a fraction threshold $l_{res}$: when the ratio $l_{res}$ particles have weight less than $a$, the probability distribution is resampled. This follows the Liu West algorithm~\cite{Liu:2001cz}, and here we adopt the Qinfer defaults, $a=0.98, l_{res}=0.5$. Resampling simply redraws the probability distribution according to the updated weights, effectively meaning more particles in the region of higher particle weights.

\begin{figure}
    \centering
    \includegraphics[scale=0.2]{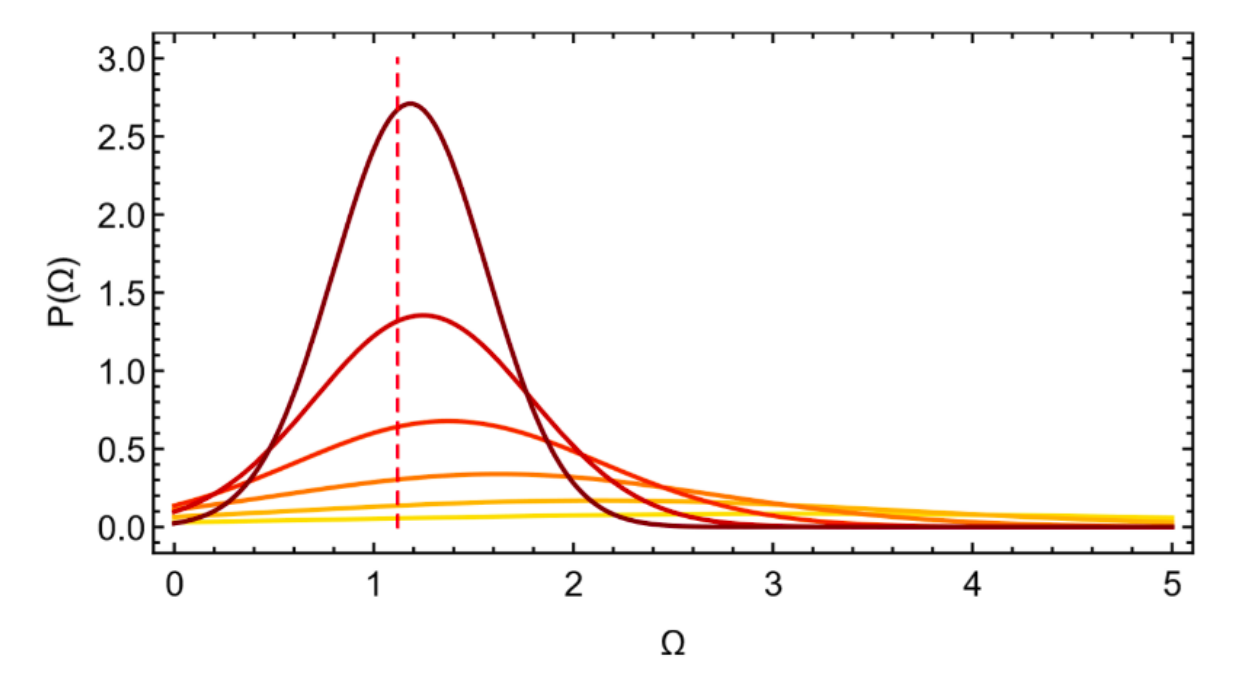}
    \caption{Evolution of parameter distribution. In yellow, initial distribution is flat across all values of $\Omega$. As the algorithm progresses (orange plots), the distribution is updating the weight of parameter values: increasing those close to the true value and decreasing those further away. Consequently, the distribution narrows around the true value (dotted red), culminating with the red distribution.}
    \label{updates}
\end{figure}

We call an iteration of $N_p$ samples is a single \emph{experiment} (or epoch), and perform $N_e$ experiments. 
In doing so, we observe the the distribution continuously narrowing around the true parameter vector, Fig.\ref{updates}. For $N$ parameters, the posterior distribution similarly converges: each dimension of the posterior corresponds to a single parameter of the Hamiltonian; the mean of each dimension gives the approximation of that parameter, which converge to $\vec{x}_0$. The uncertainty of each parameter can be estimated as the width of the corresponding parameter of the posterior distribution, $\vec{\sigma}_i$. 
When the particles are resampled (because a fraction $l{res}$ have weight less than $a$), typically the particles move closer together, converging to the particle that best represents $\vec{x}_0$. We can envision a particle cloud in high dimensional space, distributed within some \italicbold{volume} $V$; upon resampling the particles move closer together, shrinking $V$. 
We view reduction of $V$ as an indication that \gls{qhl} is learning. This effect is seen in Fig.~\ref{fig:Fig2}d. 

\subsubsection{Experiment design heuristics}\label{supp_heuristic}
Within the learning phase, \gls{qhl} is assumed to have access to the quantum system under study, and therefore be able to dictate experiments to perform in order to maximise information gained about the model. 

In this work, experiment design has two controls: input probe and evolution time. Experimentally we are constrained to the input probe $\ket{\psi} = \ket{+}\ket{+^{\prime}} = \ket{+}\frac{\ket{0} + e^{i\phi}\ket{1}}{\sqrt{2}}$. In simulation (i.e. case (i) in Analysis, main paper), we sample the input probe randomly. 
The remaining control then is the evolution time, $t$ for which to evolve the quantum system and simulate Eqn.~\ref{expected_value} to compute the likelihood for each particle in Alg.~\ref{qhl_alg}.
In order that the likelihood function returns a high value, the particle under study must mimic the system well. 
For low $t$, it is possible that this occurs coincidentally, or more likely that a coursely trained model has optimised the dominant contribution, which is more  assertive at early times (see for example Fig.\ref{fig:Fig2}e, where the spin terms' oscillatory behaviour is more strongly evident for $t<2\mu s$, and decoherence, mapped to environmental terms, are less prevelant at these times). 
Then, it is clear that the early stages of training should focus on capturing the dominant dynamics at low $t$, but to fully characterise the system, the model ought to train on higher $t$ also.

We propose a heuristic in two parts: first dynamically choose $t$ such that the model learns at a pace dictated by the information gained to date; second force the model to learn on higher times so that it must attempt to address complex dynamics. We recognise that this is not an optimal heuristic in general. 

The first phase outlined is comparable to the \emph{particle guess heuristic} provided by Qinfer. 
At each epoch of \gls{qhl} for parameterisation $\vec{x}$, two samples $\vec{x}_1, \vec{x}_2$ are drawn from $P(\vec{x})$. 
The distance is taken between these samples as the $L1$ norm, $d = \sum_{i=1}^n\abs{ (\vec{x}_1)_i - (\vec{x}_2)_i }$. $t$ is then set proportional to the reciprocal of $d$. 
Intuitively, if the volume is large, the distance between two randomly sampled particles will be large, so $t$ will be small, ensuring \gls{qhl} focuses on small $t$ when it has not yet learned the parameters to any degree of certainty. 
Then, as the volume decreases while the model learns (by resampling the posterior distribution), $d$ will proportionally decrease, so $t$ will become higher, meaning that the model is challenged to learn on higher times when it is reasonably confident of the convergence of the particles to date. 
Exponential decrease of the volume therefore corresponds to exponential increase in the times the model trains upon; or stated otherwise, if the model can reproduce dynamics for exponentially increasing $t$, we can exponentially reduce the uncertainty on the particles providing those dynamics.

\subsection{Simulations}\label{supp_simulations}
We test QMLA by simulating a chosen $\hat H_{0}$ using Phase 3 of Blue Crystal, the supercomputer of the University of Bristol, composed of 223 nodes each made of 16 $\times$ 2.6 GHz SandyBridge cores. The use of a supercomputer is justified by the need of running a large number of independent instances of the protocols, in order to have good estimation of the average performance. These simulations retain full control over the correct model form and parameters, and no experimental limitation due to setup performances, input probe states and noise. Simulations are run in Python through the \gls{qmla} framework, which relies on Qinfer~\cite{granade2017qinfer}.

QMLA instances can be run in parallel, and within individual instances, QHL on distinct models can be run in parallel, as well as BF calculations. The  bottleneck is the calculation of $|\bra{\psi} e^{-i \hat{H}^{\prime} t} \ket{\psi}|^2$ due to the exponentiation of $\hat{H}^{\prime}$. Trusted quantum simulators could compute these quantities in polynomial time as opposed to the exponential requirement doing so classically. \par 
In this work we focus on 2-qubit Hamiltonians, represented by matrices of dimension $4\times4$, whose exponentiation typically takes $t_H \sim 0.5 ms$ using \texttt{scipy.linalg.expm} in Python. QHL involves sampling of $N_P$ particles in a single epoch, over $N_E$ epochs, so each QHL instance requires $N_PN_E$ Hamiltonian exponentiations. Pairwise BF calculations involve a further $N_e$ epochs on each model in the pair, i.e. a cost of $2N_EN_P$ Hamiltonian exponentiations. 
After all layer champions are determined, a parental collapse rule involves pairwise BF between parent/child models, i.e. $N_{\mu}-1$ BFs. After parental collapse, there are a final set of $N_C \leq N_{\mu}$ layer champions, from which the overall champion is to be determined. 
The number of layers $N_{\mu}$, number of models per layer $N_m(\mu)$ (and corresponding number of BF per layer $N_{BF}(\mu)$), and number of available processes $p$ give the expected runtime for a single QMLA instance, Eqn. \ref{qmla_expected_time}

\begin{equation}
    \label{qmla_expected_time}
    \begin{split}
    T \sim t_H \times  & \left( \sum_{\mu} \ceil*{\frac{N_m(\mu)}{p}} N_P N_E \ + \sum_{\mu} \ceil*{\frac{N_{BF}(\mu)}{p}} 2 N_P N_E \ + 2 \ceil*{\frac{N_{\mu}-1}{p}} N_P N_E \ + 
    2 \ceil*{\frac{N_{C}}{p}} N_P N_E 
    \right) 
    \end{split}
\end{equation}

In the studied case, $t_H=0.5ms, N_{\mu}=9, N_E=1000, N_P=3000, p=6$ ($N_m(\mu), N_{BF}(\mu)$ can be read from the DAG in Fig. \ref{fig:Fig2}f), resulting in $T \sim 20 hrs$. Increasing $N_P, N_E$ typically improves the learned parameterisations for each model which should lead to stronger models winning more frequently, at the price of increasing the computational time required for the \gls{qmla} instance to run.

\subsection{Results}
Here we offer a discussion on the outcomes of the \gls{qmla} instances reported in the main text.  

\subsubsection{Parameter Inference}
\begin{figure}
    
    \centering
    \begin{tabular}{@{}c@{}}
        \centering
        \includegraphics[width=0.7\textwidth]{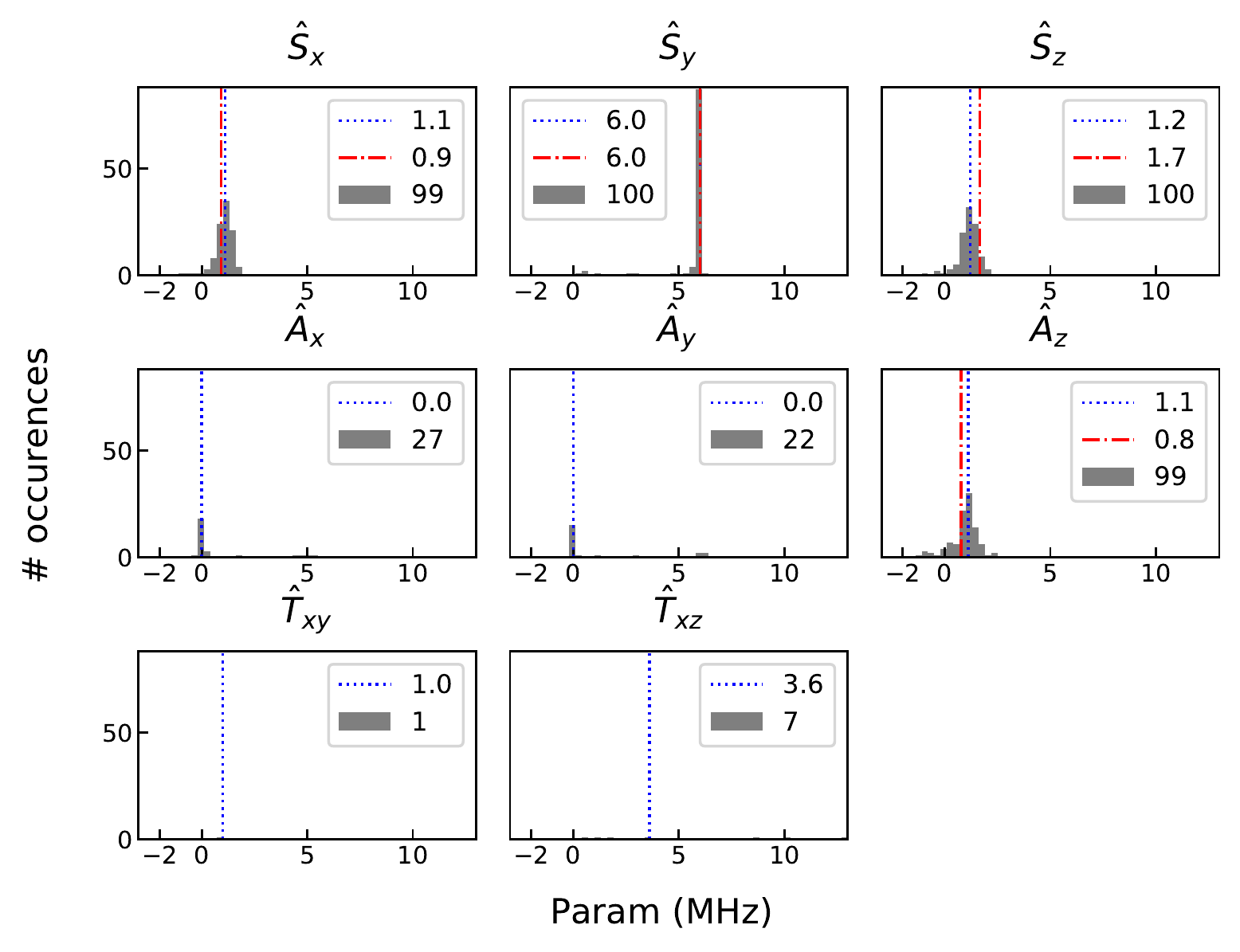}
    \end{tabular}
    \\ \small \textbf{a,} Simluated QMLA instances. Red dotted lines show the true parameters.
    
    \centering
    \begin{tabular}{@{}c@{}}
        \centering
        \includegraphics[width=0.7\textwidth]{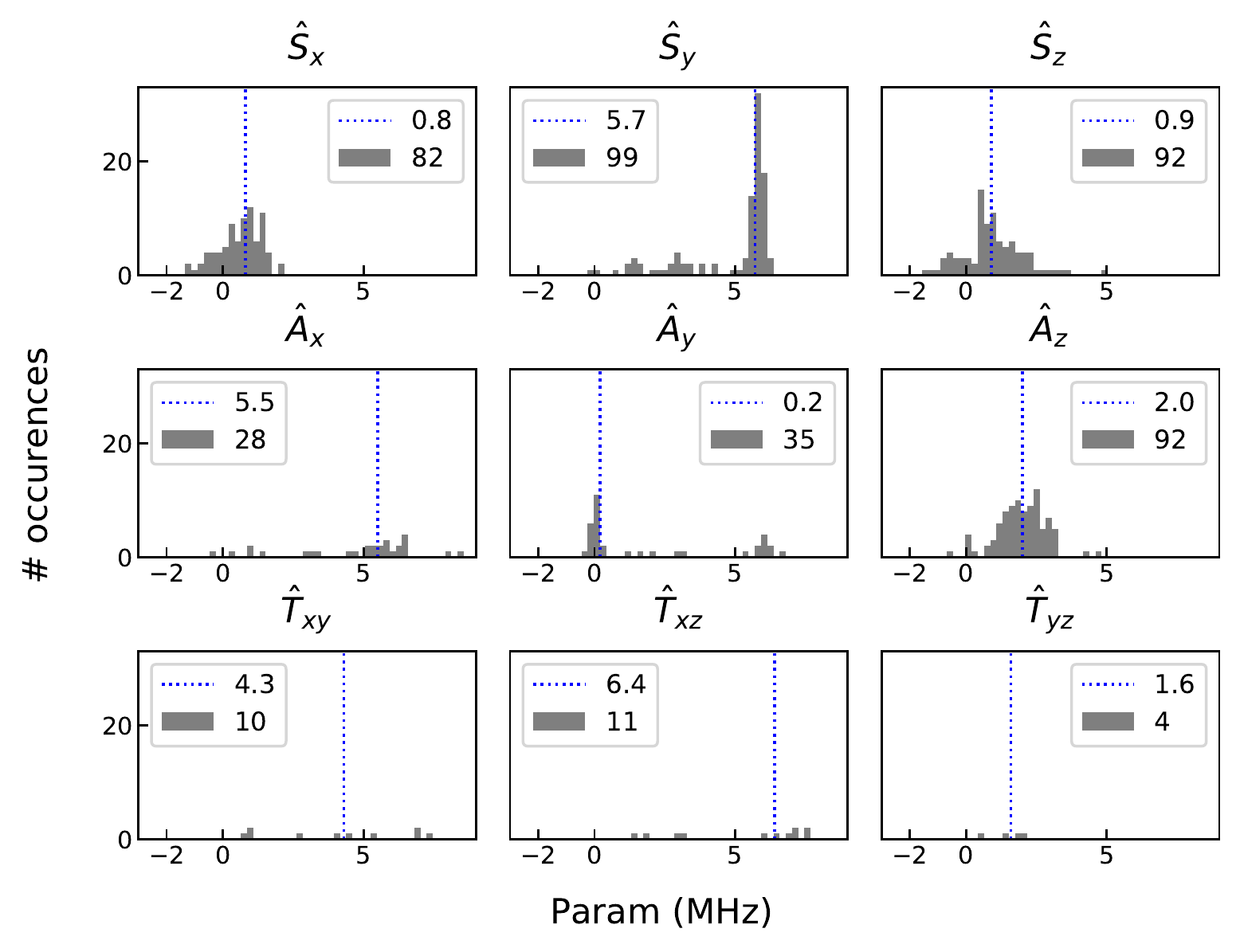}
    \end{tabular}
    \\
    \small \textbf{b,} Experimental QMLA instances.
    
    \caption{Histograms for parameters learned by champion models on (\textbf{a}) simulated  and (\textbf{b}) experimental data. Blue dotted lines indicate the median for that parameter; grey blocks show the number of models which found the parameter to have that value and the number listed in the legend reports how many champion models contained that term.
    }
    \label{learned_params_fig}
\end{figure}

Fig.~\ref{learned_params_fig} shows the values of the parameters learned by each champion model for case \textit{ii} and \textit{iii} listed in Analysis, as well as the true parameters for the simulated case. 
These also allow us to see how many champion models contain each term. 
The combination of the frequency of terms and the confidence in parameter values allows us to draw conclusions about the physics of the NV centre electron spin. 

For instance, Fig.~\ref{learned_params_fig}b shows, from \Sy, that the spin rotates in the $y-$axis with a frequency of $\sim5.7MHz$. 
We can likewise infer coupling with the environment, albeit with less precision, e.g. coupling with the $z-$axis in $\{0,4MHz\}$.

We perform one final step once \hp is nominated by a given \gls{qmla} instance. 
If some parameters probability distribution are found to be within one standard deviation from 0, \gls{qmla} proposes a reduced model, $\hat{H}^{\prime}_{r}$, which inherits all the model terms and parameters from \hp, ignoring those with seemingly negligible contribution. 
We then compute the Bayes factor, $\mathcal{ B}(\hat{H}^{\prime}_{r}, \hp)$; if this strongly favours the reduced model ($\mathcal{B}>100)$, we prefer the simpler model, resetting $\hp \gets \hat{H}^{\prime}_r$. 
The threshold is stringent enough that any non-negligible contribution which improves the predictive power will not be discarded, though interestingly in many cases \gls{qmla} favours over-fit models, where the over-fitting terms are found to have almost negligible parameters.
In Fig.~\ref{learned_params_fig}a, for instance, terms $\hat{A}_x, \hat{A}_y$ are preserved in 22 and 29 champion models, with median parameter 0.
This offers insight to the user: while the found \hp \ is not completely accurate, by considering the parameters assigned to the terms present, the user can understand the significance of those terms, and whether they are impacting their system in a meaningful capacity. 

\subsubsection{Dynamics Reproducibility}
We adpopt the coefficient of determination $R^2$ as the figure of merit indicating how well a given model reproduces dynamics generated from \ho. 
In Table~\ref{r_squared_table} we list the number of times given models won QMLA instances. 

\begin{table}
    \centering
    \begin{tabular}{|c|lr|lr|}
    \hline
     Model & \multicolumn{2}{|c|}{Experiment} & \multicolumn{2}{|c|}{Simulation}    \\
     & Wins   & $R^2$   & Wins   & $R^2$   \\
    \hline
     $\hat{S}_{y,z}\hat{A}_{z}$                      & 9      & 0.8     & 1      & 0.26    \\
     $\hat{S}_{y}\hat{A}_{x,z}$                      & 2      & 0.63    &        &         \\
     $\hat{S}_{x,y,z}\hat{A}_{z}$                    & 45     & 0.86    & 61     & 0.97    \\
     $\hat{S}_{x,y,z}\hat{A}_{y}$                    &        &         & 1      & -0.54   \\
     $\hat{S}_{x,y,z}\hat{A}_{x,y}$                  & 3      & 0.81    &        &         \\
     $\hat{S}_{x,y,z}\hat{A}_{y,z}$                  & 14     & 0.83    & 10     & 0.96    \\
     $\hat{S}_{x,y,z}\hat{A}_{x,z}$                  & 6      & 0.64    & 15     & 0.99    \\
     $\hat{S}_{x,y,z}\hat{A}_{x,y,z}$                & 2      & 0.72    & 5      & 0.97    \\
     $\hat{S}_{x,y,z}\hat{A}_{x,z}\hat{T}_{xz}$      &        &         & 1      & 0.68    \\
     $\hat{S}_{x,y,z}\hat{A}_{x,y,z}\hat{T}_{xz}$    &        &         & 5      & 0.77    \\
     $\hat{S}_{y}\hat{A}_{x,y,z}\hat{T}_{xy,xz,yz}$  & 2      & 0.31    &        &         \\
     $\hat{S}_{x,y,z}\hat{A}_{x,y,z}\hat{T}_{xy,xz}$ & 4      & 0.67    & 1      & 0.32    \\
    \hline
    \end{tabular}
    \caption{QMLA results for experimental data and simulations. We state the number of QMLA instances won by each model and the average $R^2$ for those instances as an indication of the predictive power of winning models.}
    \label{r_squared_table}
\end{table}

\begin{figure}
    \centering
    \begin{tabular}{@{}c@{}}
        \includegraphics[width=0.9\textwidth]{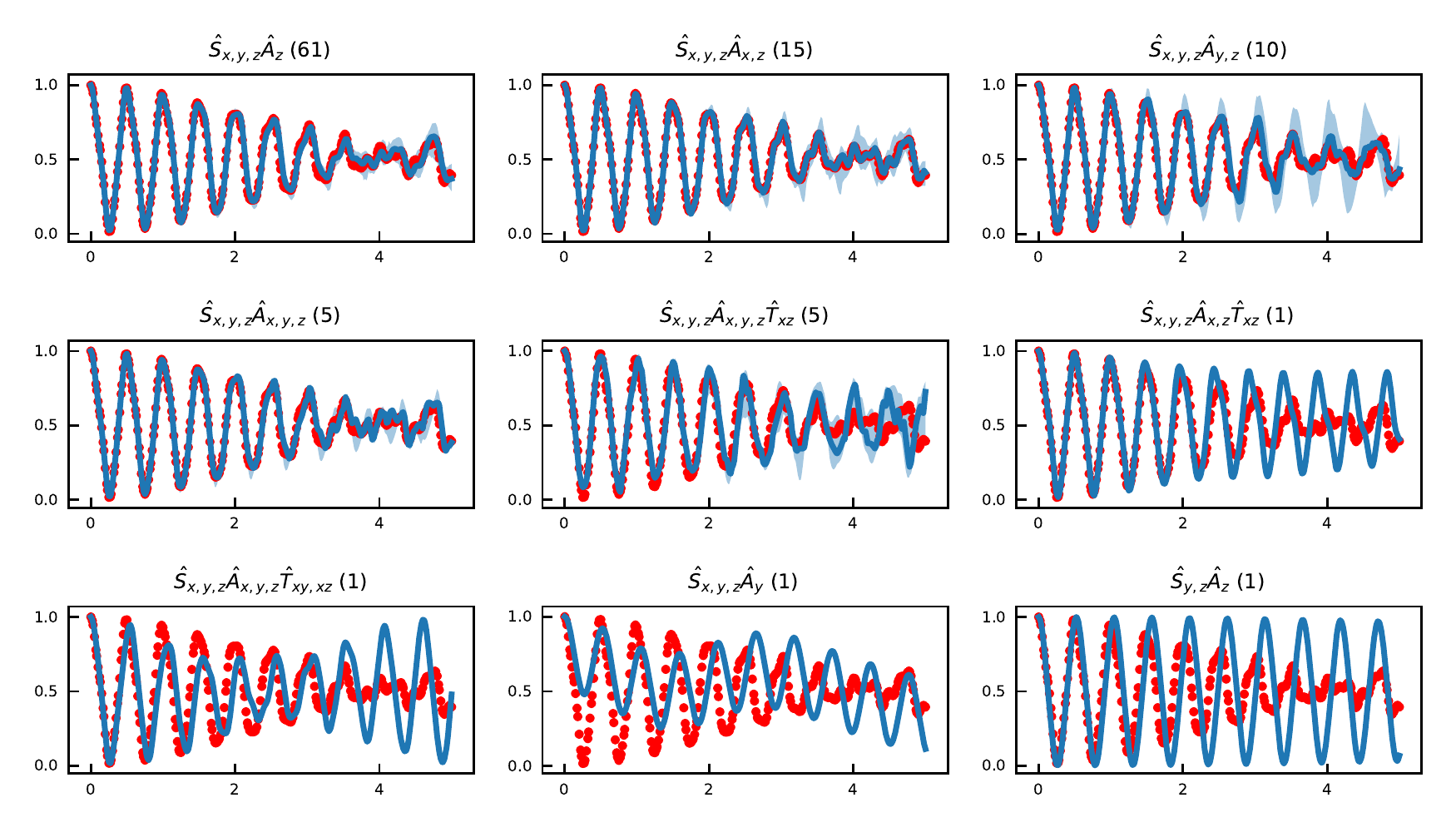}
    \end{tabular}
    \\ \small \textbf{a,} Simulated data

    \centering
    \begin{tabular}{@{}c@{}}
        \includegraphics[width=0.95\textwidth]{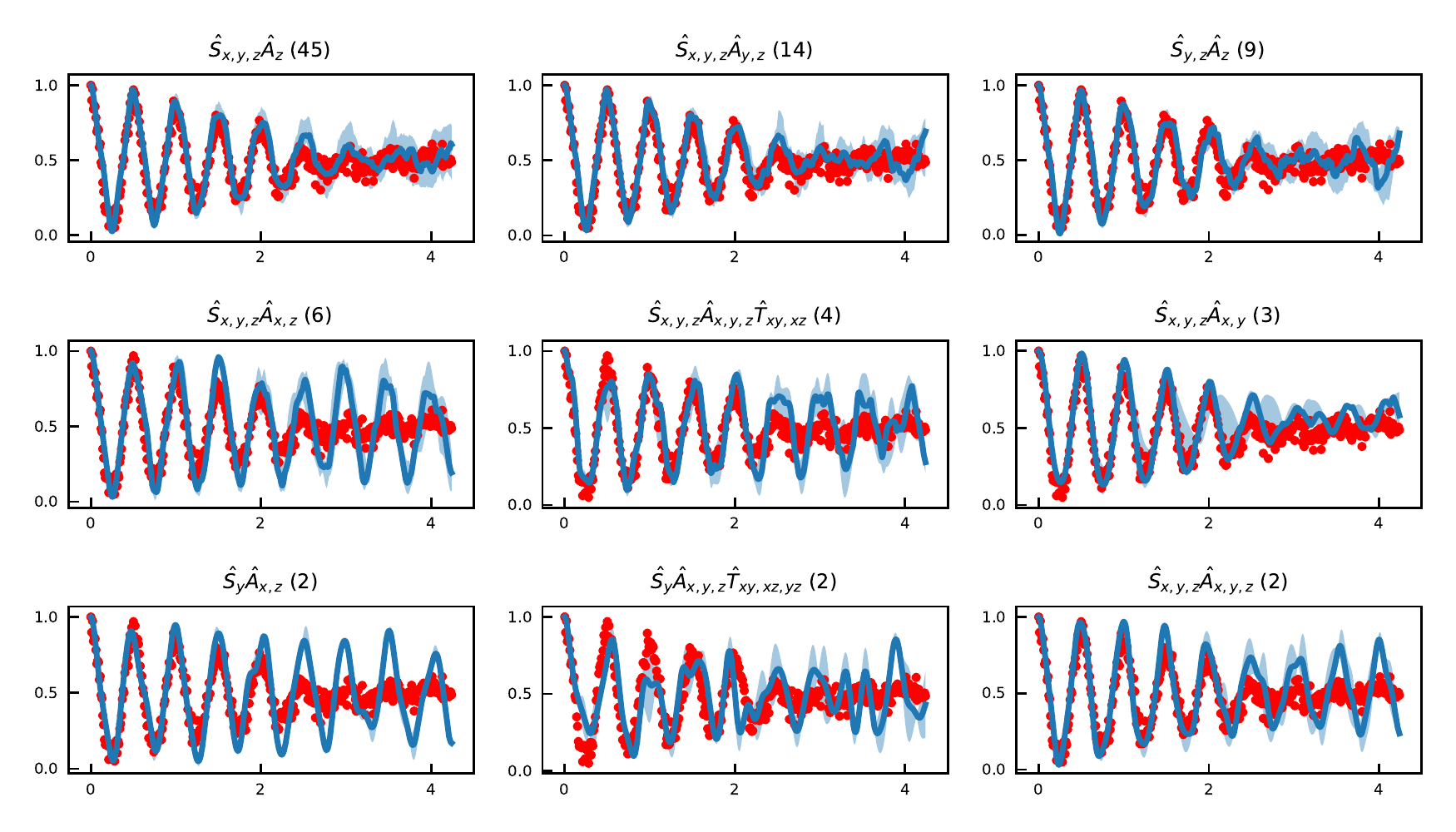}
    \end{tabular}
    \\ \small \textbf{b,} Experimental data

    \caption{Dynamics reproduced by various champion models for simulation and experiment. Expectation values are shown on $y$--axis with time on $x$--axis. Red dots give the true dynamics of \ho, while the blue lines show the reconstruction by \hp. \ \hp \ is listed on top of each plot, with the number of QMLA instances that model won in brackets.}
    \label{model_dynamics_plot}
\end{figure}

In Fig.~\ref{model_dynamics_plot}, we present the true (\ho) and reproduced dynamics (\hp) from the models listed in Table~\ref{r_squared_table} for simulated and experimental data. 
We see that, in most cases, the nominated champions models can strongly emulate the underlying system.

\subsection{Pseudocode}\label{supp_pseudocode}
We provide pseudocode for the main QMLA routine, Alg.~\ref{qmd_alg}, as well as several key subroutines for \gls{qhl}, Alg.~\ref{qhl_alg}, and the calculation of Bayes factors, Alg.~\ref{bf_alg}. Subroutines for consolidating models are given in Alg.~\ref{consolidate_alg},~\ref{parental_consolidation_alg}.
Some functionality is assumed within the pseudocode, in particular operations on the \gls{dag} to extract information such as parentage, or to prune models, though these cases are self explanatory from the given function names. 

\newcommand{\codecomment}[1]{\tcp*[1]{#1}}

\begin{algorithm}
    \caption{Quantum Model Learning Agent}
    \label{qmd_alg}

    \DontPrintSemicolon
    \KwIn{\ttt{DAG} \tcp*[1]{Directed acyclic graph, with structural and comparitive components}}
    \KwIn{\ttt{QHL()} \tcp*[1]{callable function, Alg. \ref{qhl_alg}}}
    \KwIn{\ttt{BF()} \tcp*[1]{callable function to compute the Bayes factor between $\hat{H}_j$ and $\hat{H}_k$, Alg. \ref{bf_alg} }}    
    \KwIn{\ttt{Consolidate()} \tcp*[1]{callable function, Alg. \ref{consolidate_alg}}}
    \KwIn{$Q, N_E, N_P, f(P(\vec{x}))$  \tcp*[1]{as in QHL}}
    \KwIn{$P(\vec{x}) \ \forall \vec{x} \in \vec{X}$ \tcp*[1]{probability distribution for any parameterisation $\vec{x}$ within the space of valid Hamiltonian parameterisations $\vec{X}$.}}

    \KwIn{$\mathbb{H}_0$ \tcp*[1]{set of primitive models from which to construct Hamiltonian models}}
    \KwIn{$R$ \tcp*[1]{rule(s) by which to generate new models }}
    \KwIn{$f_{T}$ \tcp*[1]{ function to check whether to terminate QMLA }}

    \KwOut{$\hat{H}^{C}$ \tcp*[1]{Approximate Hamiltonian model form}}\;
    
    $\mathbb{H}^c \gets \{\}$ \tcp*[1]{Initialise champion model set}
    $\mu \gets 0$ \tcp*[1]{First layer}
    $\mathbb{H}^{\mu} \gets R(\mathbb{H}_0)$ \tcp*[1]{Get first batch of models from growth rule and primitives}\;
    \While{
        \ttt{$f_{T}$(DAG) == False}
    }{
        
        \If{   
            $\mu$ \ttt{!= 0} 
        }{
            $\mathbb{H}^{\mu} \gets R(H_C^{\mu -1})$
            \tcp*[1]{Generate new set of models using previous layer champion as seed}
        }{}
        \For{
            $\ham_k \in \mathbb{H}^{\mu}$
        }{
            \ttt{QHL}$(Q, \hat{H}(\vec{x}_k), P(\vec{x}_k), N_E, N_P) \gets \vec{x}_k^{\prime}$  \tcp*[1]{optimal parameter set for model $\hat{H}_k$}
            $\hat{H}_k \gets \hat{H}(\vec{x}_k^{\prime})$
            
            \ttt{DAG.update($\hat{H}_k$)}
            \tcp*[1]{Update learned model on DAG within QMLA class}
        }
  
        $\hat{H}^{\mu}_C \gets$ \ttt{Consolidate($\mathbb{H^\mu}$)}
        
        $\mathbb{H}^C \gets \{\mathbb{H}^C, \hat{H}_C^{\mu}\}$ \tcp*[1]{Add layer champion to champion set} 

        $\mu \gets \mu + 1 $
    }
    $\mathbb{H}^C \gets $ \ttt{parentalConsolidation(DAG)}
    
    $\hat{H}^{C} \gets$ \ttt{Consolidate($\mathbb{\hat{H}^C}$)} \tcp*[1]{Assign global champion model as champion of champion model set}\;
    
    return $\ham^C$ \tcp*[1]{Return champion model}\;
\end{algorithm}

\begin{algorithm}
    \caption{Quantum Hamiltonian Learning}
    \label{qhl_alg}
    \DontPrintSemicolon
    \KwIn{ $Q$ \tcp*[1]{some physically measurable or simulatable quantum system to study.}}

    \KwIn{$\hat{H}^{\prime}$ \tcp*[1]{Hamiltonian model with which to  attempt reconstruction of data from $\hat{H}_0$ }}
    \KwIn{$P(\vec{x})$ \tcp*[1]{probability distribution for $\vec{x} = \vec{x}_0 $}}
    \KwIn{$N_E$ \tcp*[1]{number of epochs to iterate learning procedure for}}
    \KwIn{$N_P$ \tcp*[1]{number of particles (samples) to draw from distribution at each epoch }}
    \KwIn{\ttt{RS()} \tcp*[1]{Resampling algorithm with which to redraw probability distribution}}
    \KwOut{$\vec{x}^{\prime}$ \tcp*[1]{estimate of Hamiltonian parameters}}\;

    Sample $N_P$ times from $P(\vec{x}) \gets $ particles\;
    
    \For{$ e \in  \{1 \rightarrow N_E\}$ }{ 
        
        Sample $x_1, x_2$ from $P(\vec{x})$
        
        $t \gets \frac{1}{\left|| x_1 - x_2 \right||}$
        
        \For{
            $p \in \{ 1  \rightarrow N_P\}$ 
        }{
            Retrieve particle $p \gets \vec{x}_p$
            
            Measure Q at $t \gets d$ \tcp*[1]{datum}
            
            $| \bra{\psi} e^{-i\ham(\vec{x}_p)t} \ket{\psi} |^2 \gets E_p$ \tcp*[1]{expectation value}
            
            $Pr(d|\vec{x}_p; t) \gets l_p$ \tcp*[1]{likelihood, computed using $E_p$ and $d$}
            
            $w_p \gets w_p \times l_p $ \tcp*[1]{weight update}
        }
        \If{
            $ 1 / \sum_p w_p^2 < N_P/2$
            \tcp*[1]{check whether to resample (are weights too small?)}
        }{
            $P(\vec{x}) \gets \ttt{RS}(P(\vec{x}))$
            \tcp*[1]{Resample according to provided resampling algorithm}
            Sample $N_P$ times from $P(\vec{x}) \gets $ particles
        }
    }
    
    $\ttt{mean}(Pr(\vec{x})) \gets \vec{x}^{\prime}$\;
    
    return $\vec{x}^{\prime}$
     
\end{algorithm}

\begin{algorithm}
    \caption{Bayes Factor calculation}
    \label{bf_alg}
    \DontPrintSemicolon
    \KwIn{ $Q$ \tcp*[1]{some physically measurable or simulateable quantum system to study.}}
    \KwIn{ $\hat{H}_j^{\prime}, \hat{H}_k^{\prime}$ \tcp*[1]{Hamiltonian models after QHL (i.e. $\vec{x}_j, \vec{x}_k$ already trained), on which to compare performance.}}
    \KwIn{ $T_j, T_k$ \tcp*[1]{set of times on which $\hat{H}_j^{\prime}$ and $\hat{H}_k^{\prime}$ were trained during QHL, respectively.}}
    
    \KwOut{$B_{jk}$ \tcp*[1]{Bayes factor between two candidate Hamiltonians; describes the relative performance at explaining the studied dynamics.}}

    $T = \{ T_j \cup T_k $\} \;
    \For{$\hat{H}_i^{\prime} \in  \{\hat{H}_j^{\prime}, \hat{H}_k^{\prime}$ \} }{
        $l(D | \hat{H}_i) = 0$ \tcp*[1]{total log-likelihood of observing data $D$ given  $\hat{H}_i$} 
        \For{$t \in T$}{
            Measure Q at $t \gets d$ \tcp*[1]{datum for time $t$}
            
            Compute $| \bra{\psi} e^{-i \hat{H}_i^{\prime} t} \ket{\psi} |^2 \gets E_p$  \tcp*[1]{expectation value for $\hat{H}_i^{\prime}$ at $t$} 
            
            Compute $Pr(d | \hat{H}_i, t)$ \tcp*[1]{from Bayes' rule, using $d, E_p$}
            
            $log \left( Pr(d | \hat{H}_i, t) \right) \gets l$ \tcp*[1]{likelihood of observing datum $d$ for $\hat{H}_i^{\prime}$ at $t$}
            
            $l(D | \hat{H}_i^{\prime}) + l \gets  l$ \tcp*[1]{add this likelihood to total log-likelihood }
        }
    }
    
    $exp\left( l(D|\hat{H}_j^{\prime}) - l(D|\hat{H}_k^{\prime}) \right) \gets B_{jk}$  \tcp*[1]{Bayes factor between models}\;
    return $B_{jk}$
\end{algorithm}

\begin{algorithm}
    \caption{Consolidate}
    \label{consolidate_alg}
    \DontPrintSemicolon
    \KwIn{ $\mathbb{H}$ \tcp*[1]{set of models to consolidate}}
    \KwIn{ $N$  \tcp*[1]{number of top-ranked models to return; default 1}}
    \KwIn{ $b$ \tcp*[1]{threshold for sufficient evidence that one model is stronger}}

    \KwOut{$\hat{H}_C$ \tcp*[1]{Superior model among input set}}
     
     \For{
        $\ham_j \in \mathbb{H}$
     }{
        $s_j = 0$ \tcp*[1]{Initialise score for each model} 
     }
     
    \For{
        $\ham_j, \ham_k \in \mathbb{H}$
        \tcp*[1]{pairwise Bayes factor between all models in the set}
    }{
        $B \gets \ttt{BF}( \ham_j, \ham_k)$ \;
        \uIf{$B \gg b$ \tcp*[1]{Increase score of winning model}}{
            $s_j \gets s_j+1$
        }  
        \uElseIf{$B \ll 1/b $}
        {
            $s_k \gets s_k+1$
        }
    }
    
    $ k^{\prime} \gets \max_k \{s_k\}$ \tcp*[1]{Find which model has most points}

    return $\hat{H}_{k^{\prime}}$
\end{algorithm}

\begin{algorithm}
    \caption{parentalConsolidation}
    \label{parental_consolidation_alg}
    \DontPrintSemicolon
    \KwIn{ \ttt{DAG} \tcp*[1]{Directed acyclic graph with information about nodes' relationships/comparisons}}
    \KwIn{\ttt{BF()} \tcp*[1]{callable function, Alg. \ref{bf_alg} }}    
    \KwIn{ $b$ \tcp*[1]{threshold for sufficient evidence that one model is stronger}}

    \KwOut{$\mathbb{H}^C$ \tcp*[1]{Superior model among input set}}\;
     
    \ttt{DAG.surviving\_layer\_champions()} $ \gets \mathbb{H}$ 
    
    \For{$\ham_c \in \mathbb{H}$}{
        
        \ttt{DAG.parent($\ham_c$)} $\gets \ham_p$\;
        $B_{pc} \gets BF(\ham_p, \ham_c)$\;
        \uIf{ $B_{pc} > b$}{
            \ttt{DAG.prune($\ham_c$)}
        }
        \uElseIf{$B_{pc} < 1/b$}{
            \ttt{DAG.prune($\ham_p$)}
        }
    }
    
    \ttt{DAG.surviving\_layer\_champions()} $\gets \mathbb{H}$\;
    return $\mathbb{H}$

\end{algorithm}

\subsection{Scalability and Generalisation}\label{supp_scalability}
The backbone of QMLA is to perform QHL, for each explored model $\hat H_i(\vec{x}_i)$, for a variable number of epochs $\Xi$, obtaining cumulatively a set of observed data $D_j$ from the quantum system, and updating the parameters $\vec{x}_i$ accordingly (Fig.~\ref{fig:Fig1}b,e). 
When launching QHL instances, see sectin \ref{supp_qhl} , we assume that the QMLA is able to access key experimental controls in the system to characterise: the evolution time $\tau$ and the initially prepared probe state $\lvert \psi\rangle_{\textrm{sys}}$. As in QHL, we also require access to a trusted (quantum) simulator to (efficiently) test trial Hamiltonians forms $\hat H_i$ and parameterisations $\vec{x}_i$ \cite{Granade:2012kj, Wiebe:2014qhl, wang2017qhlexp}.
However, so far we focused on characterising a closed system, entirely described by its full Hamiltonian $\hat{H}^*_{\textrm{glo}}$, on a Hilbert space $\mathcal{H}_{\textrm{glo}}$, by preparing input (and measuring output) states: $\ket{\psi}_{\textrm{glo, in (out)}} \in \mathcal{H}_{\textrm{glo}}$. 
This need not be the case for characterising an unknown quantum system, that might be open and therefore affected  by the contribution of an environment in $\mathcal{H}_{\textrm{env}}$, so that the effective Hamiltonian can be written as \cite{breuer2002}: $\hat{H}^*_{\textrm{glo}} = \hat{H}^*_{\textrm{sys}} + \hat{H}^*_{\textrm{env}} + \hat{H}^*_{\textrm{int}}$, with $\hat{H}^*_{\textrm{int}}$ describing the interaction between system and environment. 

In order to address this more general task, we introduce the following approach. First, in the rest of this Chapter we assume that we are able to prepare states $\ket{\psi}_{\textrm{sys}}$ in the quantum system independently from its environment, so that input states are separable, of the form:
\begin{equation}
    \ket{\psi}_{\textrm{glo, in}} = \ket{\psi}_{\textrm{sys}} \otimes 
    \ket{\phi}_{\textrm{env}}.
    \label{3:eq:separableinput}
\end{equation}
Now, in most cases the environment will not be directly accessible,
so that likelihood functions as in Eqn. \ref{bayes_rule} are in general not achievable, and a preliminary partial trace over the environmental degrees of freedom needs to be taken into account. 
The likelihood for the Bayesian inference is thus here modified as:
\begin{equation}
    \mathrm{Pr}(d | \hat{H}^*_{\textrm{glo}}, \tau) = {\tr \left[  \rho_{\textrm{glo}}(\tau) \left(\ket{d}_{\mathrm{sys}}\!\bra{d}_{\mathrm{sys}}\otimes 1_{\rm env}\right) \right]},
    \label{3:eq:qmlalikelihood}
\end{equation}
with $\ket{d}_{\mathrm{sys}}$ the chosen measurement basis for $\mathcal{H}_{\textrm{sys}}$, and
\begin{equation}
    \rho_{\textrm{glo}}(\tau) = \exp(-i \hat{H}^*_{\textrm{glo}} \tau) \dyad{\psi_{\textrm{glo,in}}}{\psi_{\textrm{glo,in}}} \exp(i \hat{H}^*_{\textrm{glo}} \tau) 
\end{equation} 
the global density matrix evolved under each $\hat{H}^*_{\textrm{glo}}$.

As the algorithm progresses, the credible region for the parameters, starting from a suitable prior distribution, will be reduced in volume $V$ \cite{granade2017qinfer}. 
An efficient way to halt the QHL subroutine for $\hat{H}_j$ is to assign $\Xi_j$ dynamically, by checking periodically for saturation in the reduction of $V_j$, as a signature of a converged Hamiltonian learning instance. 
Once all models for the latest entertained layer have converged, a \textit{consolidation} phase starts (Fig. \ref{fig:Fig1}c).
In order to preserve the Bayesian framework in our protocol, we adopt here for a meaningful pairwise $(\hat H_i, \hat H_j)$ comparison between models the \textit{Bayes factor} (BF):
\nomenclature{BF}{Bayes factor}
\begin{equation}
    \mathcal{B}_{ij} = \exp [\ell(D_{ij}|\hat H_i) - \ell(D_{ij}|\hat H_j)],
    \label{3:eq:bayesfactor}
\end{equation}
where $D_{ij} = D_i \cup D_j$ is the joint pairwise dataset, and 
\begin{equation}
    \ell(D_{ij}|\hat H_i) = \sum_{d \in D_{ij}} \log 
    \Pr \big(d | H_i(\vec{x}_i) \big)
    \label{3:eq:cumulloglikelihood}
\end{equation}
is the cumulative log-likelihood.
The reason to adopt a logarithmic likelihood is here that likelihoods are prone to numerical instabilities as the number of sample measurement grows 
\cite{granade2017qinfer}, so that adopting a log-difference instead of a ratio can help reducing artefacts.
BFs are known to be a statistically robust measure for comparing the predictive power of different models, whilst favourably weighting less structure to limit overfitting~\cite{kass1995bayes}, and have been successfully used in the context of resolving multi--modal distributions in \cite{Granade:2016txa}.
The resulting $\mathcal{B}_{ij}$ are stored as a \textit{comparative} Directed Acyclic Graph (cDAG) 
\nomenclature{cDAG}{comparative Direct Acyclic Graph}
representation across the same nodes, where the edges' directionality maps the sign of $\mathcal{B}_{ij} -1$, pointing towards the model favoured by statistical evidence (Fig. \ref{fig:Fig1}c). 
If all BFs composing the graph were computed using the same dataset, i.e. $D = D_{ij} \forall (i,j)$, generating a DAG would be immediately granted by Eqn. \ref{3:eq:bayesfactor}. 
This is not necessarily the case when $D_{ij} \neq D_{jk}$, which is most likely to happen if we run an independent instance of QHL per model. However, as we expect the graph to converge to a DAG as soon as enough statistical evidence is collected (i.e. once $ \dim( D_{ij}) \gg 1$  $\forall (i,j)$, as this should make differences across datasets negligible), we explicitly exclude cycles of ambiguous interpretation in the graph,  by removing the edge which minimises $|\mathcal{B} -1|$ for each accidental cycle.
Doing so and comparing systematically all model pairs leads naturally to the selection of a \textit{layer champion}, corresponding to the node with highest indegree.  

An \textit{exploration} stage follows, whereby a new layer $(\mu+1)$ is generated. 
Generating layers that progressively increase 
the complexity of the models entertained is a key feature of QMLA, as it affects directly the \textit{interpretability} of the models. 
Indeed, we have already introduced recent works that aimed at the characterisation of e.g. quantum states via NN \cite{Carleo2017,Meinhardt2017,iten2018}. This strategy, when targeting an ideally black--box quantum device, has the limitation of employing implicitly \textit{dimensionality reduction} schemes, to map the interactions characterising the system onto the possibly smaller set of weights between the neurons within the NN.
The closest equivalent of this approach, within our Bayesian framework, would be to assume the dynamics of the quantum system to be entirely described via the special unitary group $SU(n)$, with $n$ known, or lower--bounded via the adoption of a \textit{dimensional witness} \cite{Brunner2008}.
Then, QMLA would be posed the task to learn which elements of the generator basis (e.g. the set $\Gamma_n$ of generalised Gell-Mann complex $n \times n$ matrices $\gamma_n$) to include in the model, and with which parameters. That is, QMLA shall explore among subsets $\Gamma_n' \in \Gamma_n$ models of the form $ \sum_{\hat{\gamma}_i \in \Gamma_n'} g_i \hat{\gamma}_i$ for the best Hamiltonian to reproduce measurements on the system. 
In this way, the new layer $\mu+1$ would simply e.g. include an additional $\gamma_n$ from the set.

Now these kinds of mapping, if effective, are hard to be checked against intuition (or more quantitative predictors) that a human researcher might have or acquire about a system. As worries arise for the impact of artificial intelligence in the crisis of reproducibility affecting science \cite{Hutson2018}, our intention is here to design an automated protocol bootstrapping the human characterisation process of a system, rather than replacing it. 
In a sense, a model learnt by a device to reproduce another device's behaviour, that cannot be used to distillate general properties nor understanding of the first, 
advances us from an ``untrusted black box'' to a ``trusted one'', without ultimately solving the characterisation problem.
Therefore, we decided in favour of a protocol exploring models constructed from a reduced set of primitives, that are easily interpretable, and namely single--qubit Pauli operators $\hat{\sigma}_{\alpha} \in \Sigma_P$, i.e. $ \alpha \in \{I,x,y,z\}$ with $\hat{\sigma}_I \equiv \hat{I}$.
In fact, it is known that any $n$--dimensional Hamiltonian can be written as the sum of tensor products of $\sigma_{\alpha}$ \cite{Bruus2004}: 
\begin{equation}
    \hat{H}^* = \sum_{i,\alpha} 
    h_{\alpha}^i \hat{\sigma}^i_{\alpha} +
    \sum_{i,j,\alpha,\beta} 
    h_{\alpha\beta}^{ij} \left( \hat{\sigma}^i_{\alpha} \otimes
    \hat{\sigma}^j_{\beta} \right)
    +
    \ldots
    +  \sum_{i,j,\ldots, n, 
    \alpha,\beta, \ldots \nu} 
    h_{\alpha\beta\ldots\nu}^{ij\ldots n}
    \left( \hat{\sigma}^i_{\alpha} \otimes
    \hat{\sigma}^j_{\beta}
    \otimes \ldots \otimes
    \hat{\sigma}^n_{\nu} \right)
    \label{3:eq:Paulidecomp}
\end{equation}
with Roman indices labelling the $i$--th subsystem on which operator $\sigma^i$ acts. A generic layer $\mu$ can then be interpreted as encompassing potentially all truncations of Eqn. \ref{3:eq:Paulidecomp}, that include exactly $\mu$ terms (if the sDAG started with a root encompassing one term models).
Most importantly, $k$--sparse, row--computable Hamiltonians,
i.e. those Hamiltonians whose matrix has at most $k$ non--zero elements per row or column (the locations of said elements being retrievable with an efficient classical algorithm),  $\mathbb{H}_{\textrm{kcomp}}$ have been shown to admit a decomposition of the form in Eqn.  \ref{3:eq:Paulidecomp}, with at most $\mathcal{O}(k^2)$ terms \cite{Berry2007, Childs2011}. The class $\mathbb{H}_{\textrm{kcomp}}$ is surprisingly wide, as it covers the cases of electronic Hamiltonians in quantum chemistry, as well as lattice model Hamiltonians (such as Ising and Heisenberg lattices \cite{Peruzzo2014}). In all these cases, we can thus expect the depth of the \textit{ideal} sDAG to grow polynomially in $k$, \textit{independently} from the size of the system $n$, an observation which is key in estimating the worst case scaling of QMLA.

In general, we assume $n$ not known in advance, even if dimensional witnesses might bootstrap the guess for the root model(s) of the sDAG. 
Therefore, QMLA is designed to preferentially introduce models $\hat H_j$ in the \textit{spawned} layer $(\mu+1)$ that embed additional terms as linear combinations of appropriate tensor products of primitives $\hat h_i \in \Sigma_P$ (see Fig. \ref{fig:Fig1}f), exploring the same Hilbert space dimension $n_{\mu}$ as the consolidated layer $\mu$. 
However, if $\mu$ has saturated the maximum number of independent operators allowed by $n_{\mu}$ (or by user-defined rules), QMLA will generate a layer where $n_{\mu+1}=n_{\mu}+1$.
Now, even if we admit the system Hamiltonian $\hat{H}^* \in \mathbb{H}_{\textrm{kcomp}}$,  the exploration phase is still affected by a major scalability issue: if the nodes of the ideal sDAG might scale polynomially, this is not true for a sDAG that explores brute--force also other potentially valid models. This is most easily seen by studying a specific case, e.g. truncating the expansion in Eqn. \ref{3:eq:Paulidecomp} to only include pairwise interactions, i.e. the first non--trivial terms, and also the most crucial for a wide variety of systems. The number of all possible terms $\Theta$ that can be generated in this instance scales as $\mathcal{O} \binom{n}{2}$, i.e. the sDAG depth grows combinatorially with the system's size.  Additionally, also the maximum width of the DAG grows combinatorially, as the generic layer $\mu$ includes models with any of $\mathcal{O} \binom{\Theta}{\mu}$ terms' combinations.
This intractability is well known in the field of structure learning via graphical models, and the most direct way to deal with it is the introduction of a \textit{greedy} exploration phase \cite{russell2016ai, sra2012}. That is, local optimisation is favoured against global optimisation, to keep the size of the global search DAG manageable.  

In QMLA, the greediness in the search is imposed via the \textit{inheritance} rule: all models in layer $(\mu+1)$ expand upon the Hamiltonian form of their common parent node that is the champion node in $\mu$ (Fig. \ref{fig:Fig1}a\&f).  . 
Eventually, after a dimensional matching if $n_{\mu+1} \neq n_{\mu}$. E.g. terms with the operator $\hat{\sigma}_z^1$ will be interpreted as $\hat{\sigma}_z^1 \otimes \hat{I}^2$.
This greediness dramatically reduces the global number of models considered in the exploration.
Adopting again the pairwise model, the overall number of nodes explored in the worst case
is expected to be downsized:
\begin{equation}
    \mathcal{O} \left[ \prod_{\mu=1}^{\Theta} \binom{\Theta}{\mu} \right] \rightarrow \mathcal{O} \left[ \sum_{\mu=1}^{\Theta} \binom{\Theta}{\mu} \right]
    \label{3:eq:worstcasecomplex}
\end{equation} 
when the inheritance rule is adopted. The huge improvement exemplified in Eqn. \ref{3:eq:worstcasecomplex} is clearly not enough to make QMLA tractable already for relatively small instances, and highlights the discrepancy between the expected polynomial scaling of terms and a search through a combinatorially growing models' space, which evidently goes beyond the hypothesis of $\hat{H}^* \in \mathbb{H}_{\textrm{kcomp}}$. However, the inheritance rule becomes a powerful heuristic, when combined with aprioristic knowledge about the system to characterise. 
Continuing upon our example, one might make the reasonable hypothesis that nearest neighbours interaction are to be taken into account first, and that the index labelling is representative of neighbouring conditions. In this case, the first layer of the DAG chooses one out of $n$ choices, and the overall models explored in the worst--case is $\mathcal{O} \left[\sum_{\mu=1}^n (n-\mu+1) \right]= \mathcal{O} (n^2)$. One can use the last layer of the learnt DAG to then progressively explore alternative hypotheses like index inversions, or longer--range interactions, without incurring straightforward in the intractable scaling as in Eqn. \ref{3:eq:worstcasecomplex}. 

Now that issues in the scaling of the sDAG nodes have been discussed, we are left with the scaling of the number of edges in the cDAG.
If we were to compare each model $\hat H_i$ against all others explored, the cDAG would be a \textit{complete} graph, whereby the number of edges, each corresponding to a costly BF calculation, grows combinatorially with the sDAG depth \cite{Bollobas1998}. 
In order to avoid generating overall a complete graph, the nodes within a layer form complete subgraphs of the cDAG, but the comparison among different layers occur in the QMLA only via their respective champions (Fig. \ref{fig:Fig1}). 
In this way, all nodes other than the champion node for a given layer are essentially discarded from the learning process, as they stop being compared with any other competitive model. 
Additionally, if the comparison between the parent champion and a child champion $\hat{H}_c$ leads to $\mathcal{B}_{pc} \gg $, a \textit{collapse} rule reallocates the sub-tree rooted at $\hat{H}_c$ (if any) to $\hat{H}_p$ as a parent node, discarding the whole layer of $\hat{H}_c$. This collapse rule also aims at mitigating overfitting that might arise from the greediness introduced with the inheritance rule, as terms provisionally inherited might prove superfluous once the model is expanded, and would be discarded invoking a new instance of the inheritance rule after the collapse, against the new parent node $\hat{H}_p$.  
Inheritance rules can be seen as an extreme \textit{pruning} rule for all the other nodes in $\mu$ (Fig. \ref{fig:Fig1}d), similarly to what adopted for a graphical model exploration in \cite{Granade:2016txa}. 
Therefore, another way to mitigate the greediness of the approach is to discard only those particularly unsuccessful models $\hat{H}_r$, that have: $\mathcal{B}_{rj} < b \; ,\; \forall j$ in the same layer and $b$ a user--defined threshold. Intuitively, this kind of stratagems increases the explorative nature of the QMLA, increasing the likelihood of approximating the global optimum, at the expense of the overall cost of the procedure.  

In conclusion, the cumulative knowledge of the QMLA is represented as a multi-layer DAG (see Fig.\ref{fig:Fig1}c), combining a sDAG tracking the generative process of new models and a cDAG that embeds the information about how effective a given model is in replicating the experimental dataset, compared to neighbouring nodes in the cDAG.
The pseudocode for the QMLA protocol is given in SM~\ref{supp_scalability}, along with its subroutines.
The efficiency in the QMLA's exploration of the models' space is intrinsically dependent from the extent to which an approximate solution to the characterisation problem can be deemed acceptable, and therefore how greedy towards local optimisation are the choices performed at each generation stage in the process.
This trade--off resembles the limitations in the accuracy of the predictions that can be attained via graphical classification modelling \cite{russell2016ai, Murphy2012}, versus their human interpretability \cite{bishop2006}.

\begin{table}[!ht]	
\begin{center} \centering{
  \begin{tabular}{ccc}
  \hline
  \begin{tabular}{c}
  $\dim (\hat{H})$ \\ 
  (qubits)  
  \end{tabular}  &  Operator form  & Abbreviation  \\
  \hline
  \hline
    1      &  $\hat{S}_{\alpha}$    &   S$_{\alpha}$      \\
    1      &  $\sum_{i \in \{\alpha,\beta\}} \hat{S}_i$    &   S$_{\alpha, \beta}$      \\
    1      &  $\sum_{i \in \{x,y,z\}} \hat{S}_i$     &   S$_{x,y,z}$      \\
    2      & $\sum_{i \in \{x,y,z\}} \left(\hat{S}_i \otimes \hat{\mathbbm{1}} \right) + 
        \hat{S}_{\alpha} \otimes \hat{I}_{\alpha}$  
        &   $\mathrm{S}_{x,y,z}\mathrm{HF}_{\alpha}$       \\
    2      & $\sum_{i \in \{x,y,z\}} \left( \hat{S}_i \otimes \hat{\mathbbm{1}} \right) + 
    \sum_{i \in \{\alpha,\beta\}} \left( \hat{S}_{i} \otimes \hat{I}_{i} \right)$  
        &   $\mathrm{S}_{x,y,z}\mathrm{HF}_{\alpha, \beta}$       \\
    2      & $\sum_{i \in \{x,y,z\}} \left( \hat{S}_i \otimes \hat{\mathbbm{1}}  + 
    \hat{S}_{i} \otimes \hat{I}_{i} \right)$  
        &   $\mathrm{S}_{x,y,z}\mathrm{HF}_{x,y,z}$       \\
    2      & $\sum_{i \in \{x,y,z\}} \left( \hat{S}_i \otimes \hat{\mathbbm{1}}  + 
    \hat{S}_{i} \otimes \hat{I}_{i} \right)  +
        \hat{S}_{\alpha} \otimes \hat{I}_{\beta} $
        &   $\mathrm{S}_{x,y,z}\mathrm{HF}_{x,y,z}\mathrm{T}_{\alpha \beta}$       \\
    2      & $\sum_{i \in \{x,y,z\}} \left( \hat{S}_i \otimes \hat{\mathbbm{1}}  + 
    \hat{S}_{i} \otimes \hat{I}_{i} \right)  +
        \sum_{ij \in \{\alpha \beta, \gamma \delta\}} \left( \hat{S}_{i} \otimes \hat{I}_{j} \right) $
        &   $\mathrm{S}_{x,y,z}\mathrm{HF}_{x,y,z}\mathrm{T}_{\alpha \beta, \gamma \delta}$       \\
    2      & $\sum_{i \in \{x,y,z\}} \left( \hat{S}_i \otimes \hat{\mathbbm{1}}  + 
    \hat{S}_{i} \otimes \hat{I}_{i} \right)  +
        \sum_{ij \in \{xy, xz, yz\}} \left( \hat{S}_{i} \otimes \hat{I}_{j} \right)$
        &   $\mathrm{S}_{x,y,z}\mathrm{HF}_{x,y,z}\mathrm{T}_{xy, xz, yz}$       \\
  
  \hline
  \end{tabular} }\end{center}
  \caption{List of all models explored in the QMD implementation, along with the corresponding number of qubits and the abbreviation used in the text.}
\end{table}

\subsection{Stability analysis for QHL}\label{supp_stability}
The principal question that needs to be addressed is the stability of Bayes factor analysis under the sequential Monte-Carlo approximation.  While stability has broadly been shown in the past for the problem of quantum Hamiltonian learning, the problem of using Bayes Factor analysis to compare alternative families of models is not directly covered by those results.  Here we show that, provided that both the prior distribution and the likelihood function are sufficiently smooth functions of the Hamiltonian coefficients.

Let $\epsilon \le x/2$ and $\delta \le y/2$ then
\begin{align}
\left| \frac{x+\epsilon}{y+\delta} - \frac{x}{y}\right|&\le \left| \frac{x+\epsilon}{y+\delta} - \frac{x+\epsilon}{y}\right| + \left| \frac{x+\epsilon}{y} - \frac{x}{y}\right|\nonumber\\
& \le \frac{|\epsilon|}{|y|} + |x+\epsilon| \left|\frac{1}{y} - \frac{1}{y+\delta} \right|.\nonumber\\
& = \frac{|\epsilon|}{|y|} +\frac{|x+\epsilon|}{|y+\delta|} \frac{|\delta|}{|y|}\nonumber\\
&\le \frac{|\epsilon|}{|y|} + 3\frac{|x|}{|y|} \frac{|\delta|}{|y|}.
\end{align}
The Bayes factor for two models is defined to be of the form, given data $D$ is observed under experimental settings $E$, is
\begin{equation}
\mathcal{B} := \frac{P(D|M_1;E)}{P(D|M_2;E)}
\end{equation}
We therefore have that if $P(D|M_1)$ is computed within error $\epsilon$ and $P(D|M_2;E)$ is computed within error $\delta$ such that $\delta \le P(D|M_2;E)/2$ and $\epsilon \le P(D|M_1;E)/2$ then
\begin{equation}
\Delta\mathcal{B} = \left|\mathcal{B} - \frac{P(D|M_1;E) +\epsilon}{P(D|M_2;E) +\delta} \right| \le \frac{|\epsilon| + 3 \mathcal{B} |\delta|}{P(D|M_2;E)}.\label{eq:epsbd}
\end{equation}
The overall error in the probability can arise from multiple sources but here we will focus on the error in the Bayes factors that arises from the use of a finite particle approximation to the probability density.  Specifically,
\begin{equation}
P(D|H;E) = \int P(H) P(D|H;E) \mathrm{d}H\approx \tilde{P}(D|M;E) := \frac{1}{N_{\rm part}} \left(\sum_{j=1}^{N_{\rm part}} P(D|H_j;E) \right),
\end{equation}
for an ensemble of Hamiltonian models $H_j$ drawn from the prior distribution $P(H)$. Here we use the convention that $\int f(H) \mathrm{d}H = \int f(H(\vec{x})) \mathrm{d}x^{d}$ where $\vec{x}\in \mathbb{R}^d$.
We then have from Chebyshev's inequality that with probability greater than $1-1/k^2$ the error in this Monte-Carlo approximation is
\begin{equation}
\left| P(D|H;E) - \tilde{P}(D|H;E) \right| \le k \sqrt{\frac{\int P(H) P^2(D|H; E) \mathrm{d}H - \left(\int P(H) P(D|H; E) \mathrm{d}H \right)^2 }{N_{\rm part}}}\label{eq:var}
\end{equation}
  Now using the parameterization $H= \sum_{k=1}^d x_k \hat{h}_k$ for $x_k \in [-L,L]$.  Similarly, let $P(D|M;E) = \int P(H) P(D|H; E) \mathrm{d}H$ we then have from the mean-value theorem that there exists a point in the domain of integration such that $P(\sum_k [\vec{\xi}(\mu)]_k \hat{h}_k) = P(D|M;E)$.   we therefore have from Taylor's theorem that 
\begin{equation}
P(D|H;E) \le P(D|M;E)  + \max\|\nabla P(D|\vec{x};E)\| |\vec{x} - \vec{\xi}(\mu)|.\label{eq:pert}
\end{equation}
Equations~\eqref{eq:var} and \eqref{eq:pert} then imply that
\begin{align}
\left| P(D|M;E) - \tilde{P}(D|M;E) \right| &\le k \sqrt{\frac{P(D|M;E)\max\|\nabla P(D|H;E)\||\vec{x} - \vec{\xi}(\mu)|  }{N_{\rm part}}} \nonumber\\
&\le k \sqrt{\frac{P(D|M;E)\max\|\nabla P(D|H;E)\| dL  }{N_{\rm part}}} 
\end{align}
Thus we have that with probability at least $1-1/k^2$ that $\left| P(D|M;E) - \tilde{P}(D|M;E) \right| \le \epsilon$ if
\begin{equation}
N_{\rm part} \ge \frac{P(D|M;E) k^2 \max \|\nabla P(D|H;E)\| dL }{ \epsilon^2}.
\end{equation}

In order to make progress, let us assume that $\mathcal{B} \in \mathbb{R}^+ \setminus (1-\gamma,1+\gamma)$.  This assumption is needed in order to guarantee that there is a gap between the two possibilities.  Thus if $\Delta \mathcal{B} \le \gamma/2$ then the decision about the Bayes factor will not be qualitatively affected by the Monte-Carlo error due to finite particles.  Eqn.~\eqref{eq:epsbd} then tells us that it suffices to take
\begin{align}
    \epsilon \le \frac{\gamma P(D|M_2;E)}{4},\qquad \delta \le \frac{\gamma P(D|M_2;E)}{12\mathcal{B}}
\end{align}
which implies that if the first bound above dominates that it also suffices to take with probability at least $1-1/k^2$ the number of particles used to compute the first probability is
\begin{equation}
    N_{\rm part_1} \ge \frac{16P(D|M_1;E) k^2 \max\| \nabla P(D|H_1;E)\| dL}{\gamma^2 P^2(D|M_2;E)}
\end{equation}
Now let us further assume that for $H_j$ in both model $M_j \in \{M_1,M_2\}$ 
\begin{equation}
    \max \|\nabla P(D|H_j;E)\| \le \kappa P(D|M_j;E).
\end{equation}
It then suffices to take
\begin{equation}
    N_{\rm part_1} \ge \frac{16 \kappa \mathcal{B}^2 k^2 dL}{\gamma^2}.
\end{equation}
Similarly, we can choose
\begin{align}
    N_{\rm part_2} &\ge \frac{144 \mathcal{B}^2 P(D|M_2;E) k^2 \max\| \nabla P(D|H_2;E)\| dL}{\gamma^2 P^2(D|M_2;E)},
    \end{align}
    which is implied by
    \begin{align}
    N_{\rm part_2}\ge \frac{144 \kappa \mathcal{B}^2 k^2 dL}{\gamma^2}\label{eq:partmin}
\end{align}
Therefore if the number of particles given in~\eqref{eq:partmin} is used for both models then a qualitatively accurate decision between the two models can be made according to the Bayes factors.  The same result also holds if we only assume that $\mathcal{B} \in [1+\gamma,\infty)$.

This implies that Bayes factor calculation is stable under the SMC approximation provided that the gradients of the likelihood functions are not large (i.e. $\kappa dL$ is modest) and the promise gap $\gamma$ is at least comparable to $\mathcal{B}$.  

\section{Experimental Details}\label{supp_experimental}
\subsection{Experimental system and sample description}
The measurements were performed on a home-build confocal setup at room-temperature.
In our setup we align an external magnetic field of $11 mT$ parallel to the NV centre's axis, using Helmholtz-coils. 
This magnetic field lifts the degeneracy of the electron fine states $m= \pm 1$. 
For the experiments shown here we only use the $ m= -1$ electron state.
We perform Hahn echo measurements by using off-resonant laser excitation ($532 nm$) for initialisation and readout of the electron spin.
To manipulate the electron spin in our Hahn echo experiments, we apply a $\pi/2 - \pi - \pi/2$ MW-pulse at the resonant frequency, with varying time between the MW pulses.
We use a strip line to apply microwave signals of a few microwatts.
The pulse sequence is repeated about $3\times 10^{6}$ times for each microwave frequency to acquire statistics, and the detected photon detection times are histogrammed leading to about 300 counts in each $25ns$ bin during the laser pulse.
Single photons are collected by the confocal setup and detected by an avalanche photo diode.
The detected counts at the end of each Hahn sequence are normalised with respect to the mixed state of the initialisation pulse.

For the measurement we use an electronic grade diamond sample [110], with a 4 ppb nitrogen impurity concentration and a natural abundance ($\approx 1.1\%$) of $^{13}$C (Element6).
%



\subsection{Noise in the experimental setup}
Bayesian methods are known to exhibit intrinsic properties of noise robustness against i.i.d. and Markovian sources of noise \cite{Granade:2012kj, Ferrie:2012ip}.
QMLA partially inherits such noise resilience by adopting QHL as a subroutine for learning the parameterisation of each investigated model.
However, the role of noise in our protocol has more profound implications. 
Learning a model for the noise affecting the system is inextricable from learning a model for the system dynamics itself, in all those cases when the contribution from noise sources cannot be neglected \cite{hincks2018nvlearn, Santagati2019}.
Therefore, we expect QMLA to attempt modelling noise sources within the framework of the primitives provided in the user defined library.
Apparent overfitting might then be interpreted, in some instances, as triggered by features in the dynamics unpredicted by a noiseless model.
In the specific case of the NV centre electron spin studied in this paper, the hyperfine interaction with the nuclear spins in the bath can be in itself interpreted as a ``noise'' in the isolated system picture. 
However, its effects are so crucial that we deem necessary to expand the picture correspondingly. On the contrary, we might find increases in the complexity of the model unjustified by the minor noise effects that they attempt to reproduce. 
The adoption of Bayes factors is intended to guard against such unwanted overfitting tendencies of the QMLA, providing a solid statistical ground for deciding systematically in favour, or against, increases in the model complexity~\cite{kass1995bayes}. 
However, it is crucial to review expected sources of noise in the Hahn echo experiments, that might alter the assumed correct model $\hat{H}_{\mathrm{diag}}$.

\textit{Noise in the active electron spin control}.
The Hahn echo experiments in this paper require the implementation of three MW pulses (see also Fig. 1b main text), controlling respectively: i) the initial preparation of the electron spin in the $\ket{+}$ state, i.e. a $\pi/2$-pulse, ii) the Hahn rotation, i.e. a $\pi$-pulse, iii) the preparation of the evolved state for projection onto the measurement basis $\{\ket{0}, \ket{1}\}$, i.e. another $\pi/2$-pulse, with all rotations along the y-axis of the Bloch sphere representing the electron spin. 
Constant offsets as well as i.i.d. noise might affect all such operations. 
Accurate characterisation of the setup, effectively minimises any offset affecting these operations, whereby the initial state can be represented as:
\begin{equation}
    \ket{\psi}_{\mathrm{sys}} = (\ket{+} + \omega \ket{\psi})/\sqrt{1 + \omega^2},
    \label{eq:offprobe}
\end{equation} 
with $\ket{\psi}$ a normalised one-qubit state and $\omega \ll 1$ a single fixed parameter representing the severity of the offset.
Residual offsets in the preparation and projection operations might reduce the visibility of the first peak, or slightly alter its occurrence in time.
We partially counteract these effects by rescaling the likelihood $\mathrm{Pr(\ket{0} | \hat{H}, t})$ to the full range $[0,1]$ and adjusting the zero of the experimental time against the first peak in $\mathrm{Pr(\ket{0} | \hat{H}, t})$.
Moreover, to avoid the QMLA to use recursively a biased probe state $\ket{+}$ not matching the experimental $\ket{\psi}_{\mathrm{sys}}$, we randomise the probe prepared in the simulator adopting Eqn. \ref{eq:offprobe}, but with $\omega$ a random variable normally distributed according to $\mathcal{N} (0, \sigma_{\omega})$, with $\sigma_{\omega} \sim 0.03$ as extracted from the system characterisation.
Thus, the effect of a constant offset on the system preparation is mapped onto a less disruptive randomised noise, that can be addressed easier by the Bayesian inference schemes. 
%
Similar randomisation might also be introduced in the simulator for the final rotation ahead of the projection measurement, and in the Hahn rotation angle, however these were deemed unnecessary in the light of the simulation results.
Offsets in the Hahn angle implemented might additionally screen less effectively than in the predictions the electron spin from bath effects, however, this should effect quantitatively (i.e. the strength and components of the hyperfine interaction) more than qualitatively the results in the learning (i.e. the form of the model).
%

\textit{Readout noise}.
This source of noise originates from the optical readout (see also Fig. 1 in the main text), i.e. the $\sim 33 \%$ contrast in fluorescence detected as the final electron spin state is in the $\ket{0}$ (bright) as opposed to the $\ket{1}$ (dark) stable states.
The normalised counts $C_0 (\hat{H}, \tau, \tau')$ from the experiment are proportional to $\mathrm{Pr}(\mathrm{0| \hat{H}, \tau, \tau'} ) $, which is obtained by subtracting the dark counts (i.e. the floor of the $C_0$ signal) and mapping the effective number of surplus counts onto the full interval $[0,1]$.  
Assuming that the noise floor counts are approximately constant, then $C_0$ is affected by two sources of noise: losses due to imperfect collection efficiency, and Poissonian noise in the counts readout. 
Losses are known to impact negatively the learning capability of Bayesian inference schemes in the absence of error modelling, when the number of counts is extremely low, whereas can be safely neglected already for few hundred (average) counts \cite{Santagati2019}.
In the Hahn echo experiments shown here, we reduced the impact of losses to negligible contribution, by averaging the outcome for each $C_0 (\hat{H}, \tau, \tau')$ across $M=3000000$ repetitions. 
Thus, we are left with the Poissonian distributed counts with a variance of $\sigma_P = C_0$
This noise maps well on a binomial noise model subtending the processing of data $d \in \{0,1\}$. 
Such a noise model can be naturally implemented in the Bayesian inference process adopted here \cite{granade2017qinfer}, in absence of majority voting correction schemes \cite{Santagati2019}.

\textit{Quantum projection noise}.
Quantum projection noise is an unavoidable source of noise when performing projective measurements of a quantum system onto a set of stable states. 
In this work, the NV centre evolved state with each Hahn sequences (after tracing out environmental degrees of freedom) is a single-qubit state $\ket{\psi}_{\mathrm{sys}}'$, and the final measurement operation is a projective measurement on the computational basis $\{\ket{0},\ket{1}\}$, repeated across $M=3000000$ independent instances, leading to $C_0$ counts. 
Under these measurement parameters, ignoring losses already mentioned, the quantum projection noise can be interpreted as a binomial distribution whereby $C_0$ has a variance $\sigma^2_b (0) = M p_0 (1-p_0)$ \cite{Itano1993}, where we contracted $p_0 := \mathrm{Pr}(\mathrm{0|\hat{H}, \tau, \tau'})$. 
This noise source is then dependent upon the specific $p_0$ at the end of each evolution, similarly to the Poissonian noise, but once losses are taken into account, we can observe the upper bound $\sigma^2_b < \sigma^2_P$. 
Therefore, the same considerations already done for the optical readout noise apply here as well. 
\newline

Even if, in principle, the QMLA can be employed to perform noise modelling as well, we observe how it is here impossible to disambiguate all sources of noise.
For example quantum projection noise and optical readout are respectively state dependent and readout counts dependent, but given that here we experimentally access information about the evolved state optically and we have no independent measurement, the two contributions must be considered simultaneously. 
Similarly, systematic and random errors affecting the probe state preparation, the controlled evolution and the final rotation of the evolved state cannot be effortlessly deconvoluted in the absence of dedicated measurements. 
Such experiments can well be designed in future investigations adopting QMLA as a characterisation tool, where systematic trusting of parts of the setup leads to the characterisation of others, in the spirit of a modular QHL \cite{Wiebe:2014qhl}. 
However, this is beyond the scope of this paper, as a first demonstration of the power of the QMLA approach 
And the discussion of the noise sources ensures that the interpretation of the QMLA outcome doesn't unaccounted, significative sources of noise.

\subsection{Hyperparameterised modelling of Hahn echo experiments}
\label{sec:hypparam_spinmodel}
\begin{figure*}
    \centering
    \includegraphics[width=0.98\textwidth]{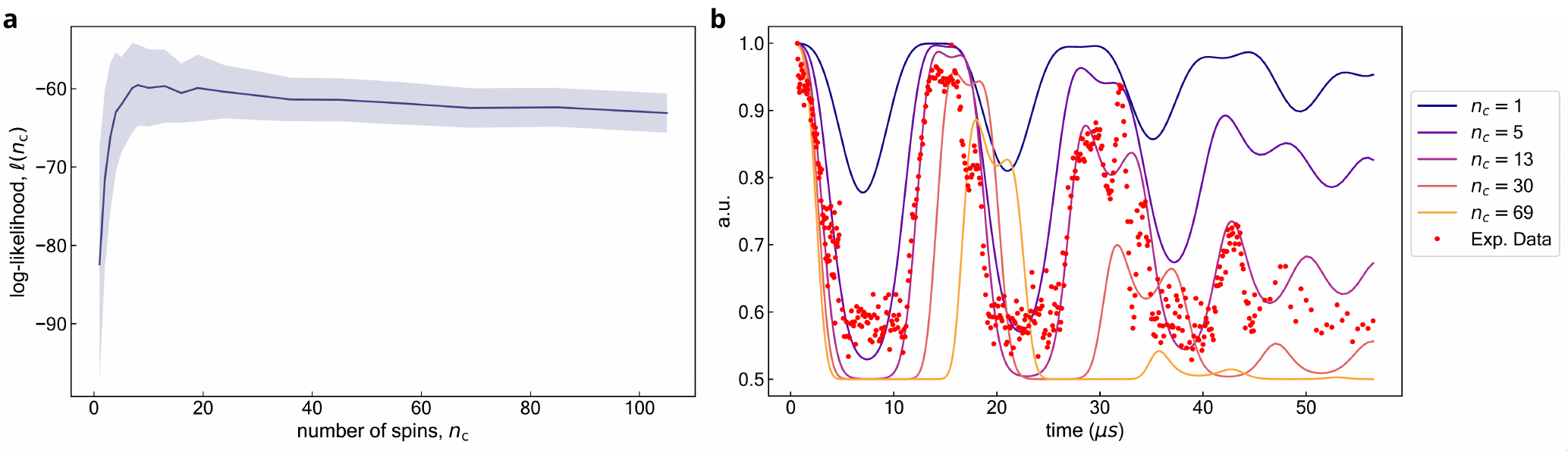}
    \caption{\textbf{a} Estimated log-likelihood for the hyperparameterised model in Eqn.~\ref{eq:prhahnsignal}, against the number of spins $n_c$. The latter is progressively increased and, for each $n_c$, the values of the hyperparameters are estimated running CLE with $1000$ particles along $100$ epochs. $ \ell (n_c)$ is estimated upon convergence and CLE repeated $150$ times. The average $ \bar \ell (n_c)$ performance is reported on the plot as a solid line, along with the 68.27\% confidence interval as a shaded area.
    \textbf{b} Plots for the expected probability $\rm{Pr}(1| \tau)$ , along with experimental data from the optical state readout (from the normalised detected PL). $\rm{Pr}(1| \tau)$ is reported color-coded for models adopting different values of $n_c$, with the hyperparameters $\{ \mathbf{B}_0, \mathbf{B}_1, \sigma_B, \omega_0, \delta_{\omega}, \sigma_{\omega}   \}$ averaged from the values learned upon convergence with the same $150$ runs as in \textbf{a}. 
    }
    \label{fig:SM_EnvironSpinLearn}
\end{figure*}

Hahn echo experiments can be interpreted according to the hyperfine interaction \cite{Rowan1965}, describing the coupling of the electronic NV spin, in state $m_s$, to the $j$-th nuclear \textsuperscript{13}C spin, combined in a four-level system $\mathcal{S}$. 
Here we summarise this interaction in terms of an effective magnetic field $\textbf{B}^j_{m_{s}}$, whereby the ground states of $\mathcal{S}$ precess at a rate $\omega_{j,0}$, whereas the excited states incur in a splitting $\omega_{j,1}$  \cite{Childress2006}. 
After the initial $\pi/2$ pulse, the nuclear and electron spin evolve freely and become progressively more entangled at a rate dictated by the hyperfine interaction, getting maximally correlated at times $\tau \propto \pi/A$ \cite{Blok2014manip}, where the Hahn signal is weakest. 
When the two spins get disentangled again, revivals can be observed in the experimental Hahn signal (see Fig.\ref{fig:SM_EnvironSpinLearn}).

Using the secular approximation and rotating frame, the target Hamiltonian in Eqn.  \ref{eq:full_Hamiltonian}, here inclusive also of the nuclei dynamics, can be written:
\begin{equation}
    \hat{H}_{\textrm{echo}} =
    \mu_e g_e B_z \hat{S}_z - \mu_n g_n B_z \hat{A}_z + A_{\parallel} \hat{S}_z \hat{A}_z + A_{xz} ( \hat{S}_z \hat{A}_x + \hat{S}_x \hat{A}_z)
\end{equation}

The dynamics of this simplified Hamiltonian, in an appropriate (electron) rotating frame, can be analytically solved~\cite{Rowan1965} to obtain the Hahn echo signal:
\begin{equation}
  \rm{Pr}(1| \tau ; \{ \mathbf{B}^j \}, \{ \omega_j \}) =  \frac{1}{2}\left(\left(\prod_jS_j\right) + 1\right), 
  \label{eq:prhahnsignal}
 \end{equation}

 for the probability of the electron spin state to be $\ket{0}$ at the end of the corresponding sequence, where the pseudo-spins:
 \begin{equation}
   S_j =  1 - \frac{|\mathbf{B}_0 \times \mathbf{B}_1^j|}{|\mathbf{B}_0|^2 |\mathbf{B}_1^j|^2} 
    \sin^2 (\omega_{0} \; \tau /2)  \sin^2 (\omega_{j,1} \; \tau /2), 
\label{eq:pseudospins}
\end{equation}
with $\mathbf{B}_0$ the external magnetic field and $\omega_{0}$ the Larmor frequency. 
Observing Eqn.~\ref{eq:prhahnsignal}, decays and revivals in the PL signal can then be interpreted respectively as beatings among the modulation frequencies $\omega_{j,1}$, and re-synchronisation when $\tau = 2 \pi / \omega_0$. 

Learning via CLE the full set of parameters $ \{ \mathbf{B}_0, \omega_0, \{\mathbf{B}_1^j\}, \{\omega_{j,1}\}  \}$, usually obtained from simulations leveraging upon aprioristic knowledge of the emitter system~\cite{Childress2006}, however, poses difficulties. 
The number of interacting \textsuperscript{13}C spins must be known in advance, and has been estimated to be potentially in the order of 1000s. 
Such a huge parameter space is computationally challenging, and might be prone to degeneracies that are known to mislead CLE methods adopting unimodal distributions.
Therefore, we construct a \textit{hyperparameterisation} of the problem, using two normal distributions $\mathcal{N}(B_1, \sigma_B)$ and $\mathcal{N}(\omega_0 + \delta_{\omega}, \sigma_{\omega})$, from where a number $n_j$ of $\mathbf{B}_1^j$ and $ \omega_{j,1}$ are drawn.
In this way, the value of $n_j$ can be estimated from experimental evidence. 
For each $n_j$ the learning can be performed against a reduced (hyper)parameter set: $ \vec{x} := \{ \mathbf{B}_0, \mathbf{B}_1, \sigma_B, \omega_0, \delta_{\omega}, \sigma_{\omega}   \}$. 

The first and crucial task is to infer from experimental data an a-posteriori estimate of the number of environmental \textsuperscript{13}C ($n_j$), required to provide an accurate description of the system dynamics, as observed from Hahn echo data. 
Thus, we adopt a bottom-up approach, where we increase linearly $n_j$, launch equivalent CLE instances for each $n_j$, and compare the agreement with experimental data adopting the usual total log-likelihood $\ell (n_j) = \sum_{E_i \in \{E\} } \log_{2} Pr( E_i | \vec{x}_{n_j} ) $. 
$\{E\}$ is here the cumulative set of experimental data against which the algorithm has trained, i.e. all those used for $\tilde{n}_j \leq n_j$. 
Increasing $n_j$ does not increase the number of parameters in the model, hence the learning process can rely on equivalent instances of CLE (see Fig.~\ref{fig:Fig3}).
However, at the same time this defies the role of using Bayes factors to prevent overestimating $n_j$.
Therefore in our algorithm we stop increasing $n_j$ once $\ell (n_j)$ converges.
The results are reported in Fig.~\ref{fig:Fig3}, where it emerges clearly how increasing $n_j$ beyond $\sim 14$ does not reproduce significantly better the experimental data, thus suggesting evidence that not less than $\sim 14$ \textsuperscript{13}C spins are effectively coupled to the electron spin. 
The phenomenological analysis does not provide at the same time an upper bound, as additional environmental spins might have a weaker hyperfine interaction that is not detected beyond noise level in this experiment.
Equivalent learning procedures applied to experiments where amplification techniques reveal weakly coupled sites might resolve additional spins~\cite{zhao2012nucspins}.
Interestingly, our estimate is well below the number of nuclear sites employed in initial simulations of Hahn echo experiments with NV-centres~\cite{Childress2006}, but agrees in order of magnitude with the number of \textsuperscript{13}C in the \textit{first-shell}, that is known to be hyperpolarisable~\cite{alvarez2015hyperpol}. 

Finally, we report in Fig.~\ref{fig:SM_EnvironSpinLearn} b the expected normalised PL signal, estimated from the model in Eqn.~\ref{eq:prhahnsignal} with $n_j = 20$, and the hyperparameters as learnt after $400$ CLE epochs, observed to be sufficient to achieve convergence of the estimated parameter values. 
Simulated data reproduce accurately experimental findings, including the revival peak positions, allowing an independent estimate of the decoherence time for this system ($T_2 = 81 \pm 3.9 \mu s $)  from the envelope of the revived PL signals.
This case also shows the increased noise robustness of the approach, compared with a standard peak-fitting methodology in estimating $T_2$, as it corrects against occasional noisy measurements (as it occurs for the peak situated at $\tau_1 = \tau_2 \simeq 32 \mu s$ (Fig.~\ref{fig:SM_EnvironSpinLearn} b).

\subsection{Finite-size effects of the nuclear spins bath and reversibility of the dynamics}
\label{sec:opensystem}

A well-established modelling of decoherence effects in open quantum systems involves the introduction of \textit{indirect measurements} occurring on the quantum system, when operations occur that effectively trace out environmental degrees of freedom, which have become correlated with the system state~\cite{breuer2002}.
In the main paper we have systematically considered models inclusive of Hamiltonian terms of the form $\sum_i \hat{S}_i \otimes \hat{B}_i$, with $\hat{S} , \hat{B}$ system and bath operators respectively, which is a standard interaction term leading to system-bath correlations~\cite{breuer2002, schlosshauer2005}. 

In the theoretical framework of open quantum systems, it is customary to model decoherence phenomena introducing a bath of $n_{\mathrm{env}}$ spins~\cite{schlosshauer2005}, or harmonic oscillators~\cite{Weiss2012}. 
The reason to adopt a bath of interacting systems is to ensure an irreversible decoherence, as usually observed in real physical systems. 
This adaptation can be seen by using Poincar{\'e} recurrence theorem. 
Assuming the state of the global quantum system starts as a separable state $\ket{\Psi (0)}  = \ket{\psi}_{\mathrm{sys}} \otimes \ket{\phi}_{\mathrm{env}}$ , the theorem states that a time $\exists \; \tilde{t}$, such that the evolved state $| \langle \Psi (0) | \Psi(\tilde{t}) \rangle  |^2 \sim 1 $, provided that the global system has a finite eigenspectrum of energies $\{E_n\}$. 
An immediate consequence is that at times 
(known as Poincar{\'e} recurrence times~\cite{breuer2002}), 
the global state can become separable again, thus leading to \textit{revivals} in the coherence of the system state. 
If we consider the simplest possible interaction term:
\begin{equation}
    \sum_b \ket{b}_{\mathrm{sys}} \bra{b}_{\mathrm{sys}} \otimes \hat{B}_b ,
\end{equation}
with $\hat{B}$ a generic operator acting on the bath and $\{ \ket{b} \}$ an orthonormal basis for the system, then the global evolved state can be represented as:
\begin{equation}
    \ket{\Psi(t)}  = \sum_b p_b \ket{b}_{\mathrm{sys}} \otimes \ket{\phi_b (t) }_{\mathrm{env}}
\end{equation}
and therefore the decoherence can be characterised in terms of the overlaps
$ M_{b,b'}(t) = |\bra{\phi_b (t) } \ket{\phi_{b'} (t) }|$ ~\cite{breuer2002}. 
Now the decay of these overlaps, and their recurrence times, depends on the size of the bath, as a richer dynamics corresponding to a higher-dimensional bath will tend to increase the recurrence times at which $M_{b,b'}(t)$ returns non-negligible.
In real systems it is empirically observed that a decay $M_{b,b'}(t) \propto \exp[-\Gamma_{b b'} t]$ occurs, which is in agreement with predictions from Markovian assumptions~\cite{chirolli2008}. 
Generally, it can be proven that the recurrence times $\tilde{t}$ for the global system scale combinatorially with the bath size $n_{\mathrm{env}}$~\cite{schlosshauer2005}.
Hence the necessity in simulations to increase the bath size for a system with a discrete spectrum, not to observe revivals as an artefact of the finite size of the bath.
This behaviour is equivalent to the known one in artificial quantum simulators, characterised by \textit{finite-size effects} in the bath degrees of freedom (e.g. ion trap simulators~\cite{Deng2005}).

In SI~\ref{sec:hypparam_spinmodel}, we have shown how the size of the bath can be inferred from \textit{Hahn echo} experiments.
In the first part of the main paper, instead, we have investigated open quantum systems dynamics making use of a single additional qubit treated as an environmental qubit, to match the maximum bath dimension attainable with our quantum photonic simulator.
However, this method leads inevitably to finite-size effects in the system dynamics, as shown in Fig.~\ref{fig:SM_EnvironSpinLearn}. 
The learning procedure can parameterise the model $\hat{H}$ opportunely to reproduce the initial decay, but being interaction terms in the Hamiltonian time-independent. 
A single environmental spin will produce Poincar{\'e} recurrence times short enough to be observed in the simulations of the reduced system dynamics. 
The standard way to address these effects is the introduction of phenomenological decay terms $\exp [-\Gamma(t)]$ in the model Hamiltonian~\cite{Granade:2012kj, Santagati2019}, to reproduce the observed decay. 
This procedure is not performed in the current work, as it would invalidate the attempt to learn the open system dynamics as an interaction with environmental qubits, preserving a time-independent model $\hat{H}$.

\end{document}